# Infrared photonics for healthcare: A roadmap for proactive and predictive health management

**Borislav Hinkov[1,*,+], Johannes Kunsch[2,*,#], Werner Mäntele[3,4,*,-], Lukasz Sterczewski[5,*,&], Ángel Sánchez-Illana[6], Jaume Béjar-Grimalt[6], Víctor Navarro-Esteve[6], David Perez-Guaita[6], Alexander Mittelstädt[4], Philippa Clark[4], Valentino Lepro[4], Sergius Janik[4], Thorsten Lubinski[4], Luis Felipe das Chagas e Silva de Carvalho[7], Hugh James Byrne[8], Filiz Korkmaz[9], Michael Kaluza[4], Mattia Saita[4], Lars Melchior[4], Alicja Dabrowska[10,11], Georg Ramer[10,11], Bernhard Lendl[10], Nathalie Woitzik[12,13,14], Klaus Gerwert[12,13,14], Peter Gardner[15,16], Hugues Tariel[17], Olivier Sire[18], Margaux Petay[10], Elisabeth Holub[10], Markus Brandstetter[19,20], Kristina Duswald[19,20], Verena Karl[19], Florian Meirer[21], Lukas Kenner[20,22,23,24], Gabriela Flores Rangel[25], Boris Mizaikoff[25], Mohamed Sy[26], Aamir Farooq[26], Liudmila Voronina[27,28], Marinus Huber[27,28], Tarek Eissa[27,28], Katharina Dietmann[27,28], Lorenzo Gatto[28], Mihaela Žigman[27,28], Joseph Rebel[27], Frank Fleischmann[27,28], Jakub Mnich[5], Jarosław Sotor[5], Bassam Saadany[29], Matthias Budden[30], Thomas Gebert[30], Marco Schossig[31], Shankar Baliga[32], Timothy Olsen[33], Christopher Harrower[33], Ivan Zorin[19], Chiara Lindner[34], Shigeki Takeuchi[35], Sven Ramelow[36,37], Paul Gattinger[19], David Stark[38], Réka-Eszter Vass[38], Killian Keller[38], Alessio Cargioli[38], Mattias Beck[38], Jérôme Faist[38], Robert Weih[39], Josephine Nauschütz[39], Julian Scheuermann[39], Jordan Fordyce[39], Johannes Koeth[39], Ka Fai Mak[40], Alexander Weigel[41,27,28], Ryszard Piramidowicz[42,43,44], Stanisław Stopiński[42,43,44], Mircea Guina[45], Jukka Viheriälä[45], Felix Jaeschke[1,45], Polina Fomina[25], Alexander Novikov[46], Viacheslav Artyushenko[46,47], Ivan Sinev[48], Nikita Glebov[48], Berkay Dagli[48], Hatice Altug[48]**

[1] Silicon Austria Labs, Villach, Austria
[2] Laser Components, Olching, Germany
[3] Uni Frankfurt, Frankfurt, Germany
[4] DiaMonTech AG, Berlin, Germany
[5] Wrocław University of Science and Technology, Wrocław, Poland
[6] Department of Analytical Chemistry, Universitat de València, Burjassot, Spain
[7] Departamento de Odontologia, Universidade de Taubaté, Brazil
[8] Physical to Life Sciences Research Hub, FOCAS, Technological University Dublin, Dublin, Ireland
[9] Biophysics Laboratory, Faculty of Engineering, Atilim University, Ankara, Turkey
[10] Institute of Chemical Technologies and Analytics, TU Wien, Vienna, Austria
[11] Christian Doppler Laboratory for Advanced Mid-Infrared Laser Spectroscopy in (Bio-)process Analytics, TU Wien, Vienna, Austria
[12] Chair of Biophysics, Ruhr-University Bochum, Bochum, Germany
[13] Center for Protein Diagnostic, Ruhr-University Bochum, Bochum, Germany
[14] betaSENSE GmbH, Bochum, Germany
[15] Photon Science Institute, University of Manchester, Oxford Road, Manchester, UK
[16] Department of Chemical Engineering, School of Engineering, University of Manchester, Oxford Road, Manchester, UK
[17] Diafir, Rennes, France
[18] IRDL, UBS, Vannes, France
[19] Research Center for Non-Destructive Testing, Linz, Austria
[20] Center for Biomarker Research in Medicine, Graz, Austria
[21] Institute for Sustainable and Circular Chemistry, Utrecht University, Utrecht, Netherlands
[22] Clinical Institute of Pathology, Medical University of Vienna, Vienna, Austria
[23] Comprehensive Cancer Center, Medical University Vienna, Vienna, Austria
[24] Department of Molecular Biology, Umeå University, Umeå, Sweden
[25] Institute of Analytical and Bioanalytical Chemistry, Ulm University, Ulm, Germany
[26] King Abdullah University of Science and Technology (KAUST), Thuwal 23955-6900, Saudi Arabia
[27] Ludwig-Maximilians-Universität München (LMU), Garching, Germany
[28] Max Planck Institute of Quantum Optics (MPQ), Garching, Germany
[29] Si-Ware Systems, Cairo, Egypt & Paris, France
[30] WiredSense GmbH, Hamburg, Germany

[31] Infrasolid GmbH, Dresden, Germany
[32] Laser Components Detector Group, Chandler, USA
[33] Omega Optical LLC, Delta Campus, 21 Omega Drive, Brattleboro, USA
[34] Fraunhofer Institute for Physical Measurement Techniques IPM, Freiburg, Germany
[35] Department of Electronic Science and Engineering, Kyoto University, Kyotodaigakukatsura, Nishikyo-ku, Kyoto 615-8510, Japan
[36] Ferdinand-Braun-Institut (FBH), Gustav-Kirchhoff-Str. 4, 12489 Berlin, Germany
[37] Humboldt-Universität zu Berlin, Institut für Physik, Newtonstraße 15, 12489 Berlin, Germany
[38] Department of Physics, Institute for Quantum Electronics, ETH Zurich, 8093, Zurich
[39] nanoplus Advanced Photonics Gerbrunn GmbH, Oberer Kirschberg 4, 97218, Gerbrunn, Germany
[40] School of Optical and Electronic Information, Huazhong University of Science and Technology, Wuhan, China
[41] Center for Molecular Fingerprinting, Budapest, Hungary
[42] Warsaw University of Technology, Institute of Microelectronics and Optoelectronics, Warsaw, Poland
[43] VIGO Photonics, Ożarów Mazowiecki, Poland
[44] LightHouse, Lublin, Poland
[45] Optoelectronics Research Centre, Tampere University, Tampere, Finland
[46] art photonics GmbH, Rudower Chaussee 46, 12489 Berlin, Germany.
[47] Art Fiber Systems, Lisa-Meitner-Str. 9, 89081 Ulm, Germany
[48] Institute of Bioengineering, EPFL, Lausanne, Switzerland

*these authors contributed equally
E-mail: [+]borislav.hinkov@silicon-austria.com , [#]j.kunsch@lasercomponents.com ,
[-]maentele@biophysik.uni-frankfurt.de , [&]lukasz.sterczewski@pwr.edu.pl

**Keywords:** infrared, health monitoring, medical diagnostics and therapy, predictive healthcare, theranostics

**Abstract**: The field of infrared (IR) photonics is currently undergoing remarkable progress, moving rapidly towards practical sensing applications demanded by medical therapy and diagnostics (theranostics). The Developments can be divided into three main categories: (i) novel devices and measurement concepts including advanced updates of classical approaches that push medical sensing into the spotlight; (ii) new demonstrations of photonic integrated circuit (PIC-)based IR devices enabling highly miniaturized sensors for point-of-care application as well as medical and wellness wearables; and (iii) technologically-mature IR demonstrators that enable first medical sensing and treatment applications.

This roadmap paper provides a consolidated overview of this highly dynamic and interdisciplinary research field with a focus on the major roadblocks that limit the widespread adoption of IR photonics in large-scale medical diagnostics. Special attention is given to the ambivalence between the molecular-level spectroscopic interpretation and a broader health-state assessment, highlighting the need for a common framework. Additionally, the paper discusses the critical importance of unified measurement standards, calibration protocols, and medical certification processes to ensure the validity of experimental results, reproducibility, and clinical trust, particularly when novel experimental techniques and AI algorithms are involved. Perspectives from major past and current contributors to application-oriented IR photonics will be provided.

Beyond current demonstrators, the roadmap emphasizes the need for energy-efficient sources and detectors, hybrid integration with electronics and lab-on-chip platforms, and the establishment of robust ethical and regulatory frameworks. These advances are essential to unlock population-scale screening, continuous monitoring through wearables, and the integration of explainable artificial intelligence. Together, these developments may accelerate the translation of IR photonics into proactive and predictive healthcare.

**Core Team (coordination, alphabetical)**:

- Borislav Hinkov (Silicon Austria Labs, Austria),
- Johannes Kunsch (Laser Components, Germany),
- Werner Mäntele (Goethe University Frankfurt & Diamontech AG, Germany)
- Lukasz Sterczewski (Wroclaw University of Science and Technology, Poland),

**Steering Advisory Team (alphabetical)**:

- Bernhard Lendl (Technische Universität Wien, Austria),

- Boris Mizaikoff (Ulm University & Hahn-Schickard Soc. for appl. Research, Germany)
- Mihaela Žigman (Ludwig-Maximilians-Universität München & Max Planck Institute of Quantum Optics, Garching, Germany)

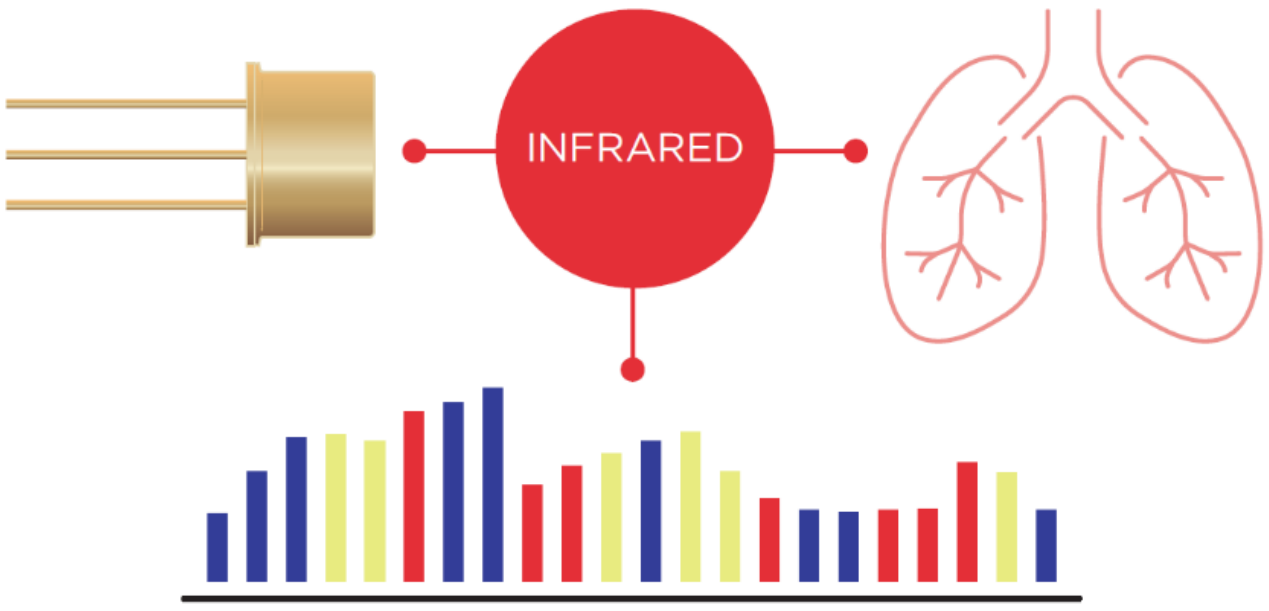

# Table of contents

## Coherent Light Sources

## Integrated photonics

Coherent Light Sources:

Integrated Photonics:

# 1. Introduction: Rationale for the IR Roadmap

**Borislav Hinkov[1,*,+], Johannes Kunsch[2,*,#], Werner Mäntele[3,4,*,-], Lukasz Sterczewski[5,*,&]**

[1] Silicon Austria Labs, Villach, Austria
[2] Laser Components, Olching, Germany
[3] Uni Frankfurt, Frankfurt, Germany
[4] Diamontech, Berlin, Germany
[5] Wrocław University of Science and Technology, Wrocław, Poland

*these authors contributed equally
E-mail: [+]borislav.hinkov@silicon-austria.com , [#]j.kunsch@lasercomponents.com ,
[-]maentele@biophysik.uni-frankfurt.de , [&]lukasz.sterczewski@pwr.edu.pl

### Status

Over the past decades, the global population has changed drastically. Life expectancy has increased significantly, while fertility rates have declined, and the population becomes clustered. This trend is most prominent in industrialized countries; however, it can be expected to be very similar in all current and future developing communities. Overall, this leads to an aging population in combination with a decrease in regional coverage [1], requiring a shift from traditional medical routines and transforming our ability for proactive and predictive health management through new strategies for our health systems. They need to include new health monitoring techniques, the implementation of digital technologies and diagnostics, and therapy through remote medicine for maintaining and improving healthcare throughout all age cohorts. Prevention, regular health screening, and early intervention are key measures for "healthy ageing". These, however, are currently burdens in our stressed healthcare systems. Transformation thus also requires new, high-technology, digitized, networked, and cost-efficient methods that cannot only be implemented in clinics, at doctors' offices, and used by experts, but also in health stations for remote diagnosis and therapy, where they can be operated by nurses and laypeople.

This Roadmap-Paper outlines and discusses in detail the potential of (mid-)infrared (MIR) spectroscopy in this transformational process. The status of medical applications is presented by a multitude of examples in the first part of the paper, followed by recent important technological progress in the field of IR technologies in the second part.

To provide context, it is useful to revisit the origins and pivotal developments in infrared science, showing how foundational discoveries inform today's innovations. The ambivalence—both promise and challenge—of IR in healthcare dates to the very discovery of infrared radiation on 11 February 1800 by Sir William Herschel. While developing filters to protect the eye during solar observations, Herschel, then Royal Astronomer, detected invisible radiation beyond the red end of the visible spectrum (the "Ultra-Red") [2], revealing the existence of IR light and underscoring its hidden potential.

Subsequent milestones continued to shape the field's evolution. In 1874, Ferdinand Braun constructed the first bulk semiconductor diode using naturally occurring lead sulfide (PbS), which is a well-known IR material [3]. Building on these advances, Jagadis Chandra Bose in 1901 patented a method for detecting radio signals with semiconductor diodes [4], demonstrating the versatility of such materials. However, the advent of vacuum tubes, which offered greater reliability, soon replaced PbS outside of photonic applications prior to its use as an IR detector [5]. Despite such shifts, IR technologies continued to progress. Notably, the first semiconductor diode lasers based on lead salts were developed in the 1960s [6], preceding even their visible-light counterparts.

It is also important to recall that Max Planck's revolutionary introduction of quantum theory in 1900 emerged from his efforts to explain black body radiation—a phenomenon describing how objects emit electromagnetic energy, primarily in the infrared region, depending on their temperature [7]. This connection between IR science and fundamental physics further illustrates the enduring significance of IR research.

By tracing these historical threads and clarifying technical concepts, this roadmap aims to bridge the gap between past discoveries and their application in today's healthcare landscape. Our objective is to guide the integration of advanced IR spectroscopy methods into medical practice, thereby fostering improved diagnostics, personalized treatment, and more efficient healthcare delivery for a healthier future.

Infrared (IR) spectroscopy has a long-standing history in molecular analysis: For about 100 years, chemists have routinely used IR spectra to identify substances or products of a synthesis. However, although applications of IR spectroscopy gradually moved into fields like biochemistry and biology over the past 50 years, medical applications are still scarce. They are mostly on an academic level but not implemented in everyday healthcare routines. The rationale of this roadmap is to identify the bottlenecks, hurdles, chances, and risks for the implementation of IR spectroscopy in healthcare based on examples from the application side as well as from the technological side.

Historically, there has been a notable disconnect between medical applications and mainstream infrared photonics, even though biomedical infrared (IR) spectroscopy has generated a substantial body of research. This divide is evidenced by the absence of biological or medical papers in a dedicated 2015 special issue on IR lasers and their applications, where none of the twenty-one articles addressed biomedical topics [8]. Such gaps highlight the need for improved interdisciplinary communication and representation. A central objective of this roadmap is to bridge these fields, foster collaboration, and integrate diverse perspectives to address today's complex biomedical challenges.

These days, there is a new dynamic and a new quality in the community. This became especially clear at the Plenary Industry Sessions of the Optica Sensing Congress 2024 (15–19 July, Toulouse, France), where more than two-thirds of the program focused on infrared and healthcare. These efforts have continued since, with this roadmap aiming to transform healthcare as a first result.

Other current developments demonstrating the increasing dynamics in the field of IR include the recent submission of another complementary IR roadmap paper to the journal *Advances in Optics and Photonics* by J. R. Meyer et al., entitled "Midinfrared Semiconductor Photonics" [9]. Their paper focuses specifically on infrared semiconductor devices and technology. Another important milestone in the field was achieved with the launch of the "protecting.health Global Initiative" by Nobel laureate Ferenc Krausz in Hong Kong on 07/11/2025, "*with the aim to jointly create the basis for transforming healthcare: toward personalized, precise prevention*" (of noncommunicable diseases, i.e., such that do not spread from person-to-person) [10].

Besides those important initiatives, it is important to select terminology that resonates across disciplines and with stakeholders for broader scientific communication and outreach. While 'Infrared Fingerprinting' is well-established within the scientific community, its association with forensic science may hinder wider acceptance. 'Infrared omics' offers a more inclusive and forward-looking descriptor for global adoption, though it may require further explanation for non-specialists. In contrast, the concept of the 'spectroscopic twin'—a digital representation derived from IR spectral data—aligns closely with the emerging 'digital twin' paradigm in healthcare. Here, a spectroscopic twin serves as a virtual model of a patient's biochemical and physiological state, supporting personalized diagnosis, monitoring, and treatment planning. This terminology connects biomedical IR technologies with broader trends in digital healthcare innovation, facilitating understanding among policymakers and investors. This roadmap seeks to unify the language and vision of biomedical infrared research. It aims to overcome historical barriers, encourage interdisciplinary dialogue, and stimulate innovation by providing a common framework for collaboration. Ultimately, these efforts will accelerate the development and adoption of infrared-based technologies in healthcare, leading to improved patient outcomes and more efficient medical practices.

## Infrared Basics

IR spectroscopy represents vibronic transitions, e.g., changes in bond lengths and angles, and covers a wide range of the electromagnetic spectrum. It relies on the fact that the dipole moment of a bond varies in strength (and direction) with changes in bond length and angle. Homonuclear molecules (such as $N_2$, $O_2$) are thus not IR active, as are atomic ions. The near-infrared range (NIR), approximately 780 nm to 2500 nm, reflects "overtones" of vibrations ("harmonics") that originate from the non-harmonic potential wells. Overtones in the NIR are strongly overlapping in a relatively narrow spectral region. They are at least 10-100 times weaker than the fundamental vibrations and, in addition, depend on temperature due to the occupancy of higher vibrational levels. For the quantitative determination of analytes in a complex matrix like a biofluid, the MIR appears to be superior to the NIR spectral range [11].

The MIR, approximately 2.5 to 20 μm, reflects transitions of almost all molecules relevant in biological systems and thus will be in the focus of this roadmap. In this region, a clear "band-to-bond" assignment is possible in many cases, at least for small molecules: it is thus common to use the term "localized vibrations". For historical reasons, the wavenumber scale (parallel to the wavelength scale) is frequently used. The far-infrared range (FIR), approx. 20–1000 μm, is shared with the THz community and rather reflects collective than localized modes. It appears to be less suitable for diagnostic purposes, at least in the medical field.

Fundamentally, MIR spectroscopy exhibits an extremely high specificity for molecules, their conformation, molecular interactions, and their matrix, seemingly making it an ideal tool for biomedical analysis. While this specificity is demonstrated by quantum chemical calculations for small molecules, the same theoretical approach generally fails for large molecules, e.g., biopolymers at molecular mass above 10 000 kDa. Hybrid methods combining quantum chemistry for active centers of proteins and molecular dynamics calculations for the rest of the molecules have shown promising results in the calculation of IR spectra [12]. However, this theoretical approach generally fails for structures like cells and tissues, where IR spectroscopy has accumulated a wealth of experimental evidence [13].

In contrast to the high specificity, the sensitivity of MIR is rather low. Transition moments for vibrational levels are typically 10–100 times weaker than for electronic levels, thus often preventing MIR from the analysis of ppb levels and below (unless procedures for enrichment or specific spectroscopic techniques are employed). However, the specificity for concentrations down to ppm outweighs this deficit.

A common representation of the IR signatures of molecules is a look-up table summarizing the vibrational modes or a spectrum indicating which bond can be found at a certain spectral position. Again, this may work for small molecules and has been used for decades by chemists for identification. However, in this paper, we prefer the alternative picture that complex molecules should not be spectroscopically identified by their unique absorption bands, but rather by their full "spectroscopic DNA" – a vector of all absorption band coordinates. Nevertheless, it can be helpful to indicate spectral regions of relevance as a first guideline for certain applications. (Fig. 1)

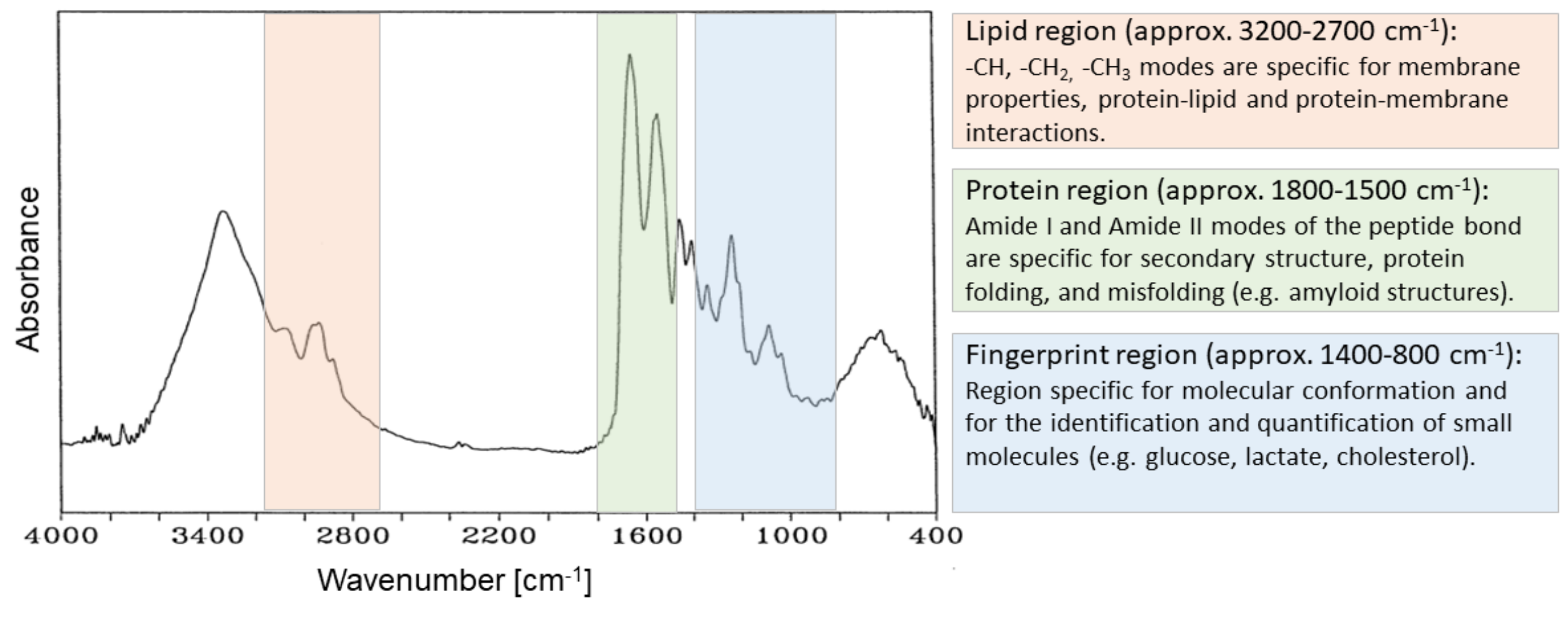

**Figure 1.** IR spectrum originating from a protein in a biological membrane. The entire spectral range can be used for analysis. The shaded spectral regions are characteristic for lipids forming a biological membrane (red), for proteins revealing conformation and interactions (green), and for biologically and medically relevant small molecules (blue).

### Current and future technical challenges

A first technical challenge for "medical IR spectroscopy" is the choice of the light source. All spectroscopic techniques require photons in a well-defined energy range. For IR spectroscopy, photons with energies ($E$) in the 50–400 meV range are required, which can originate in the simplest case from thermal sources or more advanced from coherent laser sources. The energy bandwidth ($\Delta E$) (resolution) required is not as critical as for IR spectroscopy of gases with sharp lines since molecules of interest in "medical IR spectroscopy" typically exhibit broad bands due to hydrogen-bonding and matrix heterogeneity, with half-widths of 5–10 cm$^{-1}$ for liquid and solid samples. Another criterion is the flow density of available photons (the power of the measuring light): Infrared spectrometers using thermal sources in combination with filters or grating monochromators deliver at most several 100 nW. This is sufficient for many applications and has the advantage of using non-coherent and non-polarized light, with low noise and high long-term stability. Other applications require much higher photon flows in the mW range that can only be delivered by lasers. The aspect of IR sources for medical applications will be detailed in the second part of this roadmap paper, with a focus on technologies.

A second challenge for medical IR spectroscopy is the efficient coupling of the measuring beam to the sample, i.e., the "sample interface". These can be conventional transmission cells in the case of *ex-vivo* applications (e.g., for the analysis of body fluids), attenuated total reflection (ATR) technologies for *ex-vivo* and for skin *in vivo* analysis, reflection technologies (e.g. for skin analysis), or direct measurement of the absorption process in tissue by photoacoustic or photothermal methods. Beyond conventional optics, fiber optics can be used as convenient sample access, e.g., for *in vivo* studies, or photonic integrated circuits (PICs) can serve as miniaturized devices for *ex vivo* and, in the future, potentially *in vivo* studies of body fluids. Each medical IR spectroscopic application must find a suitable sample interface for high reproducibility and convenient use (e.g., in the case of *in vivo* applications).

A third challenge is the detection of the photons after interaction with the sample. The technologies for the detection of IR photons have advanced much in the past decades. The historical IR detectors, so-called thermopiles that detect the heat generated by absorption of IR photons, have greatly improved and have been complemented by pyroelectric detectors that generate charge separation because of IR absorption and heat generation. The latter detectors, mostly working at room temperature, can be mass-fabricated and are standard in IR spectrometers. While these are typically termed "thermal" detectors because the initial effect used for detection is heating, "quantum detectors" made from various materials use the absorption of IR photons for the generation of charge carriers ("photoconductive") or charge separation ("photovoltaic" detectors or "quantum detectors"). Those quantum detectors may need to be cooled down to cryogenic temperatures and rely on the manifold of available semiconductor materials that allow covering the entire spectral region of interest (approx. 2–20 µm). General criteria for the selection of IR detectors are wavelength response, sensitivity, detectivity $D^*$ (a measure for detection calculated from sensitivity and noise level), response time, linearity and homogeneity, but also the operating temperature and, last but not least, the price [14]. The aspect of IR detection for medical applications will be detailed in the section on technologies.

In summary, key components like uncooled detectors and lasers are now made from rugged, well-understood materials and give a new push to classical instruments like Fourier transform infrared (FTIR) spectrometers. (M)IR Photonic Integrated Circuits (PICs), which are compact, chip-scale devices that integrate multiple, active and passive, photonic elements and functions on a single chip, are currently being developed for potential wearable applications, promising to unlock the full potential of MIR technology for everyday and personal healthcare applications. Scalability is another issue that has been learned through historic lessons in the community during the development of the technological basis, and that is addressed by highly scalable PIC approaches.

### Challenges for the integration into healthcare practice

For good reasons, healthcare practice is based on routine procedures that are well-established and have undergone numerous and continuous tests and critical evaluations. No patient would like to learn shortly before a critical surgical intervention that a novel, previously not used technique will be applied for the first time. Implementing IR spectroscopy into the medical routine thus requires thoughtful considerations before technological advances can be harnessed to improve patient well-being and reduce healthcare costs. Only a fraction of the IR applications for healthcare reported in this roadmap have been validated in clinical settings. In some applications, for the first time, infrared and medical data from large cohorts have been correlated, leading to reliable, robust, and reproducible exploratory results.

As the following step, IR spectroscopic procedures intended for healthcare use and the instruments they are based on require official clearance by authorities in the EU or by the FDA in the US. This often is a time-, labor-, and costly process, also depending on the risk classes (I–III) that define the potential risk for the patient. It is strictly recommended that this step be considered already in the early phase of the development, typically when device prototypes are developed based on the experience with lab setups. The example of X-rays, where the first medical application was reported only a few weeks after Röntgen discovered the "X-rays" on November 8, 1895, in Würzburg, Germany, can serve as a warning.

With rare exceptions, the focus of MDs, fortunately, is on diagnosis and therapy, while their background in physics and spectroscopy is typically rather basic. The more important thing is the integration of clinicians and general practitioners with their views and needs into the development of devices and their validation. Another group to integrate is healthcare managers and health insurance companies (the "payers"). If these groups can be convinced that IR-based methods in healthcare are not just a "nice-to-have" but a real "added benefit", the implementation can become a success.

## Concluding remarks

This roadmap demonstrates, through many examples for applications, that IR technologies have reached a level of maturity that calls for integration into healthcare. IR technologies can provide an added value for routine settings (such as body fluid analysis: "lab-on-the-desk"), for specialized diagnostics applications (like early Alzheimer's diagnosis), or for non-invasive diagnostics (such as breath analysis or blood glucose measurement), to name just a few examples (Fig. 2). Moreover, IR technologies can be used for large-scale healthcare screening and phenotyping (such as determining the risk for cancer) by creating a "spectroscopic twin" of the patient that potentially enables precise personalized prevention of noncommunicable diseases.

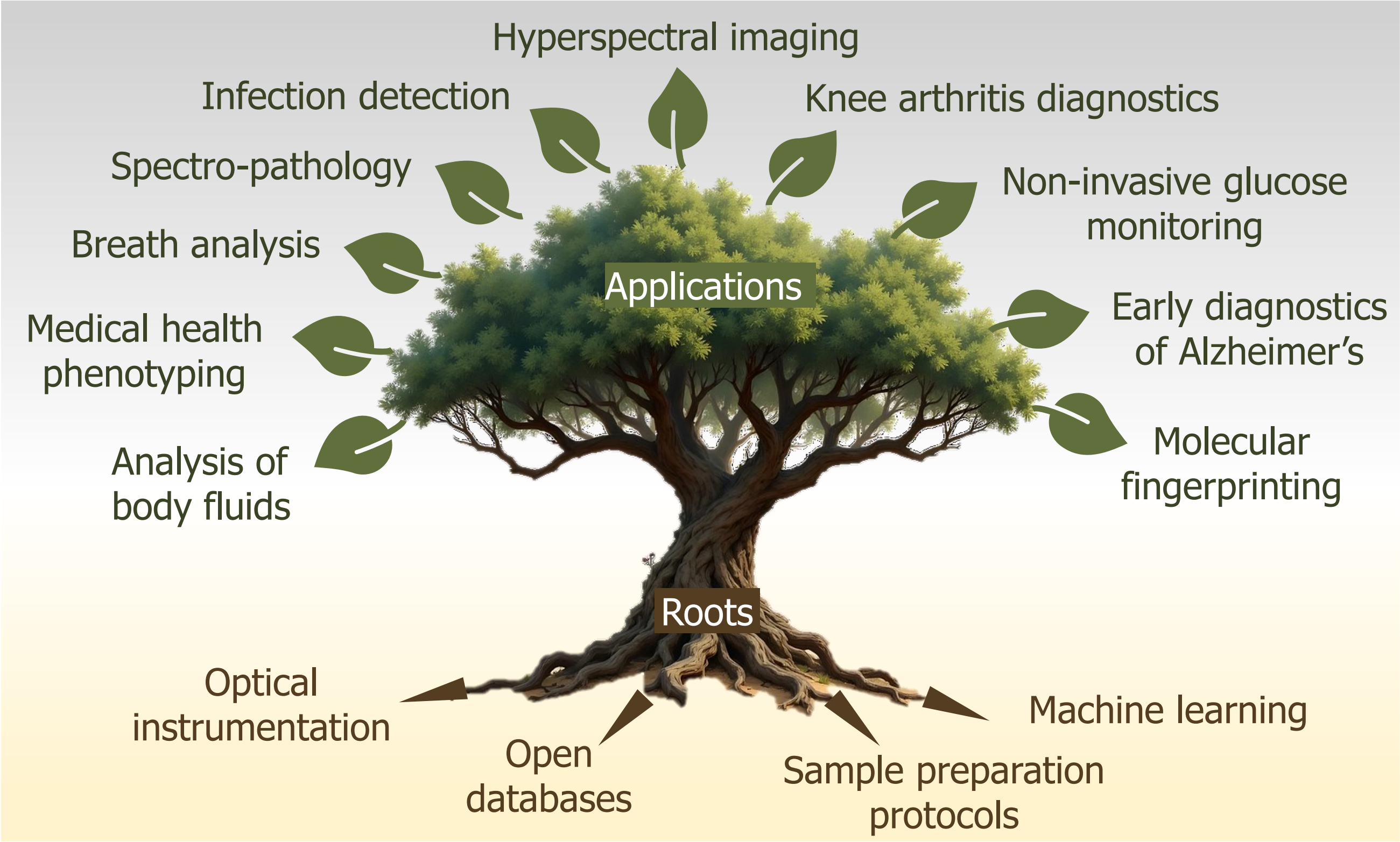

**Figure 2.** This tree symbolizes the interplay of technological developments, information technology, and IR applications in healthcare.

The benefits of these technological advancements and initiatives can be seen as follows:

- Enabling earlier detection of health deviations, leading to more healthy years of life for those affected
- Providing non- or low-invasive diagnostics for certain diseases that will improve patient compliance in prevention and treatment
- Enabling reagent-free diagnostics, leading to economic benefits for society
- Personalized precision prevention initiatives that may help reduce the burden on healthcare systems by enabling targeted interventions and improving efficiency.
- Strengthening economic power through an extended active lifespan as individuals remain healthier and productive for a longer time
- Improving medical service for existing and upcoming "medical deserts" (i.e., regions of low or no healthcare coverage – term adapted from French "*déserts médicaux*" [1]) by deploying portable and wearable infrared technologies, allowing remote diagnosis and monitoring, even by non-experts.
- Addressing the problem of existing and upcoming shortage of general practitioners through point-of-care devices, which facilitate rapid assessments and reduce the frequency of specialist appointments.
- Reduction in medical complexity by utilizing technologies that combine multiple diagnostic capabilities in a single device, supporting a "one catches many" point-of-care philosophy.

## Acknowledgements

The authors would like to thank Dr. Mihaela Žigman (Ludwig-Maximilians-Universität München & Max Planck Institute of Quantum Optics, Garching, Germany) for constructive feedback and comments.

# 2. Introduction to medical applications of infrared spectroscopy

**Werner Mäntele[1,2,*], and Johannes Kunsch[3,#]**

[1] DiaMonTech AG, Boxhagener Strasse 82A, 10245 Berlin, Germany
[2] Institut für Biophysik, Goethe-Universität, Max-von-Laue-Strasse 1, 60428 Frankfurt/Main, Germany
[3] Laser Components, Olching, Germany

E-mail: *maentele@biophysik.uni-frankfurt.de ; #j.kunsch@lasercomponents.com

## Status

Infrared (IR) spectroscopy as an analytical tool has more than 100 years of history. Based on the knowledge that IR radiation is somehow specific for materials, first IR spectrophotometers were developed and became standard equipment in analytical chemistry. These dispersive instruments (i.e. equipped with filters, with a grating or a prism for wavelength selection) were notoriously insensitive and slow, yet sufficient to identify materials. Sample preparation for dry substances was routine in KBr pellets or nujol mulls and in thin-layer cuvettes for organic liquids. No one expected information from aqueous solutions, not to mention from biological materials. IR was fully in the hands of organic and inorganic chemists: the "Chemistry-Era" of IR spectroscopy [1].

The situation changed when Fourier transform infrared (FTIR) spectrometers for the mid-IR (MIR) were developed around 50 years ago (coming from the far-IR where requirements for mechanical precision of the interferometer were lower). The multiplex advantage of the interferometric technique reduced acquisition time for spectra more than 10–100-fold (and thus allowed averaging to increase sensitivity). Gradually, scientists started to dare the analysis of biopolymers (e.g. proteins) and biological membranes with ample water present. Still, the low spectral emission of the thermal sources common for dispersive and FTIR instruments was considered a limiting factor for some applications. We can nevertheless speak of a "Biochemistry-Era" of IR spectroscopy that slowly started in the 70s and 80s of the last century.

Attempts to develop laser techniques for more powerful MIR sources were manifold, such as the $CO_2$ or the CO laser, but with a limited spectral range. Early tunable semiconductor lasers, such as the lead salt laser, turned out to be a dead end because of the unreliable material system: too complicated to use at cryogenic temperatures. The situation changed completely when the first Quantum Cascade Lasers (QCL) were developed in the 90s. More on this in the "Technology Part": relevant for certain applications are output powers of milliwatts and higher, tunability, and beam quality. Finally available: an IR light source for biomedical applications that are "hungry for photons". A low but steadily increasing number of scientific papers on IR applications in medicine allows us to speak about an upcoming "Medicine Era" of IR spectroscopy and start the implementation of IR spectroscopy in healthcare [2, 3].

## The way we look at infrared spectra

An interesting aspect of IR spectroscopy, making its way from chemistry through biochemistry to medicine, is the way spectra are analyzed. In the "Chemistry-Era", spectra were analyzed by analogy to reference spectra collected in handbooks or in an "IR Atlas": Absorption bands present/absent at certain wavelengths, band intensities were "very strong", "strong", "medium" or "weak" and had or had no shoulders. Let us label this approach as "phenomenological 1.0". Quantitative analysis of spectra was almost unknown. Among other reasons, this had to do with sampling techniques: Concentrations of samples in a KBr pellet or a nujol mull (nujol is a mineral oil with low IR absorbance) are hard to determine. Sample cells with defined optical path length and sampling methods for complex biological materials had to be developed, allowing semi-quantitative and quantitative *in vitro* and *in vivo* studies. Today, mostly transmission cells with path lengths in the 10–50 µm range (also as flow cells) or attenuated total reflection (ATR) cells are standard. This forms the basis for quantitative analysis of, e.g. body fluids reported here as an example for applications.

With substantial input from theoretical chemists, a more thorough understanding of quantum chemistry, molecular mechanics, force fields, and bond strengths has led to a more detailed understanding of IR spectra. Band positions and amplitudes can be calculated even for biopolymers. Let us label this approach to the understanding of IR spectra as "molecular mechanical".

With the development of sophisticated data evaluation procedures and computational power, a "phenomenological 2.0" analysis of IR spectra became possible, which somehow resembles the classical analysis of spectra in the "chemistry era", except being much more sensitive and specific. It does not rely on the "molecular mechanical" view, but on fingerprints of molecules that manifest in the spectra, even with weakest contributions. Spectra of complex biomedical samples such as body fluids, membranes, cells, and tissue, *ex vivo* or *in vivo*, can be analyzed with chemometric methods, neural network analysis, artificial intelligence, and others, as shown in Fig. 1 [4, 5]

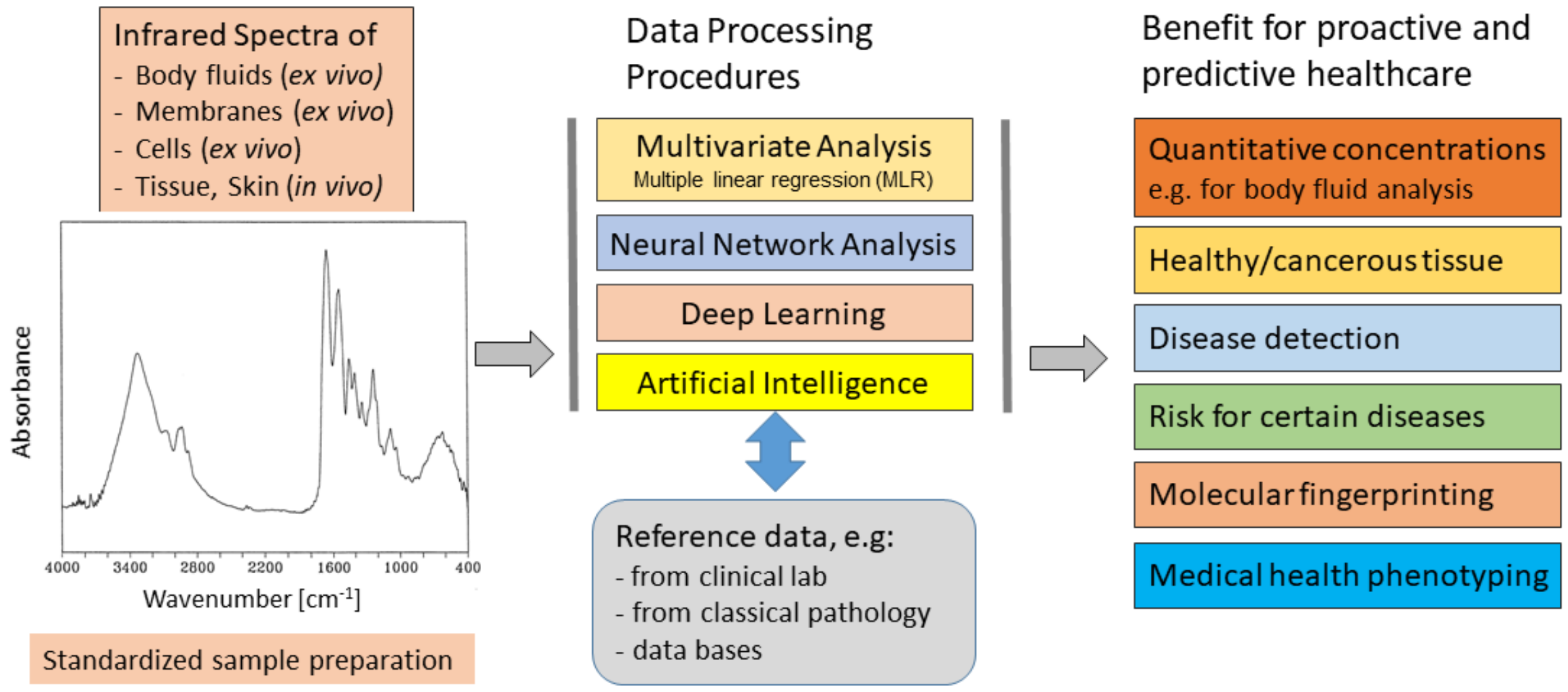

**Figure 1:** Flowchart of "phenomenological 2.0" applications also called "Infrared omics" or "spectroscopic twin".

This approach is meanwhile established and requires standardized sample preparations, well-defined IR instrument conditions, and a solid data processing strategy with robust calibration and validation procedures. Reference data may come from the clinical laboratory, from pathologists, or from conventional diagnostic methods. Examples are provided in this application section of the roadmap for the analysis of body fluids, tissues and skin.

With the advent of infrared microscopes, either coupled to an FTIR instrument or to QCL sources, spatial information is available in addition to molecular information. This represents a valuable tool, e.g. in the diagnostics of cancerous tissue and the decision between healthy/cancerous for surgical removal [6]. For this specific IR technology, the name "spectropathology" has been coined.

### Advances in science and technology to meet challenges

Medical doctors are generally not interested in recording IR spectra but prefer dedicated instruments that directly provide relevant health information. Footprints of IR-based instruments need to be small (e.g. for the operating room or the doctor's practice). Compact FTIR instruments with thermal sources will be one option for spectroscopy, QCL-based instruments; another option for applications requiring higher IR power (e.g., for the non-invasive glucose analysis reported in this section). Some applications aim to handheld systems that would also require laser-based IR sources. Affordability is a relevant issue in all health systems. High-end FTIR or laser instruments are difficult to implement in healthcare, unless they offer unique and unprecedented information. It is clear that further users – MDs as well as nurses and health assistants – must be involved at an early stage whenever a new IR technology is to be implemented. It will also be necessary to include health insurances – the payers – at an early stage and to convince them about the added value as compared to established techniques.

A relevant barrier in the implementation of IR technologies in healthcare is clearance as a medical device, i.e., by CE or FDA approval. Only a few developments in the field of biomedical IR spectroscopy have managed to pass this costly and manpower-consuming barrier, although they could provide clinical validation (a must for the clearance). Any instrument that provides diagnostic or therapeutic (theragnostic) information is, in principle, a medical device. It is recommended to be aware of this hurdle and prepare for it even at the early stage of development.

### Concluding remarks

In summary, IR spectroscopy has undergone multiple metamorphoses since chemists established it for material identification many decades ago. These metamorphoses have been possible with considerable input from physicists, engineers, as well as data and computer scientists. Many pilot studies have demonstrated that infrared spectroscopy can provide valuable information in proactive and predictive healthcare. It is time that "**Infrared goes to clinics and point of care**".

### Acknowledgements

We have to thank many colleagues – among them infrared experts as well as medical doctors – for numerous discussions on the pros and cons of biomedical IR spectroscopy, whether encouraging or discouraging. When I started my IR career 50 years ago by recording a protein spectrum in aqueous solution on a dispersive instrument, very few people would believe that this was of any use (WM).

# Molecules:
# 3. Re-thinking clinical IR: How measurement mode and sample preparation shape diagnostic performance

**Ángel Sánchez-Illana[1,*], Jaume Béjar-Grimalt[1], Víctor Navarro-Esteve[1], and David Perez-Guaita[1,#]**

[1] Department of Analytical Chemistry, Universitat de València, Burjassot, Spain
E-mail: *angel.illana@uv.es, #david.perez-guaita@uv.es

### Status

Mid-infrared (MIR) spectroscopy entered clinical research well before the advent of Fourier Transform IR (FTIR) instrumentation. Between the late 1940s and 1950s, seminal studies demonstrated MIR analysis of tissue and blood on IR-transparent substrates and even direct measurements of complex biofluids using 60 µm liquid cells [1]. These early approaches differed markedly in measurement geometry and sample preparation, yet all pursued the same goal: extracting relevant chemical information from real clinical specimens and minimal or inexistent sample processing.

Today, clinical MIR studies coexist across multiple measurement strategies, including transmission measurements of liquids, and attenuated total reflectance (ATR)-based approaches using internal reflection elements (IREs) applied to both liquid samples and dried deposits [2], transflection measurements [3,4], and emerging evanescent-wave photonic platforms [5,6]. Despite this diversification, the aspiration for direct analysis or minimal sample processing remains central to most MIR-based clinical strategies.

Rather than advocating a single optimal measurement strategy, we argue that the diagnostic potential of clinical MIR spectroscopy is fundamentally shaped by the co-design of measurement setup and sample processing, aligned with the analytical demands of the clinical application. Most approaches aim to remove components of the sample matrix, thereby isolating the analytes of interest and enhancing the sensitivity and selectivity of diagnostic or quantification models (Fig. 1). The key, therefore, is to improve model performance by enhancing the quality of the information acquired. The most basic and widely used preprocessing step is the removal of water, the dominant component of liquid biopsies, by forming dry films on transmission or ATR substrates. Other strategies selectively target specific sample fractions, such as proteins, cells, or lipids, depending on the disorder under investigation.

In fact, the preprocessing of clinical biopsy samples is ubiquitous and depends on the intrinsic nature of the sample (e.g., blood, plasma, or serum). Representative examples include liquid–liquid microextraction with organic solvents, which has been employed to enhance the quantification of lipidic parameters in serum and to detect drugs in urine at the ppm level [7,8]. Protein precipitation strategies and ultrafiltration workflows have also been reported, either to selectively remove specific molecular weight fractions from serum to obtain more detailed spectra of metabolic components, or to extract proteins from urine for subsequent analysis [9,10]. Collectively, these less frequently adopted approaches highlight how modest, purpose-driven sample preparation can unlock analytical capabilities that are difficult, or even impossible, to access through direct analysis alone.

### Current and future challenges

A first challenge in integrating sample pretreatment into clinical MIR workflows stems from the unavoidable introduction of spectral artefacts during preprocessing. Preconcentration strategies (e.g., drying, filtration, or extraction) amplify not only diagnostically relevant signals but also contaminants introduced during sample handling, such as plasticizers, glycerol residues from filtration membranes, or polymeric components leached from consumables. While often negligible under direct analysis, these contributions can dominate spectra after preconcentration, compromising model robustness and transferability across devices and sites.

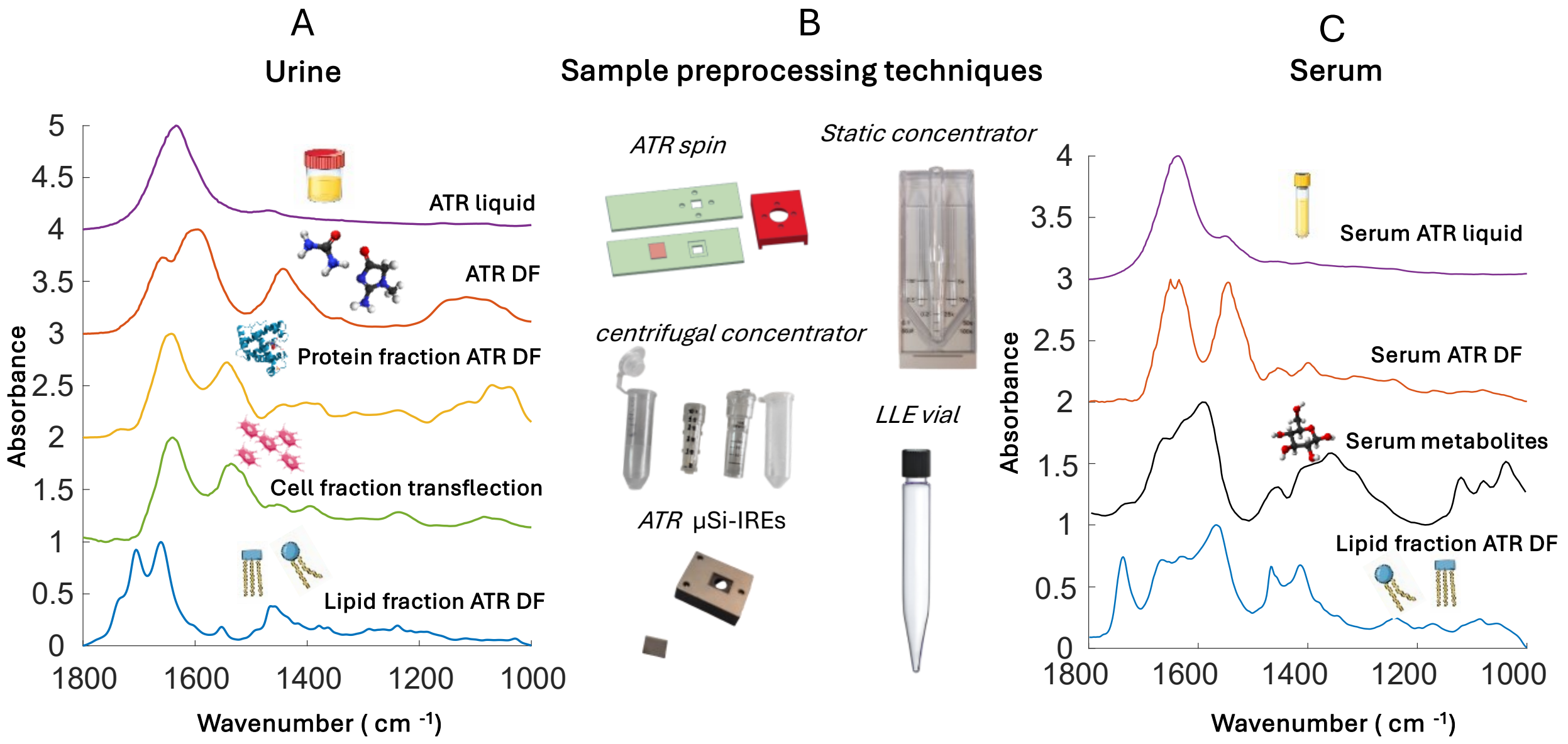

**Figure 1.** MIR Spectra of urine (A) and serum (C) can be acquired using different measurement strategies and sample-processing techniques (B), providing information on different molecules depending on the clinical application. Direct measurement of serum and urine results in spectra with elevated contributions from water (violet), whereas spectra obtained after drying (red) show clearer features of biochemical molecules. Furthermore, protein extracts obtained using centrifugal and static protein preconcentrators (yellow) clearly display spectral bands associated with urine proteins. A similar approach can be used to remove serum proteins and enhance signals from minor metabolites (black). Cell fractions can also be studied using integrated cytocentrifugation with ATR-spin devices (green), and liquid–liquid microextraction (LLE) can be used to investigate lipid fractions (blue).

Within internal reflection elements (IREs)-FTIR–based clinical workflows (e.g., ATR), this challenge is closely linked to the widespread use of dried films, which are particularly attractive due to their simplicity and their ability to boost sensitivity through elimination of water and preconcentration into the ATR crystal (limits of detection in the low mg/L range have been reported). In this configuration, microgram-level amounts of material are sufficient to generate high-quality spectra, such that even modest preconcentration factors can shift spectral composition toward analytically irrelevant species. Well-documented examples include glycerol contamination following ultrafiltration [11], plastic-derived signals after high-factor liquid–liquid extraction [7]. In parallel, dry-film formation introduces additional physicochemical effects, such as coffee-ring–driven spatial heterogeneity, which can further modulate the effective interaction between the sample and the evanescent field. Although detailed imaging studies have shown that quantitative analysis remains possible within well-defined linear regimes and controlled deposition geometries [10], these effects highlight that preconcentration in ATR dry films must be treated as a design choice rather than a neutral sensitivity enhancement. As emerging high-sensitivity IR approaches based also in dry film sampling, such as nanomechanical IR spectroscopy, push detection limits into the pg range [12], both contamination and morphology-related effects are expected to become even more limiting if not explicitly controlled, posing additional challenges for clinical translation and calibration transfers between quantification models.

More importantly, the portability, simplicity, and, ultimately, the near-patient testing capabilities of IR spectroscopy, which constitute some of its most defining and attractive features, may be compromised by the introduction of additional processing steps. The translational bottleneck is that such workflows often require additional steps (centrifugation, solvent handling, phase separation, timed incubations), increasing time-to-result and operator dependence, two factors that directly erode PoC, bedside, and IVD feasibility. The key challenge, therefore, is to develop “smart” pretreatment strategies that preserve the analytical advantages of selective fractionation while remaining simple, fast, and standardized enough for non-expert use and regulated deployment.

### Advances in science and technology to meet challenges

Several emerging advances are making MIR workflows with coupled sample pretreatment realistic in clinical settings by enabling true integration between sample handling and the optical interface. On the handling side, 3D printing, microscale concentration or filtration devices, selective chemical or biochemical platforms, and microfluidics provide controllable ways to fractionate and deliver samples; on the optical side, ATR with microstructured IREs, transflection geometries, and guided-wave photonic platforms (waveguides) offer interfaces that naturally accept such controlled deposition or flow.

As a concrete example of versatile optical interfaces designed for potential integration with sample processing, microstructured silicon IREs (μSi-IREs) have recently gained attention. They are wafer-fabricated, multiridge ATR substrates that enable affordable, potentially disposable, and system-integrable evanescent-wave measurements, with structure and performance trade-offs (e.g., ridge angle, footprint, and alignment) [13]. These IREs are therefore well matched to 'sample-to-surface' pretreatment concepts, where controlled deposition and enrichment are designed into the sampling step. Accordingly, μSi-IRE platforms have been coupled to 3D-printed sampling interfaces for direct blood serum measurements in cancer triage [14], and have also supported 'ATR-spin' 3D printed modules that integrate cytocentrifugation with direct cell deposition (from cell cultures, blood, or urine) onto the IRE [3,15].

Beyond IREs, evanescent-wave MIR photonics extends the same co-design philosophy to fully engineered optical paths, notably waveguides, for which multiple clinically oriented sensing concepts have been proposed in recent years [6,16]. In parallel, ongoing efforts in MIR plasmonic amplification have matured into SEIRAS (surface-enhanced infrared absorption), where tailored nanophotonic substrates provide near-field enhancement and have already been demonstrated in complex biofluids [17]. **Transflection substrates such as silver-based slides, well established** in IR microscopy experiments, could also be used in combination with sample processing naturally, as recently demonstrated for the analysis of cells from biofluids [3,4].

Beyond these optical substrates, there is still considerable untapped room for built-in pretreatment driven by microfluidics and surface chemistry. For instance, the recent 'spectIR-fluidics' concept illustrates how customizable, fully enclosed microchannels can be integrated with high-sensitivity on-chip ATR using standard FTIR instrumentation [18]. In parallel, chemical and chemical-biology functionalization can turn the ATR surface into a selective extractor, enabling structure-sensitive readout, for example, through reversible immuno-IR formats that capture target proteins from body fluids directly on the crystal [19]. More generally, tailored coatings (e.g., hydrophobic polymers on ATR waveguides) can preconcentrate classes of analytes and exclude interferents. This approach is well established in aqueous sensing and readily transferable to clinical matrices [20].

### Concluding remarks

Whereas most current efforts focus on improving MIR-based diagnostic models through increasingly sophisticated deep learning algorithms, there remains a valuable opportunity to enhance the quality of information obtained via optimized sample treatment. Clinical applications of MIR spectroscopy stand at the intersection of analytical potential and practical feasibility. As we have argued, the analytical gains afforded by sample pretreatment come with physical and chemical consequences that must be consciously minimized. Advances in integrated optics, disposable sampling interfaces, microfluidics, and controlled deposition strategies demonstrate that it is possible to make preprocessing part of the measurement, rather than an external workaround. These developments help MIR spectroscopy move toward workflows that maintain point-of-care simplicity while providing deeper information by targeting specific sample fractions (or even applying data fusion across them), which dramatically increases the potential to uncover selective spectral markers, enabling more specific and sensitive diagnostic tools. Ultimately, clinical MIR does not need to forgo sample preprocessing; instead, it can incorporate it into tailored, standardized, reproducible, and regulated platforms, accelerating its translation from proof-of-concept studies to trusted clinical tools.

**Acknowledgements**

This study has been funded by Instituto de Salud Carlos III (ISCIII) through the projects PI23/00135, PI23 and co-funded by the European Union. Authors also acknowledge the financial support from the projects RYC2019-026556-I, PID2023-148947OB-I00 and CNS2023-145528 funded by MCIN/AEI/10.13039/501100011033. V.N.E. acknowledges the financial support by the grant PRE2021-098833 from MICIN/AEI/10.13039/501100011033 and the FSE+. Á.S.I. acknowledges the support of grant JDC2022-049354-I funded by MCIN/AEI/10.13039/501100011033 and by the "European Union NextGenerationEU/PRTR".

# 4. IR Analysis of body fluids: Reagent-free point-of-care and bedside applications

**Werner Mäntele[1,2], Alexander Mittelstädt[1], Pip C. J. Clark[1], Valentino Lepro[1], Sergius Janik[1] and Thorsten Lubinski[1]**

[1] DiaMonTech AG, Boxhagener Strasse 82A, 10245 Berlin, Germany
[2] Institut für Biophysik, Goethe-Universität, Max-von-Laue-Strasse 1, 60428 Frankfurt/Main, Germany

E-mail: werner.maentele@diamontech.de, maentele@biophysik.uni-frankfurt.de

### Status

Analysis of body fluids such as blood or urine form the basis of most diagnostic processes in medicine. Typically, samples are taken and transferred to a clinical central lab or to a service provider, thus resulting in delays – up to days - for the initiation of therapeutic steps. In addition, laboratory analysis is predominantly based on chemical or biochemical reagents and requires frequent calibrations, both being relevant cost factors. In the case of blood, typically venous blood samples are taken by trained personnel, at a volume of several mL. This also involves a lot of consumables (vials, syringes, tubes) that end as potentially infectious waste.

Most analytical procedures of samples rely on a spectroscopic measurement in the visible or UV range. They use either direct measurement of pigment absorption (e.g. in the case of haemoglobin or bilirubin) or indirect measurement of a dye or enzyme reaction with subsequent spectroscopic detection. Detection and quantification in the mid-infrared spectral range from approximately 3 to 20 µm is not currently used in clinics, even though the constituents of body fluids have highly specific fingerprint bands in this range because of their molecular structure.

The potential of mid-IR spectroscopy (MIR) of body fluids for a quantitative analysis using IR spectra of blood plasma analysed by partial-least-square (PLS) methods was realized as early as 1989 [1]. Since then, many improvements have been reported. Spectra of whole blood, blood plasma, and urine samples have been recorded in transmission or attenuated total reflection (ATR) spectroscopy and analysed by multivariate data analysis, the latter leveraging a range of different statistical and machine learning approaches including artificial neural network models [4–8]. Some of these studies were motivated by glucose analysis and have yielded a good predictive accuracy for glucose concentration in whole blood or blood plasma samples which is sufficient for clinical applications.

A common feature of all MIR studies on body fluids published so far is that the number of samples in the calibration set is relatively small (typically <100) and that the range of the analysis covers the span of physiological rather than extreme pathological values. To develop a method which can be applied in routine clinical analysis and which enables stable prediction even for extreme combinations of pathological values, a larger number (>1000) of carefully selected calibration samples was used, spanning the full range for each parameter and following specific disease profiles [9]. Reference analysis was also optimized, especially in the case of urine samples where clinical analysis does not fully rely on clear concentration values for certain parameters, but on statements like "not present", "above limit" or "below limit". The concept of blood and urine analysis by MIR spectroscopy was later extended to dialysis fluid with the intention to provide a real-time monitoring of detoxification for critical patient cases and a personalized haemodialysis [10].

In all publications reported so far, infrared analysis can provide a minimum of 8-10 blood parameters at clinical precision (cf. Fig. 1) with a minimum of blood (some µL), without reagents, and within minutes. These blood samples can be obtained by finger pricking instead of venous blood extraction, do not need trained personnel, and they are much less critical for Intensive Care Units patients that require frequent blood analysis. It is also evident that with a broader basis of reference samples, additional parameters such as HDL,

LDL, creatinine, lactate, c-reactive protein (CRP) and others are within reach. All this can be reached without the use of reagents.

The situation for urine analysis is similar. A significant number of renal parameters can be deduced from IR spectra. Urine has been analysed using dried samples on IR windows, liquid samples in thin-layer cuvettes or ATR cells (for a recent review, see [11]). Ultimately, a flow ATR cell directly combined to a urinal is conceivable that could represent a simple and efficient tool for quick renal screening.

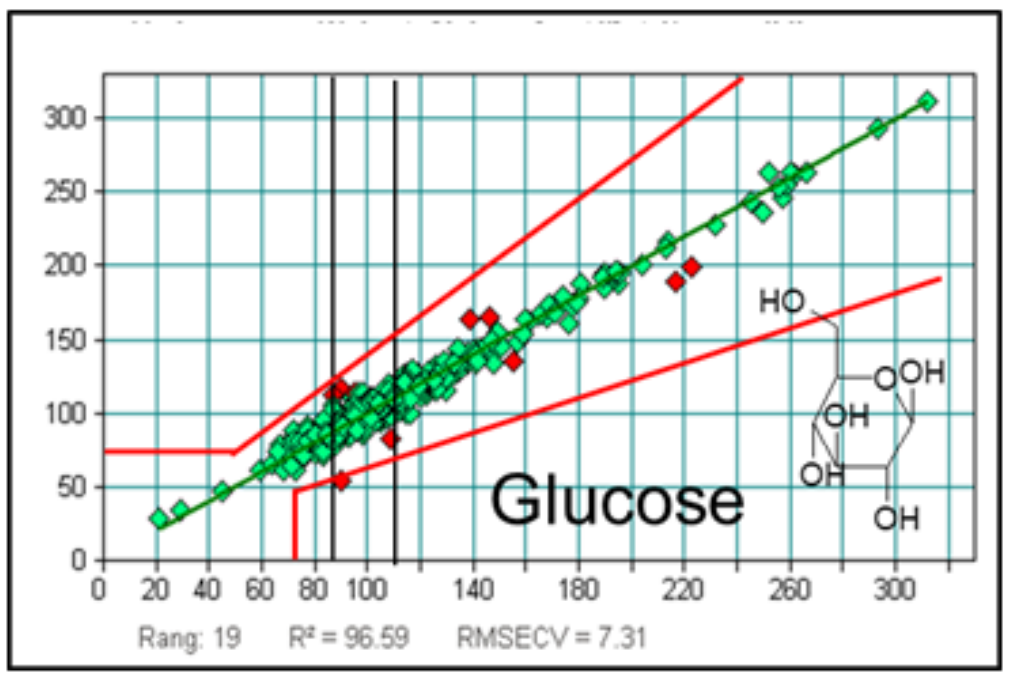

| Blood parameter | Range [mg/dL] | Standard Error [mg/dL] | Clinical Precision |
|---|---|---|---|
| Glucose | 20 – 340 | ± 7,6 | 🙂 |
| Urea | 10 – 220 | ± 4,4 | 🙂 |
| Cholesterol | 35 – 360 | ± 15 | 🙂 |
| Triglycerides | 0 – 1700 | ± 19,7 | 🙂 |
| Total Protein | 2000 – 11000 | ± 322 | 🙂 |
| Albumin | 1000 – 6500 | ± 148 | 🙂 |
| Hemoglobin | 7000 – 18000 | ± 234 | 🙂 |
| Immunglobulin | 200 – 2200 | ± 122 | 😐 |

**Figure 1.** Examples for MIR-based blood analysis (data taken from ref. [9]). **Left:** Cross validation of whole blood glucose. Reference values from clinical analysis are plotted on the horizontal axis; values from IR analysis on the vertical axis. Clinical precision is within that of blood glucose self-measuring devices that work with a test strip. **Right:** Whole blood parameters analysed by MIR spectroscopy on the basis of a calibration with 1000+ reference samples. IR analysis of all parameters except immunoglobulin meets clinical needs.

### Current and future challenges

One challenge for the analysis of blood is sample preparation. Whole blood samples should be immediately analysed after extraction, since blood clotting or sedimentation of red and white blood cells may lead to false results. Alternatively, plasma samples can be prepared that are free from cellular components (and consequently miss haemoglobin) but can be stored for longer. For whole blood samples, standard operating procedures with a strict timing are recommended since red blood cells, over time, metabolize glucose.

A second challenge is the initial calibration procedure. Reliable reference values can usually be obtained from well-run clinical laboratories. This implies more stringent calibration in the clinical labs as well as the willingness for repeated measurement of the same sample to improve reference accuracy. In the process of establishing a robust calibration matrix, care has to be taken to identify e.g. blood samples with a large span of parameters: realistic predictions for “pathologic” samples can hardly be obtained with a calibration from “normal” samples. Once a robust calibration matrix has been established, it can be used on different versions of IR instruments and transferred to new (“naïve”) instrumentation.

Further challenges are for the algorithms that lead from the raw IR spectra to medically usable data. In view of the complexity of body fluids as a matrix, univariate analysis, i.e. selection of one characteristic IR wavelength, if necessary with a reference wavelength, is not applicable. In general, the IR spectra of blood and urine constituents are broadly overlapping and thus require multivariate analysis. Established mathematical tools such as principle component regression (PCR) and partial least-square regression (PLS) are available and can be complemented with artificial neural networks or other artificial intelligence-based analysis. Care has to be taken in the development of the user interface to avoid ballast – medical personnel or laypersons are not interested in spectra or in complex math.

A hurdle for the implementation of reagent-free point-of-care and bedside applications for body fluid analysis is approval as a medical device, i.e. CE or FDA clearing. This is a relevant, time- and labour consuming step, although the measuring device need not be in direct contact to the patient and acts as a “laboratory device”.

### Advances in science and technology to meet challenges

The instrumental challenges for IR-based body fluid analysis can be met by standard FT-IR equipment combined with e.g. ATR cells and by standard multivariate analysis software. Nevertheless, stable thermal IR sources, compact interferometers, stable sample interfaces (sample cells) and sensitive IR detectors are required for dedicated medical IR instruments, ideally with a small footprint. With upcoming tuneable mid-IR quantum cascade lasers (QCLs), or with QCL arrays, handheld devices are conceivable. With further improvement of laser sources, an increase in sensitivity can be expected that potentially enlarges the measurable blood or urine parameters to those at lower (below ppm) concentrations.

### Concluding remarks

In summary, the feasibility of reagent-free mid-IR analysis of body fluids has been demonstrated, and current as well as future technologies will improve sensitivity as well as applicability (and affordability). IR spectroscopy *per se* is limited to molecules, and does not provide information on atomic ions (e.g. $Cl^-$, $Ca^{2+}$, $K^+$). However on molecules and molecular ions (such as e.g. $PO_4^{3-}$), the wealth of information that can be quickly obtained in one measurement, without any reagents, from tiny amounts of samples, and can complement and (partially) replace clinical chemistry analysis. Point-of-care and bedside applications (“lab-on the desk”) represent an attractive alternative for healthcare centres and doctor’s offices, and offer new opportunities for healthcare in “medical deserts”.

### Acknowledgements

The authors would like to thank the entire team at the Biophysics Institute at Frankfurt University for the fundamental work in the initial phase of this project.

# 5. Infrared spectroscopy of biofluids for clinical and point-of-care diagnostics: bridging biochemistry and pathology

**Luis Felipe das Chagas e Silva de Carvalho[1], Hugh James Byrne[2]**
[1]Departamento de Odontologia, Universidade de Taubaté
[2]Physical to Life Sciences Research Hub, FOCAS, Technological University Dublin, Aungier Street, Dublin D02 HW71, Ireland.

E-mail: luisfelipecarvalho@hotmail.com; hugh.byrne@tudublin.ie

## Status

Infrared (IR) spectroscopy provides a non-invasive, label-free, and rapid analytical method for probing the biochemical composition of biological samples. In clinical practice, biofluids such as blood, plasma, saliva, and urine contain a wealth of molecular information that reflects physiological and pathological states. Unlike morphological or imaging-based diagnostic methods, which primarily capture structural abnormalities, IR absorption spectroscopy captures the vibrational signatures of functional groups in biomolecules, yielding biochemical fingerprints that correlate with disease.

Fourier-Transform Infrared (FTIR) and Attenuated Total Reflectance (ATR-FTIR) are the dominant techniques for biofluid analysis due to their minimal sample preparation, high reproducibility, and compatibility with sample volumes as low as microliters. Miniaturized optics, portable mid-infrared sources, and chemometric algorithms have accelerated translation from research to bedside and Point-of-Care (PoC) applications [1–5].

Recent studies demonstrate the ability of IR spectroscopy to detect subtle biochemical shifts linked to diabetes, malignancies, hepatic injury, renal dysfunction, and sepsis. Its integration into pathology workflows enables early disease detection, monitoring of therapeutic response, and predictive risk assessment. Combined with AI-driven data analysis, IR spectroscopy offers quantitative, objective biochemical information complementing conventional histopathology, providing a foundation for predictive and personalized healthcare [6–9].

The idea of bringing FTIR spectroscopy into medical/dental offices and hospitals for direct use by healthcare professionals is becoming increasingly closer to reality. The fact that the technique can provide very rapid results for various diseases can accelerate the diagnostic process and greatly assist in patient treatment [10].

## Current and future challenges

Despite extensive evidence of diagnostic potential, the clinical implementation of IR spectroscopy faces critical challenges [11–13]. Biological variability, sample heterogeneity, and water absorption complicate spectral interpretation. Overlapping vibrational modes of proteins, lipids, and carbohydrates can mask disease-specific biomarkers. Thus, standardization of sampling, pre-processing, and spectral normalization remains essential for reproducible diagnostics.

Another barrier lies in regulatory and clinical validation. To achieve medical-grade reliability, large-scale, multi-center studies must correlate spectral signatures with clinical outcomes across diverse populations. Integration into existing laboratory workflows also requires automated preprocessing pipelines and intuitive interfaces accessible to clinicians.

Additionally, clinical education must evolve, as few pathologists and clinicians are trained in vibrational spectroscopy. Overcoming the perception gap between optical spectroscopy and “gold-standard” biochemical assays will depend on robust validation, improved instrumentation stability, and interpretability of AI-assisted models.

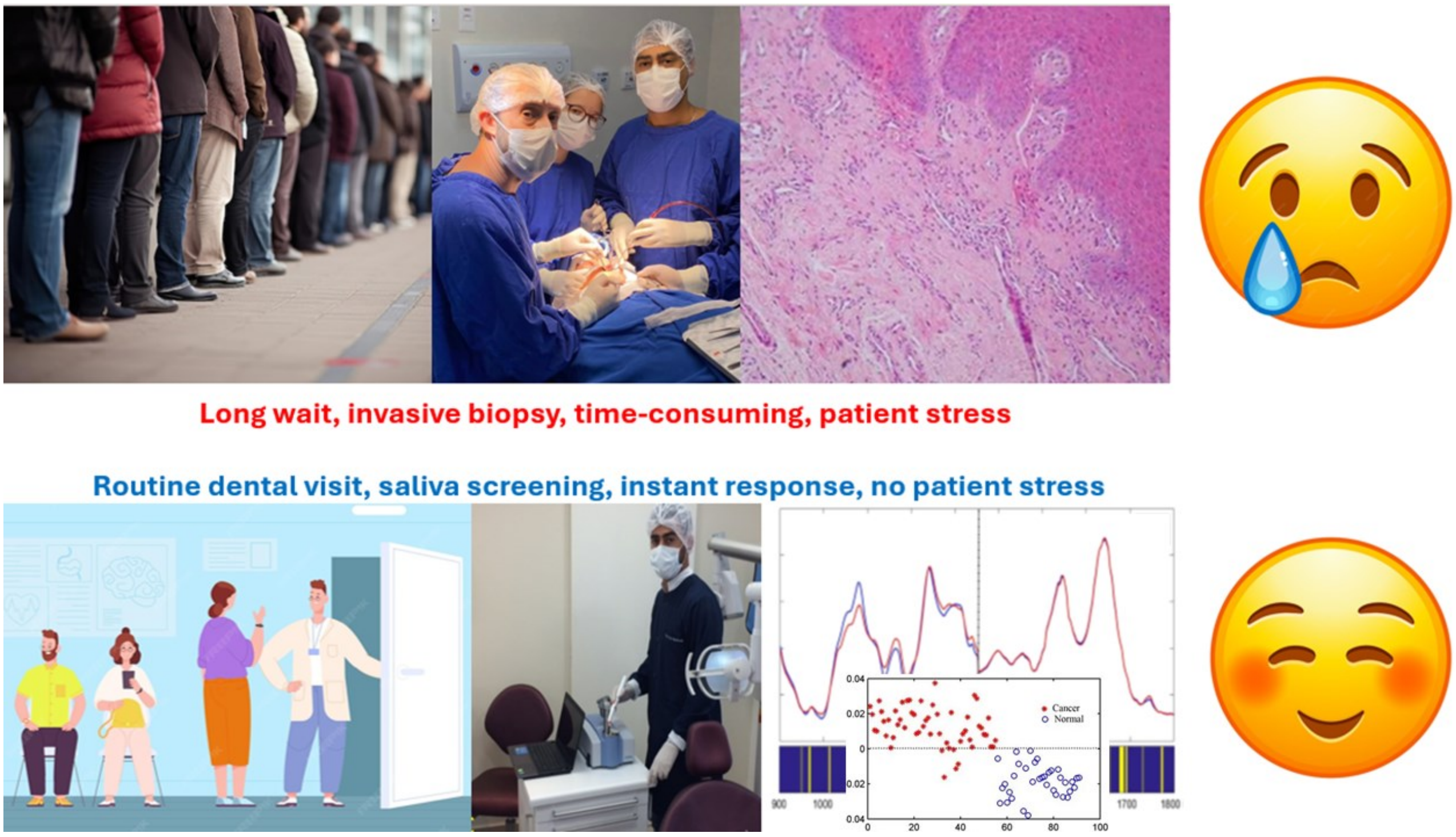

**Figure 1.** The upper path represents the current approach, which is time-consuming and often incurs unnecessary costs. Chair-side spectroscopic screening as part of a routine dentist visit, speeds up the process significantly, reduces costs, and identifies cases which require follow up biopsy.

## Advances in science and technology to meet challenges

Technological advances are rapidly addressing many of these limitations. Developments in quantum cascade lasers (QCLs), broadband supercontinuum sources, and micro-electromechanical systems (MEMS) have enabled compact, high-performance mid-IR instruments suitable for clinical environments [14].

The development of robust, portable, and efficient systems, such as the ALPHA II ATR system from Bruker, should facilitate the entry of this technology into clinical settings. It is also worth noting that the system features an ATR crystal that can be heated, allowing liquid samples to be dried and analyzed more rapidly [15].

Machine learning and deep learning frameworks now drive spectral classification, biomarker discovery, and disease prediction, achieving accuracies comparable to standard clinical assays [16]. Integration with microfluidics and disposable substrates enhances reproducibility and contamination control, critical for routine hospital use [17].

Hybrid modalities, combining IR with Raman spectroscopy or Optical Coherence Tomography (OCT), enable complementary biochemical and structural readouts, expanding diagnostic scope [18]. Future directions include automated data pipelines integrated with electronic health records for predictive analytics, paving the way toward digital pathology guided by spectroscopic biomarkers.

## Concluding remarks

Infrared spectroscopy of biofluids represents a paradigm shift in clinical diagnostics, providing biochemical insight invisible to traditional morphological techniques. The convergence of optical engineering, computational analytics, and clinical validation is transforming IR spectroscopy into a reliable Point-of-Care tool. Continued interdisciplinary collaboration will be key to achieving routine adoption in pathology laboratories and real-time hospital monitoring, ultimately advancing proactive and predictive healthcare. The knowledge of the technology by clinicians (from various areas) is of fundamental importance for its real advancement as a new diagnostic modality.

## Acknowledgements

Luis Felipe das Chagas e Silva de Carvalho is funded by Fundação de Amparo à Pesquisa do Estado de São Paulo (FAPESP – 2017/21827-1) and by Conselho Nacional de Pesquisa e Desenvolvimento (CNPq - 406761/2022-1).

# 6. Urine analysis using FTIR spectroscopy: clinical successes and challenges toward standardization

**Filiz Korkmaz**

Biophysics Laboratory, Faculty of Engineering, Atilim University, Ankara, Turkey

E-mail: filiz.korkmaz@atilim.edu.tr

## Status

Infrared spectroscopy has long been used in chemistry and materials science to probe molecular vibrations, but its application to biological fluids is more recent. Over time, Fourier transform infrared (FTIR) spectroscopy—especially when combined with attenuated total reflectance (ATR)—became attractive in biomedical contexts for its minimal sample preparation and non-destructive nature. Urine, as a complex biofluid containing salts, urea, creatinine, proteins, metabolites, and potential pathological biomarkers, has traditionally been analyzed by biochemical assays and separation techniques (e.g., chromatography, mass spectrometry). While those classical methods are well-established and quantitative, they often target only a few analytes per assay. In contrast, FTIR enables rapid, reagent-free acquisition of holistic biochemical fingerprints, capturing multiple functional group vibrations simultaneously and highlighting subtle spectral differences that may point to potential biomarkers. A recent review comprehensively summarized the applications, strengths, and limitations of vibrational spectroscopy in urine analysis [1]. In the urinary stone field, FTIR has been a long-standing standard for identifying crystalline deposits [2]. More recently, ATR-FTIR spectroscopy has been successfully employed for diagnostic purposes, including diabetes, diabetic kidney disease, urinary tract infections, renal hyperfiltration, prostate cancer and autism spectrum disorder [3-8]. Consequently, FTIR urine analysis is now recognized as an emerging complementary method to biochemical assays, offering a rapid, holistic profiling approach that can detect metabolic alterations and disease-related signatures.

## Current and future challenges

Despite these advances, FTIR urine spectroscopy faces several challenges before it can be fully integrated into clinical diagnostics. A major limitation lies in sensitivity and detection thresholds: many urinary biomarkers occur at micromolar or nanomolar levels, where dominant matrix components such as water, urea, and salts obscure weaker spectral features. The resulting spectral complexity and overlapping bands make interpretation difficult. Moreover, biological and inter-individual variability further complicates spectral analysis, since urine composition is strongly influenced by diet, hydration, circadian rhythm, and renal function. Pre-analytical variables—including sample storage, evaporation, and pH fluctuations—can also distort baselines and alter spectral intensities. Because urine is water-rich, water absorption remains a significant obstacle unless careful drying or dehydration methods are applied. Reproducibility, throughput, and standardization of ATR contact pressure and optical geometry also require careful control. Another persistent barrier is quantitative calibration: FTIR spectra provide relative intensities, necessitating chemometric calibration and external validation against gold-standard methods. A lack of standardized preprocessing and normalization protocols across laboratories continues to limit reproducibility. Additionally, machine learning models—although powerful—are sometimes viewed as 'black boxes' that hinder clinical acceptance. Large-scale, multi-center validation studies are therefore essential to demonstrate reproducibility and diagnostic reliability. Without regulatory validation from authorities such as the FDA or EMA, widespread clinical adoption will remain slow.

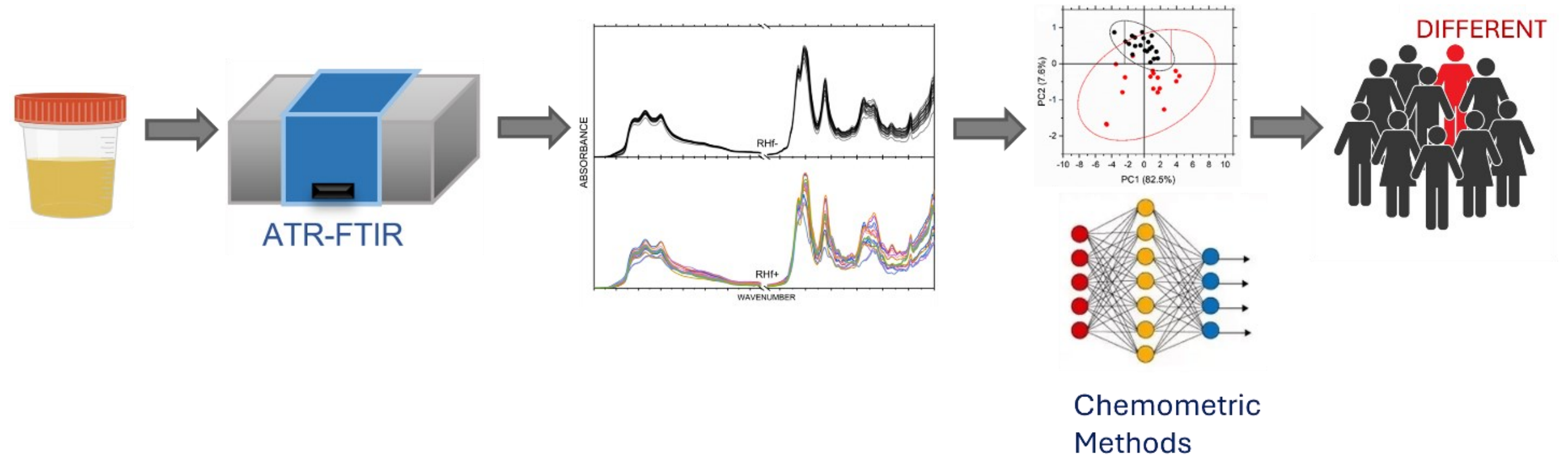

**Figure 1.** Urine analysis workflow using ATR-FTIR.

## Advances in science and technology to meet challenges

Several technological and methodological advances are helping to overcome these barriers. Evaporation of excess water in urine sample improves sensitivity for low-abundance metabolites [9]. Developments in detector technology, such as cooled focal plane array (FPA) sensors, enhance signal-to-noise ratios. Optimized ATR crystal materials (e.g., diamond, germanium) and variable-angle configurations improve reproducibility and penetration depth. On the computational side, recent advances in chemometrics, deep learning, and transfer learning allow more precise extraction of subtle spectral patterns and improved classification accuracy. Robust spectral preprocessing, feature selection, and calibration transfer techniques enhance cross-instrument comparability. Miniaturized, field-portable spectrometers—enabled by quantum cascade lasers and Micro-Electro-Mechanical Systems (MEMS)-based interferometers—are opening opportunities for point-of-care urine screening. Efforts toward standardization, spectral libraries, and automated analytical pipelines are also ongoing to improve reproducibility. The development of an artificial urine protocol improved control over sample variability [9]. Together, these examples highlight how thoughtful experimental design and data integration are transforming FTIR spectroscopy from a laboratory tool into a clinically relevant analytical technique.

## Concluding remarks

FTIR spectroscopy of urine represents a promising, non-invasive, and reagent-free analytical platform that complements conventional biochemical methods. Although limitations related to sensitivity, overlapping bands, and biological variability persist, continuous progress in instrumentation, sampling strategies, and computational data analysis is steadily improving the method's diagnostic potential. With harmonized protocols, transparent modelling approaches, and rigorous clinical validation, FTIR-based urine analysis is well-positioned to become a valuable tool for early disease detection and metabolic monitoring in modern healthcare.

# 7. Non-invasive blood glucose monitoring

**Werner Mäntele[1,2], Michael Kaluza[1], Sergius Janik[1], Mattia Saita[1], Lars Melchior[1], Thorsten Lubinski[1]**

[1] DiaMonTech AG, Boxhagener Straße 82A, 10245 Berlin, Germany
[2]Institut für Biophysik, Goethe-Universität, Max-von-Laue-Strasse 1, 60428 Frankfurt/Main, Germany

E-mail: werner.maentele@diamontech.de

### Status

Diabetes as a disease that affects worldwide about 530 million people for which there is no known cure at present. [1]. An important step in the management of diabetes is the frequent and precise measurement of blood glucose levels (BG) and the adaptation of food intake, physical activity, medication or insulin injection. For around 40 years, the measurement of blood glucose implied pricking the skin for a drop of blood and measuring BG level with a test strip, a painful and uncomfortable invasive method and thus not applied closed-meshed enough. Recently, minimally invasive sensors have been introduced that use a flexible needle in the interstitial fluid (ISF) of skin for a quasi-continuous glucose monitoring (CGM). These sensors are particularly appreciated by Type-1 diabetes patients in spite of frequent failures, allergic reactions and high costs [2].

For almost three decades, researchers have tried to use photonic techniques for a fully non-invasive glucose monitoring (NIGM) in sweat, saliva, tear fluid, urine and interstitial fluid (ISF). Glucose appears in these body fluids, but there can be inadequate correlations with BG as well as delays with respect to dynamic up's and down's of BG level. Among these proxies, ISF has been identified as an optimal fluid that can be measured in skin at depths below about 20-50 µm. The top layer of skin, the *stratum corneum*, contains some glucose, however, not representative for BG level. A direct transdermal photonic measurement of glucose in the ISF thus requires (i) penetration of the *stratum corneum* towards deeper layers and (ii) specific identification of the glucose molecule on the basis of spectral fingerprints [3].

Glucose in water is colourless, i.e. glucose has no electronic transitions in the VIS. Further to the UV, there are only unspecific electronic transitions. In the MIR at wavelengths between approx. 8 and 11 µm, however, there is a vibrational fingerprint of glucose arising from –C–O– stretching and –O–H– bending modes that can be the target of MIR and Raman spectroscopy. These MIR fingerprint bands give rise to vibrational overtones in the near IR (NIR) in the 1200–2000 nm spectral region. These NIR overtones are 2–3 orders of magnitude weaker than the MIR fundamentals. Although this is partly compensated by the higher penetration depth of NIR light, a reliable detection of glucose has not been accomplished yet (for a review, see Ref. [4]).

### Current and future challenges

The challenges of a transdermal measurement of ISF glucose by MIR spectroscopy are (i) sufficient penetration depth in skin, (ii) the complexity of skin layers that requires a highly specific spectroscopic technique and (iii) a sensitive detection of glucose signals, aiming to the diabetes-relevant glucose concentration range from approx. 50 mg/dL to 400 mg/dL. IR spectroscopy can meet these challenges. A penetration depth of 60–120 µm in skin (depending on the water content) is realistic for IR radiation in the 8–11 µm range. However, transmission measurements are not feasible due to the lack of sufficiently thin skin parts. Measurements that use attenuated total reflection (ATR) technology can be excluded as well, since the evanescent wave penetrates only by 5–10 µm into the skin. Several groups have analysed IR radiation backscattered from skin. Although they could demonstrate glucose patterns in the backscattered light, it appears that only top and very shallow layers are probed by this technique [5, 6].

In order to circumvent this drawback, several groups have proposed a photoacoustic measurement of ISF glucose by excitation of skin with pulses from a quantum cascade laser (QCL) at a several IR wavelengths in the glucose fingerprint and using air-filled photoacoustic (PA) cells in contact with skin with a microphone for detection. Dependent on the design of the PA cell, it can be used in resonance by adapting the QCL pulse frequency, which substantially increases sensitivity [7-9]. A substantial drawback of detection with an air-filled PA cell is the acoustic impedance in the skin-air and air-microphone interface. Alternatively, time-gated PA spectroscopy was used [10].

An alternative to PA measurements is the photothermal (PT) signal that originates from heat deposition at the glucose molecule following vibrational relaxation [11,12]. The generated PT signal migrates to the skin surface, much slower than the PA signal. By placing an internal reflection element (IRE) in contact with the skin surface, the heat signal transfers to the IRE and locally (at the skin spot where the QCL beam enters) alters the optical parameters of the IRE, e.g. its refractive index. This can be interpreted as a temporary (because of the pulsed excitation) thermal lens. This thermal lens is then read out by a second laser beam from a red or NIR laser diode, whose deflection is recorded with a position sensitive detector (PSD). The deflection signal is directly related to the glucose concentration in skin [11].

MIR-based glucose measurement has been validated in clinical tests with an accuracy comparable to CGM devices [13,14]. DiaMonTech has developed a table top (D-Base) device and is finalizing a handheld device (D-Pocket).

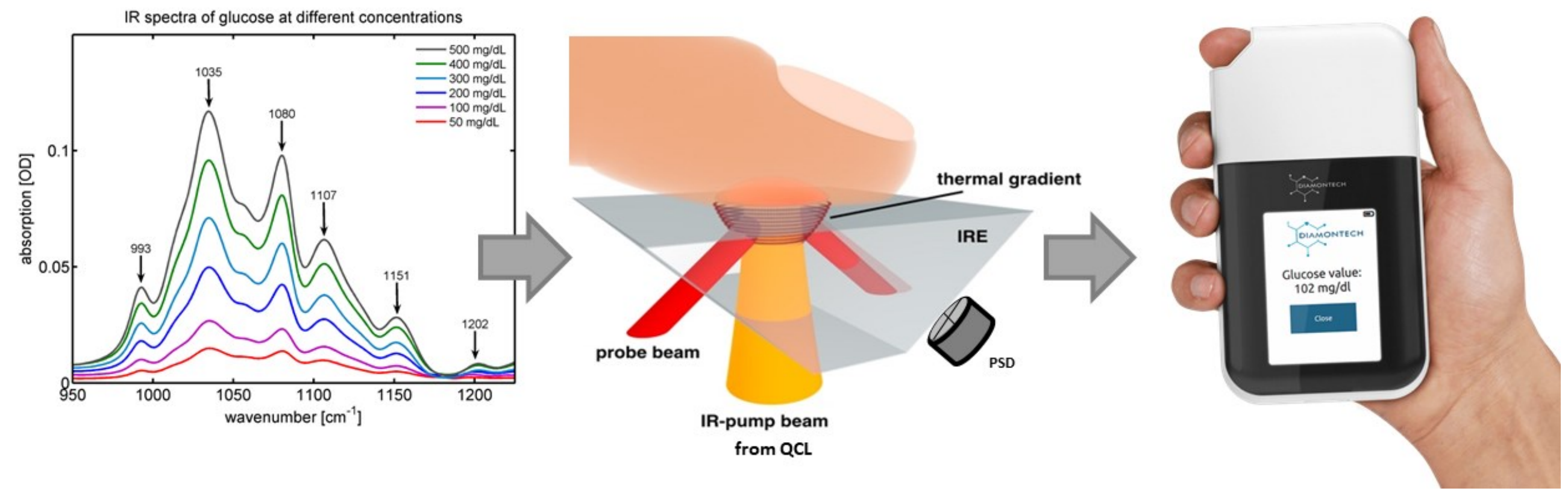

**Figure 1.** Non-invasive transdermal glucose monitoring: A specific fingerprint of glucose in the MIR is excited by a pulsed QCL, either scanned across the spectrum or at 10-15 discrete wavelengths from a QCL array. Vibrational excitation of glucose molecules in skin leads to a heat gradient in skin that couples to an internal reflection element (IRE), and generates a transient thermal lens, probed by a second red laser diode. The deflection is detected by a position-sensitive photodiode (PSD). Based on the amplitude of deflection, the glucose concentration is calculated.

## Advances in science and technology to meet challenges

An essential step in MIR based NIGM is the development of IR laser sources. The development of QCLs over the past 20 years has strongly advanced this field. External cavity tuneable lasers (EC-QCL) can cover a spectral range of >300 $cm^{-1}$, sufficient for IR spectral fingerprints e.g. from glucose or lactate. Alternatively, individual QCL emitters can be combined to arrays that form the basis of compact, handheld and affordable sensors. The average output power of these QCL is typically several mW, more than sufficient for *in vivo* applications, complies with safety standards for medical and consumer use. Further miniaturization of sources and detection will lead towards smartwatch applications.

## Concluding remarks

MIR-based transdermal skin spectroscopy has been demonstrated as an approach to solve the problem of a true non-invasive glucose measurement. This will allow close-meshed, painless, and comfortable glucose measurements and open new perspectives for a better management of diabetes.

## Acknowledgements

The authors would like to thank the entire team at DiaMonTech team for their contributions in R&D and the team at Frankfurt University for the fundamental work in the initial phase of this project.

# 8. Mid-IR laser-based dispersion spectroscopy: A new highly sensitive and robust avenue for liquid sensing

**Alicja Dabrowska[1,2], Georg Ramer[1,2] and Bernhard Lendl[1]**

[1] Institute of Chemical Technologies and Analytics, TU Wien, Vienna, Austria
[2] Christian Doppler Laboratory for Advanced Mid-Infrared Laser Spectroscopy in (Bio-)process Analytics, TU Wien, Vienna, Austria

E-mail: bernhard.lendl@tuwien.ac.at

### Status

The unique properties of tunable mid-IR lasers enable the development of sensing platforms that advance liquid-phase spectroscopy beyond the limits of established absorption techniques. Traditional absorption spectroscopy, governed by Beer's law, quantifies the attenuation of light intensity through an absorbing sample but neglects the wave characteristics of light.

By exploiting the coherence and well-defined polarization of modern mid-IR quantum cascade laser sources within interferometric architectures such as the Mach Zehnder interferometer (MZI), wavelength-dependent phase shifts induced by anomalous dispersion (wavelength dependent refractive index changes) across the analyte's absorption band can be measured with high precision. By placing a sample and a reference in the two arms of the MZI a compact spectrometer can be built and the phase shift between the two arms can be resolved by precise movement of a piezo-actuated mirror and placed in one arm of the MZI, which is then directly proportional to the dispersion spectrum of a sample (see Fig. 1) (1,2). In fact, using the MZI both amplitude and phase information, hence both the imaginary as well as the real part of the complex refractive index of liquid samples can be acquired, providing a more comprehensive understanding of the sample's optical properties for the purpose of its identification and quantification (3).

Compared to intensity-based absorption methods, dispersion spectroscopy offers several critical advantages that might make the difference for a successful application of mid-IR measurement of biomedical samples. First, it operates over a larger dynamic range, as phase shifts accumulate linearly with optical path length or increasing concentration, in contrast to the exponential attenuation of intensity (4). Second, the technique maintains constant sensitivity across a broad spectral range and extended pathlengths, even in matrices such as water with a strongly varying absorbance across the mid-IR spectral region. This makes it particularly well suited for studying aqueous and biological samples, where the linearity of conventional detectors poses a limitation. Finally, phase-based detection intrinsically suppresses certain types of intensity noise, commonly present in pulsed QCL sources, and baseline instabilities that limit the precision of intensity-based methods (5–7).

### Current and future challenges

Despite the potential of laser-based dispersion spectroscopy, there are several challenges that must be addressed to further improve the technique's sensitivity and applicability. Detecting subtle variations in the complex spectra of body fluids requires exceptional sensitivity often down to or exceeding levels of $2x10^{-5}$ AU (8). The current spectrometer already achieves sensitivity comparable to that of high-end FTIR instruments, enabling detection of refractive index changes as small as $6x10^{-7}$ RIU and has been tested for its suitability to monitor both static and dynamic samples in its dispersion mode in the protein as well as carbohydrate spectral regions (2,9). Nevertheless, the phase-resolved approach due to its unique advantages and different signal-to-noise ratio (SNR) characteristics holds substantial potential for further improvements. A potential limitation of this technique might be scattering particles present in the liquid sample.

The sensitivity of the system is fundamentally constrained by the uncertainties in interferometric distance measurements, including laser phase noise and refractive index fluctuations of the surrounding medium (e.g., air). Additionally, the accuracy of phase estimation (here mechanical movement of the piezo-mirror) remains a critical factor in reducing measurement uncertainty (10–12). As laser power and pathlengths contributing to signal increase are essentially constrained by the complete sample (water) absorption and logarithmic scaling of intensity with pathlength on permissible pathlength, the focus must shift to minimizing noise via phase-sensitive interferometric detection. This represents the most promising route for enhancing the SNR in dispersion spectroscopy without compromising measurement accuracy. Finally, developing a robust and portable analyzer suitable for operation outside specialized laboratory environments represent another important step toward real-world applications (13).

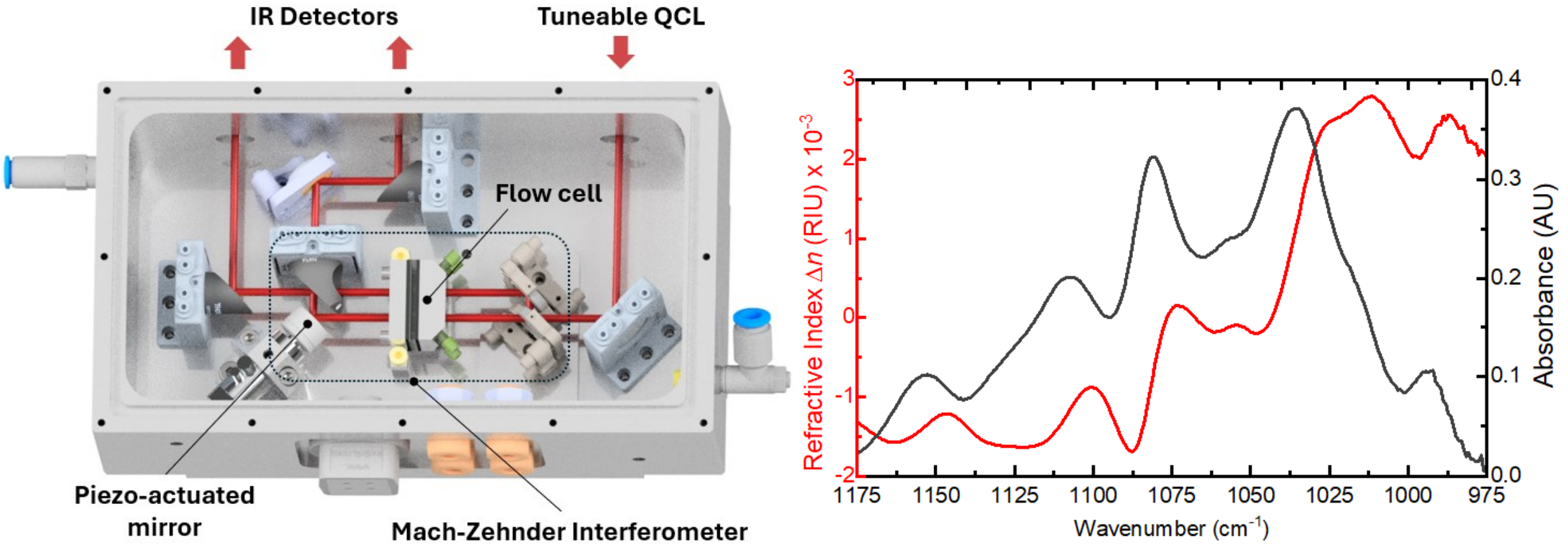

**Figure 1.** Mid-IR laser-based dispersion spectrometer for liquid-phase analysis (left) and an exemplary mid-IR absorption and dispersion spectrum of 10 mg $ml^{-1}$ glucose in aqueous solution measured by the spectrometer in a transmission mode at a pathlength of 170 μm (right).

## Advances in science and technology to meet challenges

Addressing the challenges in laser-based dispersion spectroscopy will require advancements in both hardware and software. On the hardware side, laser sources with higher monochromaticity and reduced phase noise are essential. Continuous-wave EC-QCL or multiplexed DFB-QCL arrays targeting specific parts of the spectrum could enhance measurement precision. The use of fast optical phase shifters and modulation schemes may further support stable and unambiguous phase readout (14,15). In addition, reducing the sensor footprint through the integration of mid-IR photonic integrated circuits (PICs) with microfluidics could lead to more compact, low-cost and ruggedized systems that enable real-time analysis of small-volume biological samples, capitalizing on the on-going efforts of expanding silicon photonics into the fingerprint region for sensing applications (16–18). On the software side, progress is being made in developing advanced algorithms for the interpretation of complex-valued data. Machine learning and artificial intelligence methods, complementing established chemometric tools, can help optimize the extraction of meaningful information from complex refractive index data and enhance the predictive power of dispersion spectroscopy for various applications, particularly in biomedical diagnostics (19,20).

## Concluding remarks

Laser-based dispersion spectroscopy represents a promising advancement in mid-IR sensing, offering significant improvements over traditional absorption-based techniques. By focusing on phase shifts rather than intensity changes, one can take full advantage of sensitivity of interferometry, different SNR characteristics, and higher permittable sample thicknesses to unlock higher sensitivities required for complex biological samples. While the technique has already demonstrated substantial capabilities, further advancements in noise mitigation, and data analysis will be key to realizing its full potential in various scientific and biomedical applications.

### Acknowledgements

The authors acknowledge the funding from Horizon Europe research and innovation program under grant agreement no. 101093008 (M3NIR) and the Austrian Federal Ministry of Labour and Economy, the National Foundation for Research, Technology and Development and the Christian Doppler Research Association.

# 9. Biophysical infrared spectroscopy as a diagnostic paradigm: Structural biomarkers for preclinical Alzheimer's and Parkinson's disease

**Nathalie Woitzik[1,2,3], Klaus Gerwert[1,2,3]**

[1] Chair of Biophysics, Ruhr-University Bochum, Bochum, Germany
[2] Center for Protein Diagnostic, Ruhr-University Bochum, Bochum, Germany
[3] betaSENSE GmbH, Bochum, Germany

E-mail: klaus.gerwert@rub.de

## Status

Infrared spectroscopy has evolved into a highly sensitive approach for detecting early molecular changes in human biofluids. Advances in quantum cascade lasers (QCLs), microfluidics, and machine-learning–assisted spectral analysis have enabled a transition from fundamental biophysical measurements toward clinically deployable diagnostics. The method provides label-free access to molecular fingerprints, capturing changes in protein secondary structure, lipid composition, and metabolic states with high structural specificity (Barth and Zscherp 2002).

Neurodegenerative diseases constitute a strong application area, as both Alzheimer's disease (AD) and Parkinson's disease (PD) are driven by pathogenic protein misfolding long before symptom onset. Amyloid-β (Aβ) aggregation in AD begins one to two decades before clinical manifestation (Selkoe 2001, Jucker and Walker 2013, Jack et al 2018). Similarly, α-synuclein (α-Syn) misfolding underlies early Parkinsonian pathology (Spillantini et al 1997, Brundin et al 2010). These pathogenic folding transitions generate disease-characteristic amide-I absorption signatures, enabling minimally invasive early detection in biofluids. In both AD and PD, progressive amide-I frequency downshifts ($\approx 1650 \rightarrow 1625\ cm^{-1}$) report the increasing formation of β-sheet–enriched neurotoxic aggregates, as conceptually summarized in Figure 1.

At Ruhr University Bochum, the Chair of Biophysics has established infrared structural biomarkers for such misfolding processes. Using an immuno-infrared sensor, Aβ misfolding—rather than peptide concentration—was shown to yield superior diagnostic and prognostic value (Nabers et al 2016). The misfolding signal correlates with CSF biomarkers and amyloid PET imaging and appears already in preclinical stages (Nabers et al 2018). More recently, the infrared spectral (iRS) platform detected α-Syn misfolding in CSF with high sensitivity and specificity in two independent cohorts, highlighting its applicability across synucleinopathies (Schuler et al 2025). The approach was further extended to blood measurements, demonstrating that misfolded protein conformations associated with proteinopathies can also be detected in plasma and serum, enabling accessible and scalable testing beyond CSF-based diagnostics.

The associated spin-off betaSENSE translates these developments into a fully automated clinical platform. The iRS system enables detection of pathological protein conformations, individualized AD/PD risk prediction, stratification for clinical trials through prodromal misfolding signatures, and therapeutic-response monitoring via structure-sensitive readouts. These advances represent a complete translational trajectory from biophysical discovery to clinically actionable diagnostics.

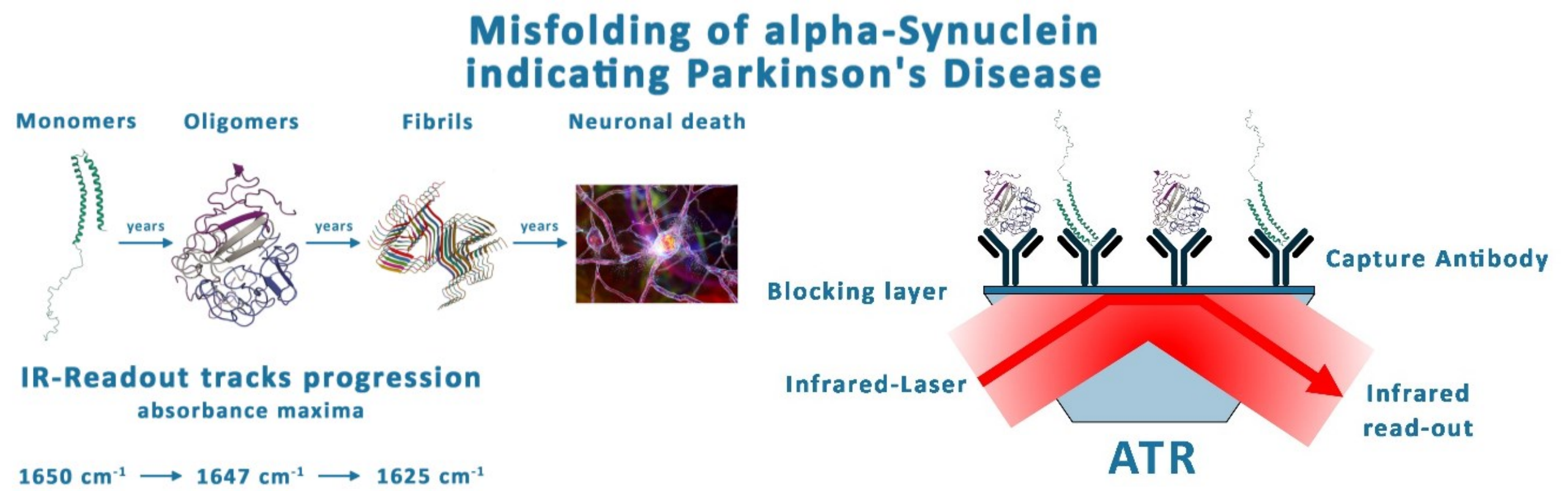

**Figure 1. Biophysical workflow for ATR-IR–based monitoring of α-synuclein misfolding and aggregation in Parkinson's disease.** α-Synuclein undergoes a conformational conversion from intrinsically disordered monomers to neurotoxic oligomers and β-sheet–dominated amyloid fibrils, driving progressive neurodegeneration and neuronal loss. Secondary-structure evolution is resolved in a label-free manner via characteristic low-frequency shifts of the amide I mode ($\approx 1650 \rightarrow 1625\ cm^{-1}$), reporting a relative increase in β-sheet population. The right-hand panel outlines an ATR sensing configuration with immobilized capture antibodies passivated by a peptide blocking layer. Interfacial structural signatures of bound aggregates are excited by coherent mid-IR laser radiation and detected through the evanescent ATR field, enabling selective and sensitive readout of α-synuclein secondary-structure fingerprints at the sensor interface.

## Current and future challenges

Despite rapid progress, several factors limit widespread clinical adoption. A central challenge is the standardization of sample handling, optical geometry, and environmental conditions. Biofluids exhibit biological variability, and even small deviations in path length, hydration, or temperature introduce spectral shifts. Automated microfluidics, reference standards, and harmonized protocols will be essential to achieve reproducible multicenter measurements.

Another major challenge concerns clinical interpretability. Although machine-learning classifiers achieve high diagnostic accuracy, clinicians require transparent links between spectral features and biochemical mechanisms. For misfolding assays, diagnostically relevant information arises from shifts in the amide-I band and alterations in secondary-structure composition, which correspond to β-sheet–enriched pathological assemblies. These spectral signatures must be translated into clinically interpretable categories—such as low, intermediate, or high misfolding—aligned with established frameworks including A/T/N-Stagingstaging for AD and MDS criteria for PD. Integration with cognitive assessments, imaging findings, and established biomarker modalities will be needed for broad clinical acceptance.

Regulatory approval requires multicenter validation, analytical performance characterization, reproducibility across instruments and sites, and demonstration of clinical utility across diverse populations. Standardized workflows, defined reporting formats, and longitudinal stability studies will be necessary for IVDR/FDA pathways.

Technical challenges remain as well. While QCL sources have become more stable and compact, long-term drift, thermal sensitivity, and system costs still limit scalability. For routine clinical operation, thermally robust, cost-efficient, and maintainable infrared engines are required.

## Advances in science and technology to meet challenges

Several technological advances are expected to improve diagnostic performance and scalability. Progress in QCL engineering will allow broader tuning ranges, higher spectral power density, and improved thermal stability. Multi-QCL integration on a single chip, combined with on-chip beam splitting and multiplexed microfluidic channels, will enhance throughput and reduce cost per measurement.

Automation of microfluidic workflows, including precise path-length control and real-time environmental compensation, will further improve reproducibility. At the computational level, next-generation spectral models will enable targeted frequency-region acquisition, adaptive measurement strategies, and improved interpretability through explainable AI.

In parallel, mechanistic integration with structural biology will strengthen the link between specific spectral motifs and pathogenic conformers. This will enable improved diagnostic specificity and facilitate therapeutic monitoring by detecting treatment-induced structural shifts earlier than clinical endpoints.

### Concluding remarks

Infrared spectroscopy is emerging as a structurally sensitive, minimally invasive diagnostic technology capable of detecting neurodegenerative pathology at preclinical stages. Through its ability to quantify protein misfolding directly, the method provides a coherent link between biophysical mechanism, molecular pathology, and clinical phenotype. With continued advances in standardization, QCL technology, data analytics, and multicenter clinical validation, infrared spectroscopy is well positioned to become a central component of predictive and preventive healthcare.

### Acknowledgements

We thank the State NRW and the Wissenschaftsrat for funding PRODI, the center for Protein-Diagnostik at the Ruhr-University.

# Cellular:
# 10. Infrared hyperspectral imaging in healthcare

**Hugh J. Byrne[1] and Peter Gardner[2,3,*]**

[1] Physical to Life Sciences, Technological University Dublin, D08 CKP1, Dublin, Ireland
[2] Photon Science Institute, University of Manchester, Oxford Road, Manchester, UK
[3] Department of Chemical Engineering, School of Engineering, University of Manchester, Oxford Road, Manchester, UK

E-mail: *peter.gardner@manchester.ac.uk

## Status

Infrared hyperspectral imaging (IR-HSI) integrates vibrational molecular spectroscopy with digital imaging, enabling the acquisition of both spatial and spectral information from biological samples. Over the past two decades, infrared microspectroscopy has evolved from a niche analytical research tool into a promising diagnostic platform for healthcare applications [1,2]. Its potential application to biopsy tissue analysis represents one of the most transformative uses of the technique, offering label-free, non-destructive molecular characterization of tissues at micron scale spatial resolution. Using an infrared micro spectrometer coupled with a focal plane array (FPA) detector, a detailed biochemical fingerprint can be obtained, reflecting variations in proteins, lipids, nucleic acids, and carbohydrates and these biomolecular changes often precede visible histopathological alterations, characteristic of disease [3].

Ageing populations and increases in cancer screening are putting extreme strain on pathology services. Traditional histopathology, which relies on staining and visual assessment, remains the diagnostic gold standard. However, it is inherently subjective and limited by staining variability and interpretive differences among pathologists [2,3]. IR-HSI complements and potentially augments these methods by offering reproducible insights into tissue biochemistry without the need for dyes or reagents. Recent developments in infrared sources, particularly quantum cascade lasers (QCLs) and machine learning algorithms, have accelerated the potential translation of IR-HSI into preclinical and clinical workflows, particularly for oncology, as tumor type, cancer grade, and stage can be determined (Fig. 1).

Despite these advances, IR-HSI remains absent in clinical practice. Barriers to adoption include the need for standardization (particularly of substrates), speed of data acquisition, robust data interpretation, and integration with existing pathology infrastructure [2]. The field remains vibrant, however, because it bridges molecular pathology and digital diagnostics, two pillars of precision medicine, potentially enabling earlier disease detection, personalized treatment decisions, and deeper biological understanding.

## Current and future challenges

Integration of IR-HSI into clinical biopsy analysis faces significant technical, computational, and translational challenges. Although numerous studies have shown that Fourier Transform IR imaging, coupled with machine learning, can identify tissue types and detect the presence of cancer, it is too slow to be used routinely in a hospital pathology laboratory [4]. A partial solution is provided by the advent of the higher power, tunable mid infrared QCLs. If the key diagnostic frequencies are known in advance, it may be sufficient to measure just those frequencies, saving a significant amount of time [5,6]. The number of frequencies required however, is a matter of debate, and will depend upon the size of the biological change being analyzed. For prostate cancer, it has been shown that there is a significant drop off in diagnostic performance between 25 and 10 frequencies, although the scan time per 1 mm core is reduced from 330 s to 196 s [7].

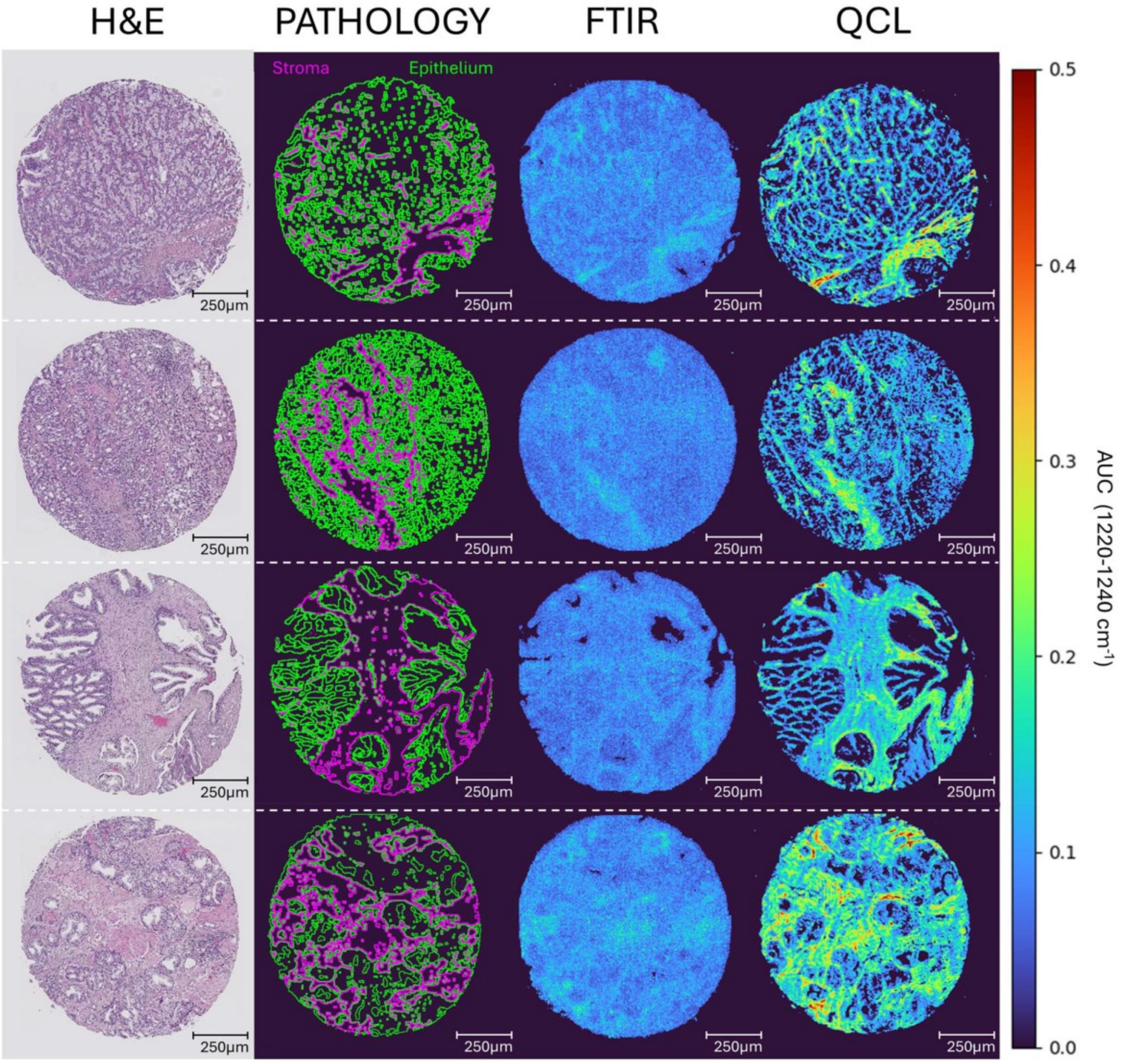

**Figure 1.** Comparison of integrated lower wavenumber spectral regions (1220–1240 cm$^{-1}$) for unprocessed data of four separate prostate cores imaged using FTIR and QCL systems, with H&E stained adjacent sections and key manually drawn pathological annotations for stroma and epithelial regions (pink and green respectively) provided for visual comparison. The scale bar shows 250 µm [12].

A second issue hampering clinical adoption is that of a suitable substrate. For best results, the tissue should be mounted on an infrared-transparent substrate, such as $CaF_2$, and the measurement made in transmission mode. However, $CaF_2$ is expensive, compared with a normal glass pathology slide. An alternative method is to use an infrared-reflecting slide, and measure in transflection mode, but this suffers from the electric-field standing-wave effect and interference issues [8]. These lead to oscillations in the spectral intensity that are difficult to correct for [5].

Regulatory acceptance and clinical validation remain major hurdles. Unlike conventional histopathology, IR-HSI-derived spectral biomarkers must demonstrate clinical utility and robustness through large-scale, multicenter trials. Establishing clinically interpretable models is complicated by the "black-box" nature of many AI algorithms used for spectral classification. This is a problem, since the EU AI Act states that AI-based models used in medical diagnoses are classed as the highest risk and will have to be completely explainable [9] This means that the black-box approach needs to be reconsidered.

Instrument cost and size could also limit accessibility. New IR imaging systems are typically in the range of £300 K, so it is important that, once an application has been identified, a full health economic study be carried out to evaluate whether such a method would be cost effective. Finally, data integration, linking hyperspectral data with genomic, proteomic, and clinical metadata, remains an open challenge, but also an opportunity for holistic disease modelling.

### Advances in science and technology to meet challenges

Addressing these challenges requires coordinated advances in optics, computation, and systems integration. On the hardware front, next-generation focal plane array detectors with enhanced mid-infrared sensitivity are enabling faster acquisition and higher signal-to-noise ratios. Crucially, quantum cascade laser (QCL)–based illumination is displacing conventional FTIR sources, providing high brightness and programmable spectral access that shortens scan times, essential for intraoperative or high-throughput pathology. Recent studies have shown that, rather than discrete frequency imaging, the QCL based systems can now scan the full fingerprint region from ~1800–950 $cm^{-1}$ over a whole tissue slide in ~20 min., making routine application possible [10].

Emerging, highly tunable QCL technologies will further accelerate progress [11]. External-cavity (EC-QCL) designs with mode-hop-free, wideband tuning and kHz–MHz sweep rates allow rapid, targeted interrogation of diagnostically salient bands (amide I/II, lipid CH stretches), supporting adaptive or compressive hyperspectral protocols that cut acquisition time and data volume [12,13]. QCL frequency combs paired with dual-comb detection promise near real-time, multiplexed spectral capture with high resolution, opening a path to snapshot or line-scan dual-comb IR-HSI that could dramatically reduce minutes-long scans to seconds [14]. Increased power and narrower linewidths should also benefit photothermal IR modalities (Specifically optical photothermal infrared O-PTIR ), extending IR-HSI to sub-diffraction contrasts and thicker specimens [15,16].

Computational innovations are equally transformative. Machine-learning pipelines that exploit task-driven band selection (aligned with agile QCL tuning) have the potential to deliver orders-of-magnitude reductions in data without sacrificing diagnostic accuracy [17].

On the standardization and data infrastructure side, open spectral libraries, calibration phantoms, and QCL-specific transfer functions could mitigate inter-site variability. Automated sample handling, humidity control, and multimodal correlative imaging (Raman, MSI, H&E) will further validate spectral biomarkers [18].

### Concluding remarks

Infrared hyperspectral imaging has matured into a powerful, non-destructive technique, capable of revealing the biochemical complexity of biopsy tissues beyond the limits of traditional microscopy. While substantial challenges remain in data standardization, computational scalability, and clinical validation, rapid advances in photonic technologies, particularly in tunable QCL technology, alongside innovations in machine learning and miniaturized instrumentation are steadily addressing these barriers. The convergence of these developments could transform IR-HSI from a research tool into a cornerstone of digital pathology and precision oncology, providing clinicians with objective molecular insights at the point of diagnosis. Its successful integration into healthcare would not only improve diagnostic accuracy but also reduce time to treatment, and a better fundamental understanding of the pathogenesis of disease, paving the way for a more personalized, data-driven model of care [19].

### Acknowledgements

PG acknowledges CLIRPath-AI and EPSRC Heathcare Technologies Network Plus grant EPSRC EP/W00058X/1.

# 11. Mid-infrared spectroscopy: Clinical utility in infection detection

**Hugues Tariel[1], and Olivier Sire[2]**

[1] Diafir, Rennes, France
[2] IRDL, UBS, Vannes, France

E-mail: Hugues.tariel@diafir.com, olivier.sire@univ-ubs.fr

## Status

Rapid detection of infectious diseases is a critical component of modern clinical practice, as early and accurate diagnosis directly influences patient outcomes, limits disease progression, and reduces the public health burden of infectious agents. Timely pathogen identification enables clinicians to initiate targeted antimicrobial therapy rather than empiric broad-spectrum treatment, thereby minimizing unnecessary drug exposure that drives antimicrobial resistance. In addition, rapid diagnostic methods support more efficient allocation of hospital resources.

Despite these needs, many infections are still diagnosed using conventional bacterial culture on agar plates. Although this approach is cost-effective and reliable, it typically requires several days to yield definitive results, thereby delaying clinical decision-making. Alternative diagnostic techniques based on surface or whole-cell recognition have been developed, including microscopy, immunoassays, nucleic acid amplification and sequencing, flow cytometry, biochemical profiling, and molecular probe–based assays targeting species-specific cellular or genetic markers. While these methods can reduce turnaround times to hours, they generally require specific reagents, markers, or labels. Metabolomics-based techniques, such as matrix-assisted laser desorption/ionization time-of-flight (MALDI-TOF) mass spectrometry, are also widely used in clinical laboratories but remain dependent on specialized infrastructure and trained personnel.

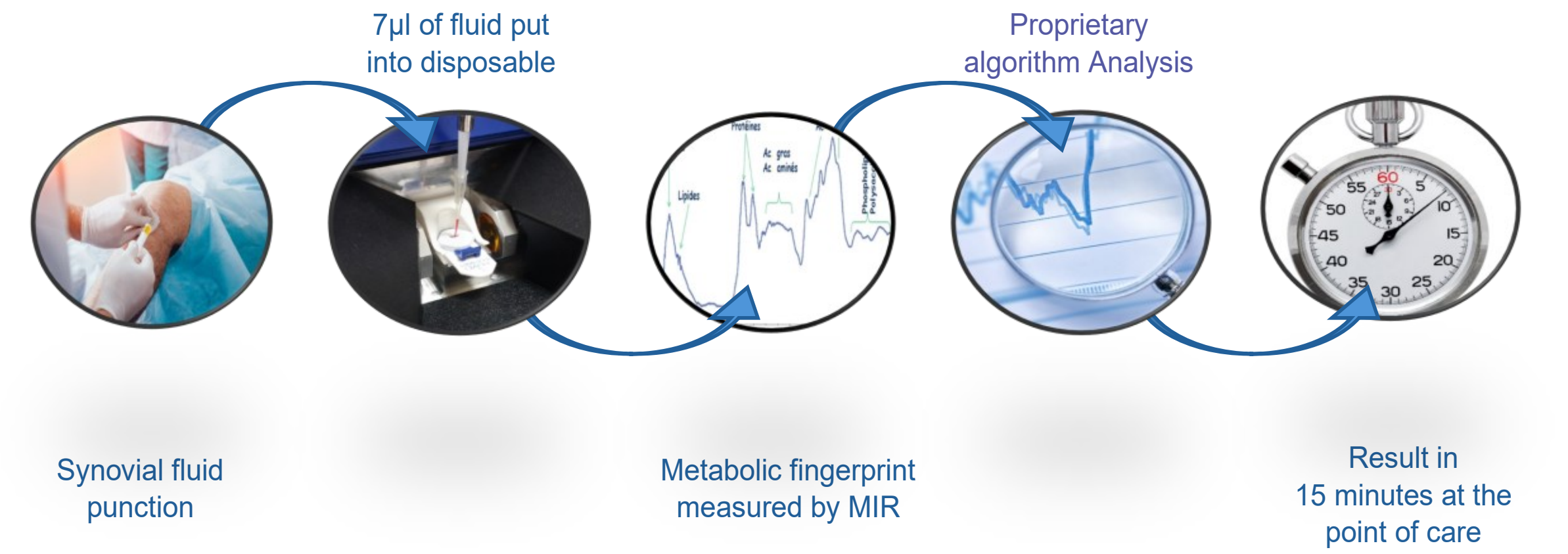

**Figure 1. MIR diagnosis process,** quick and label free, at the point of care

Mid-infrared (MIR) spectroscopy provides a rapid, label-free alternative metabolomic approach[1, 2]. MIR spectra capture a comprehensive biochemical signature of biological fluids or tissues and are often described as "metabolic fingerprints." In medical applications, these fingerprints can be analyzed using machine learning algorithms to extract clinically relevant diagnostic information [3, 4]. Moreover, the integration of MIR spectroscopy without extensive sample preparation or labeling enables rapid analysis directly at or near the point of care, facilitating timely clinical decision-making. This capability is particularly advantageous for infection detection, where delays in diagnosis can have significant consequences for both individual patients and public health.

Previous studies have demonstrated that MIR spectroscopy can detect and discriminate contaminants across multiple bacterial strains [5, 6]. Notably, it has been successfully applied to determine the terrestrial origin of fecal contamination in oysters by differentiating *Escherichia coli* serovars [7].

The Synofast solution [8] has been developed to enable the identification of septic arthritis within 10 minutes. Septic arthritis is a medical emergency that can result in irreversible joint damage or death; consequently, patients are often hospitalized and treated empirically with antibiotics while awaiting laboratory results. Synofast is CE-marked and is currently being deployed in French hospitals as a MIR spectroscopic diagnostic tool, thereby demonstrating the clinical feasibility and real-world performance of this technology. Table below summarizes the diagnostic performance observed in the derivation cohort (Synofast, 402 patients) and in an independent validation cohort (Synofresh, 307 patients).

**Table 1 – MIR synovial fluid infection test performance**

| | SYNOFAST Cohort | SYNOFRESH Cohort |
|---|---|---|
| Area Under ROCurve | 0.95 | 0.94 |
| Negative Predictive Value | 0.99 | 098 |
| Accuracy | 90% | 90% |

**Table 1:** Predictive model performance is estimated from the Area Under the ROCurve. NPV and accuracy are computed for a given threshold.

## Current and future challenges

Mid-infrared (MIR) spectroscopy detects the absorption of fundamental vibrational modes of most organic molecular bonds, generating strong and readily interpretable signals. This enables analysis of very small sample volumes without the need for exogenous labels. However, individual molecular bonds are shared by numerous biomolecules and metabolites, and their associated absorption bands are influenced by multiple factors, including the nature of the biological matrix (e.g., blood serum, urine, synovial fluid, saliva), the sampling procedure, and the thermal history of the specimen (e.g., number of freeze–thaw cycles).

In contrast, microbial MIR signatures are predominantly governed by membrane and cell wall constituents that contain species-specific molecular compounds, such as glycolipids, teichoic acids, and other serovar-defining markers. These structural components are among the most abundant cellular constituents and therefore largely dominate bacterial MIR spectra. This intrinsic lack of molecular specificity in complex biological fluids contributes to a perceived "black box" effect, which has limited the acceptance of MIR spectroscopy within the medical community, where diagnostic paradigms traditionally rely on well-defined and individually identified biomarkers. Consequently, only a limited proportion of early academic findings have translated into clinically validated diagnostic tools.

A further implication is the requirement for large patient cohorts to develop and validate MIR-based diagnostic algorithms, together with strict standardization of sample collection, handling, and storage protocols. For widespread clinical implementation, consensus on standardized practices across medical centers is essential; otherwise, the associated logistical burden and costs would be prohibitive. For example, multicenter studies in hepatopathies have demonstrated that pre-analytical variability—such as the choice of anticoagulant for blood samples (e.g., heparin versus EDTA)—can substantially impair robustness, generalizability, and predictive performance of models trained on specific datasets.

Additionally, methodological parameters related to MIR spectral acquisition, including the choice of measurement modality (transmission, attenuated total reflectance [ATR], or fiber-evanescent wave spectroscopy [FEWS]), require careful consideration. Post-acquisition mathematical correction of raw spectra is unlikely to fully compensate for systematic biases introduced by heterogeneous collection and measurement conditions.

Beyond experimental standardization, the secure handling of sensitive medical data and the preservation of patient confidentiality must be ensured, while accommodating hospital workflow constraints and enabling remote data processing and real-time transmission. In this regard, the primary challenges are regulatory rather than technical.

### Advances in science and technology to meet challenges

We identify three major axes that are expected to facilitate the dissemination and clinical adoption of MIR spectroscopy for medical diagnostics.

From a technological standpoint, the most anticipated advances concern MIR sources and detectors, with the objective of enabling more compact, affordable, and reliable platforms that can be deployed as close as possible to the point of care. Although current Fourier-transform infrared (FTIR) systems are widely recognized for their robustness and analytical reliability, their high cost and, to a lesser extent, their size remain significant barriers to deployment outside hospital settings or large healthcare facilities.

A key area of innovation lies in the development of mid-infrared light-emitting diodes (MIR LEDs), which are currently substantially more expensive than their visible-spectrum counterparts. The emergence of group IV–based components, particularly those relying on germanium (Ge) and tin (Sn), represents a promising alternative. These materials can be fabricated in standard semiconductor foundries, without the need for specialized III–V production lines, potentially enabling cost reduction and broader accessibility. In parallel, quantum cascade lasers (QCLs) are expected to provide higher output power, thereby improving signal-to-noise ratios. However, QCLs remain costly at wavelengths above 4 μm, which are required for infection-related applications. Despite this limitation, ongoing advances in MIR source technologies are also anticipated to drive improvements in detector performance, contributing to overall system optimization.

With respect to predictive modeling, the application of machine learning has enabled substantial progress through multivariate statistical approaches and advanced preprocessing of raw spectral data. In spectroscopic datasets, the ordering and continuity of variables are highly structured and convey a large proportion of the relevant information. Pattern recognition strategies based on encoding MIR spectra using spline coefficients provide a representation that closely reflects the underlying physical properties of MIR signals. Spline fitting captures the subtle curvature of typical spectra with high fidelity and allows the detection of small variations arising from changes in band overlap, intensity, and frequency shifts [9].

Finally, data mining and natural language processing approaches offer additional opportunities to improve interpretability by integrating knowledge from large-scale bibliographic resources, including more than 60,000 publications. By aggregating prior experimental evidence, these methods may help establish robust relationships between spectral variations and physiological or pathological states of biofluids, moving beyond isolated molecular bond assignments toward a more systems-level understanding.

### Concluding remarks

Mid-infrared (MIR) spectroscopy represents a promising approach for near-patient medical testing, owing to its rapid analysis time and operational simplicity. These characteristics make it particularly well-suited for clinical settings where timely diagnostic information is essential.

The establishment of well-curated infrared-linked biobanks, together with continued advances in artificial intelligence, is expected to enhance diagnostic performance while improving the biological understanding of MIR-derived signatures.

As MIR source technologies mature and follow cost-reduction trajectories comparable to those observed for focal plane arrays (FPAs) in thermal imaging, MIR-based diagnostic systems may progressively expand beyond hospital environments into primary care and, eventually, home use.

# Tissue:
# 12. Super-resolution mid-IR microscopy

**Margaux Petay[1], Elisabeth Holub[1], Bernhard Lendl[1] and Georg Ramer[1,2,*]**

[1] Institute of Chemical Technologies and Analytics, TU Wien, Vienna, Austria
[2] Christian Doppler Laboratory for Advanced Mid-Infrared Laser Spectroscopy in (Bio-)process Analytics, TU Wien, Vienna, Austria

E-mail: *georg.ramer@tuwien.ac.at

## Status

Conventional mid-IR spectral imaging is limited in its spatial resolution by the Rayleigh limit to >2.5 µm for most of the spectrum [1]. Several established techniques can circumvent this limit: optical photothermal infrared (O-PTIR), atomic force microscopy -infrared (AFM-IR), and scattering scanning nearfield optical microscopy (s-SNOM). Both O-PTIR [2] and AFM-IR [3] indirectly detect the sample IR absorption via photothermal-induced phenomena. In these two techniques, the sample is illuminated with a pulsed mid-IR source, which, when absorbed, leads to local heating proportional to the sample's optical absorption. This heating induces both thermal expansion and refractive index changes. O-PTIR uses a visible probe beam to detect local heating at a spatial resolution on the order of the probe beam wavelength [2]. Here, the sample expansion, deformation, and change in refractive index affect the intensity of the detected visible probe beam. In AFM-IR, an atomic force microscope (AFM) cantilever reads out the local surface expansion [3], improving spatial resolution down to a few nanometers [4]. Like AFM-IR, s-SNOM uses an AFM cantilever to achieve chemical imaging with a spatial resolution in the range of a few nanometers. However, unlike O-PTIR and AFM-IR, which indirectly detect the sample absorption through photothermal effects, s-SNOM detects the light scattered by the AFM cantilever oscillating in the vicinity of the sample's surface, providing both absorption and complex refractive index information [5].

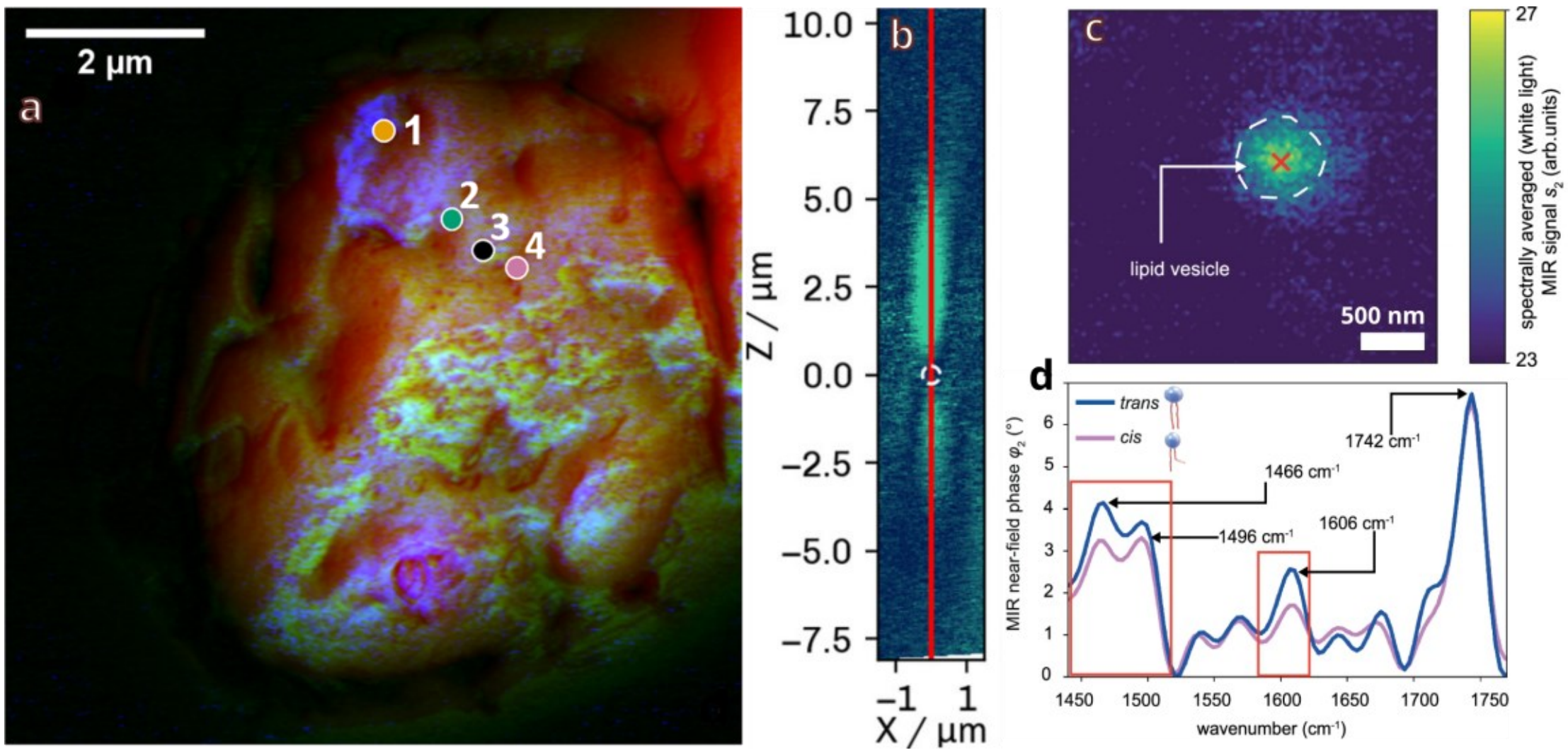

**Figure 1 a)** AFM-IR overlay image of a red blood cell combining AFM height with lipid (C=O stretching) and protein (amide I) absorptions. Reproduced from Ref. [8]. **b)** O-PTIR signal distribution around a 500-nm-sized polymer bead, showing that there is now signal at the location of the absorber and two maxima above and below. Reproduced from [13] **c)** White light s-SNOM image of a lipid vesicle recorded in water. **d)** Spectra taken at red cross marker in **(c)** at 465 nm or 365 nm. **(c)** and **(d)** reproduced from Ref. [11].

All three techniques enable label-free chemical imaging at high spatial resolution and have shown promising capabilities for biomedical applications. Offering sub-micrometer resolution and the ability to operate in aqueous environments, O-PTIR has been used to study fresh mouse biopsies [6], live single cells [2] and microcalcifications in breast cancer [7]. AFM-IR nanoscale imaging has been applied to the direct imaging of organelles, human milk extracellular vesicles, breast cancer calcifications and the detection of bacteria inside of human cells [8] (see figure 1 a), among others. While s-SNOM is mainly used to characterize plasmonic structures with prospective applications as surface enhanced Raman spectroscopy (SERS) or surface enhanced infrared spectroscopy (SEIRA) substrates in medical sensors, s-SNOM has also been leveraged to study biological specimens, including Si-particles in cancerous cells [9], radiation-induced damages in collagen [10], single proteins, bacterial cells and vesicles [11] (see Fig. 1c,d).

### Current and future challenges

Regarding the robust and reproducible implementation of all three above-mentioned technologies for biomedical applications, two main challenges remain: i) imaging/signal artifacts, and ii) low throughput capabilities. Indeed, the high spatial resolution chemical imaging capabilities of O-PTIR, AFM-IR, and s-SNOM rely on indirect, complex detection mechanisms that are prone to technique-specific artifacts and signal instabilities. O-PTIR and AFM-IR have both complex signal transduction chains [3] due to their IR absorption photothermal-based detection scheme. Consequently, in addition to the pure absorption information – with transduction efficiency depending on the sample's geometry, mechanical and photothermal properties – O-PTIR and AFM-IR signals can be impeded by technique-specific parameters, such as the tip-sample interaction for AFM-IR [12], or probe laser optical effects for O-PTIR [13,14] (see Fig. 1b). Despite these experimental considerations, O-PTIR and AFM-IR, unlike s-SNOM, provide local spectra that can be directly compared to conventional far-field reference IR spectra. Since s-SNOM indirectly probes the sample´s dielectric function through the light scattered by an oscillating tip at the sample's surface, mathematical models are required to retrieve the sample's pure absorption spectra [15]; these models are typically developed for homogeneous materials with simple geometries and not for biological materials with high local variations and surface roughness [16].

With regards to throughput, AFM-IR and s-SNOM are scanning probe microscopy techniques and are limited by their electronic feedback loop and scanning speed to accurately track the sample's surface. For a low-roughness sample, a typical single-wavelength 10 µm x 10 µm AFM-IR/s-SNOM image with a low spatial sampling rate (256 pts/line) will take about 5 minutes. If images at multiple wavelengths are required, a strategy of acquiring a full spectrum at each pixel may be beneficial, resulting in significantly longer measurements, as each spectrum takes approximately 30 seconds. In contrast, O-PTIRO-PTIR typically offers faster acquisition times due to its lower spatial resolution. For instance, a single-wavelength image of 100 µm x 100 µm (500 pts/line) will take a few minutes. Widefield techniques for O-PTIR have been described which trade spatial resolution for the ability to collect a large-scale image at once [17].

All three techniques are currently used in research laboratories for fundamental biomedical research but are not yet routinely employed for medical or clinical applications. To do so, the previously mentioned challenges will need to be addressed, sample handling/preparation developed, and standardized and automated experimental protocols implemented.

### Advances in science and technology to meet challenges

For the key challenges of optical artifacts in the O-PTIR signal, work has already started to describe and understand them [13,14]. However, it is currently unclear whether these artifacts are best minimized through experimental protocols, through different optical designs or a combination of both. Quantitative-phase-imaging-based approaches for O-PTIR signal transduction have been implemented and may see reduced artifacts; while the fluorescence-detected O-PTIR variant is not affected by standing wave artifacts (frequently observed with O-PTIR), in contrast to conventional O-PTIR it requires fluorescence labelling in samples without autofluorescence [17].

Several groups are studying thermoelastic effects in AFM-IR imaging [18]. Here, interestingly, a better understanding of the contribution of such effects to the AFM-IR signal could not only lead to improved data quality but could also enable new AFM-IR imaging modes, such as depth-resolved and surface-sensitive AFM-IR [19]. Among new AFM-IR modes for improved performance, a novel “null-deflection" AFM-IR mode has been developed, demonstrating enhanced AFM-IR imaging quality through the elimination of contact stiffness and tip amplitude artifacts [20]. However, improvements will be required to enable the technique to operate at the same throughput as conventional AFM-IR.

While technique-specific limitations are being addressed, with most notably theoretical studies improving comprehension of signal-transduction, O-PTIR, AFM-IR, and s-SNOM will all benefit from improvements in tunable mid-IR laser technology. For instance, a more stable laser emission profile would contribute to reducing power background effects in spectra, while a broader tuning range would allow for a more exhaustive description of the sample's chemistry compared to the restricted spectral range of most currently employed tunable mid-IR sources. Finally, for all three technologies, community efforts will be required to develop standardized protocols, which we consider the main remaining barrier to ensuring measurement comparability and reproducibility among research groups. This is a key step for the routine use and broad acceptance of these modalities in clinical research.

### Concluding remarks

O-PTIR, AFM-IR, and s-SNOM are three complementary techniques that offer chemical imaging beyond the optical diffraction limit, providing a samples' molecular description at the nanoscale level as well as subcellular chemical imaging capabilities, which are drawing strong interest for both fundamental biomedical research and clinical applications. To get there nonetheless, significant work is needed regarding both instrumental technological advances and the theoretical description of the physical concepts behind the detection mechanisms of each technique, for robust and artifact-free data acquisition. To ensure reproducibility, further standardization and coordination of protocols are also needed. However, once those challenges are overcome, O-PTIR, AFM-IR, and s-SNOM are promising techniques for biomedical applications, providing label-free molecular chemical information at a spatial resolution that is not achieved by any other modality.

### Acknowledgements

The financial support by the Austrian Federal Ministry of Labor and Economy, the National Foundation for Research, Technology and Development, and the Christian Doppler Research Association is gratefully acknowledged. G.R., M.P., and B.L. acknowledge funding from the Austrian Science Fund (FWF) [doi.org/10.55776/COE7]. This project has received funding from the European Union’s Horizon 2020 research and innovation program within the research project “Tumor-LN-oC” under grant agreement no. 953234.

# 13. Micro- and nanoplastics detection

**Markus Brandstetter[1,2], Kristina Duswald[1,2], Verena Karl[1], Florian Meirer[3] and Lukas Kenner[2,4,5,6]**

[1] Research Center for Non-Destructive Testing, Linz, Austria
[2] Center for Biomarker Research in Medicine, Graz, Austria
[3] Institute for Sustainable and Circular Chemistry, Utrecht University, Utrecht, Netherlands
[4] Clinical Institute of Pathology, Medical University of Vienna, Vienna, Austria
[5] Comprehensive Cancer Center, Medical University Vienna, Vienna, Austria
[6] Department of Molecular Biology, Umeå University, Umeå, Sweden

E-mail: markus.brandstetter@recendt.at

### Status

Microplastics (<5 mm) and nanoplastics (<1 μm), collectively referred to as MNPs, have been detected in multiple human biological matrices, highlighting their health relevance. Their presence across various body compartments, including the gastrointestinal tract, the circulatory system, and human biofluids indicate systemic exposure and potential translocation across epithelial barriers [1]. In vivo experiments in mice have shown that nanoparticles can cross the blood brain barrier, while other research has demonstrated cellular uptake of MNPs and translocation to daughter cells during cell division, implying a potential role in tumor progression [2]. Supporting this, a recent study on bladder cancer revealed MNPs in a notable share of bladder tissue specimens [3]. Additional studies further suggest that ingested MNPs might lead to a potential risk of activating inflammatory factors, induce oxidative stress and contribute to dysbiosis in the intestinal microbiota [4]. Leading research concludes that MNPs represent a plausible risk factor for human health and that current exposure levels warrant precautionary investigation [1].

From the very beginning of MNP research, FTIR and Raman microspectroscopy have become the workhorse for non-destructive polymer particle identification in complex matrices, complementing mass-spectrometric techniques, such as pyrolysis GC-MS. In contrast to the latter, microspectroscopic approaches provide information on particle counts, size and morphology and have already become a routine analytical tool in environmental MNP studies. The requirements for sample processing are more stringent than for mass spectrometry. Concerning this issue workflows must be developed and validated separately for different biological matrices. Recently, advanced laser-based methods have emerged, such as direct infrared, photothermal and near-field imaging. These techniques promise to collectively span diverse size ranges, resolutions, and measurement speeds. Further advances in infrared photonics and correlative approaches will be essential to move from proof-of-principle detection towards quantitative, clinically interpretable biomarkers for the MNP body burden.

### Current and future challenges

**Ultra-trace concentrations:** Due to the low concentrations of MNPs found in both biological and environmental samples, pre-concentration is often needed. Efficient removal of proteins, lipids and cellular debris without altering polymer chemistry or particle morphology remains difficult. Although chemical digestion effectively removes organic residues and isolates MNPs, the harsh conditions might alter the particles morphology slightly. Enzymatic purification protocols developed for environmental samples are promising [5] but need adaptation and validation for human derived samples [6]. **Size ranges and nanoplastics:** Established infrared spectroscopic workflows achieve reliable identification for particles above 10 μm, which critically limits the detection range towards smaller particles. Yet toxicologically relevant fractions may extend into the sub-μm regime. Raman imaging systems offer higher resolution but struggle with several key limitations like fluorescence of biological content, and low signal-to-noise ratios for nanoplastics embedded in complex matrices [7]. First successful attempts to address the sub-μm region with optical photothermal (O-PTIR) and near-field techniques were reported, but require further consolidation [7,8]. **Metrological gaps**: There is still no harmonized set of reference materials that encompass various weathering pathways and sample digestion protocols, proficiency tests, or even agreed reporting units for MNPs in human specimens (polymer mass, particle numbers, surface functionalization, etc.). Recent reviews emphasize poor comparability between studies due to differences in sampling, purification protocols, size binning and spectral post-processing. Particle number, mass, surface area and polymer-specific metrics are rarely reported together, hampering efficient risk assessment [9]. **Spectral interpretation and data overload:** High-throughput infrared and Raman imaging generate large hyperspectral datasets. Manual inspection and simple library matching are prone to bias and false positives, particularly in tissue sections where biogenic materials and additives produce overlapping bands. Polymer additives and environmental weathering can substantially alter the chemical appearance of MNPs, leading to oxidation, additive leaching and shifts in characteristic spectral bands, which further complicate spectral identification[10]. Advanced chemometrics and machine learning can help but require curated training data and transparent validation [11,12]. **Linking exposure to health effects:** While MNPs have been associated with oxidative stress, inflammation processes, altered metabolic and cellular functions [1], causality remains uncertain. Key challenges are co-exposure to other pollutants, temporally variable exposure histories and the difficulty of obtaining matched control tissues. **Translation into healthcare:** For proactive and predictive health management, analytical methods must evolve from research-grade, labor-intensive protocols towards robust, partially automated workflows with defined quality assurance. Integration into biobanking, pathology and occupational health monitoring raises additional demands on sample throughput, cost, and regulatory acceptance. Addressing these challenges will require close collaboration between environmental scientists, spectroscopists, clinicians, toxicologists and standardization bodies.

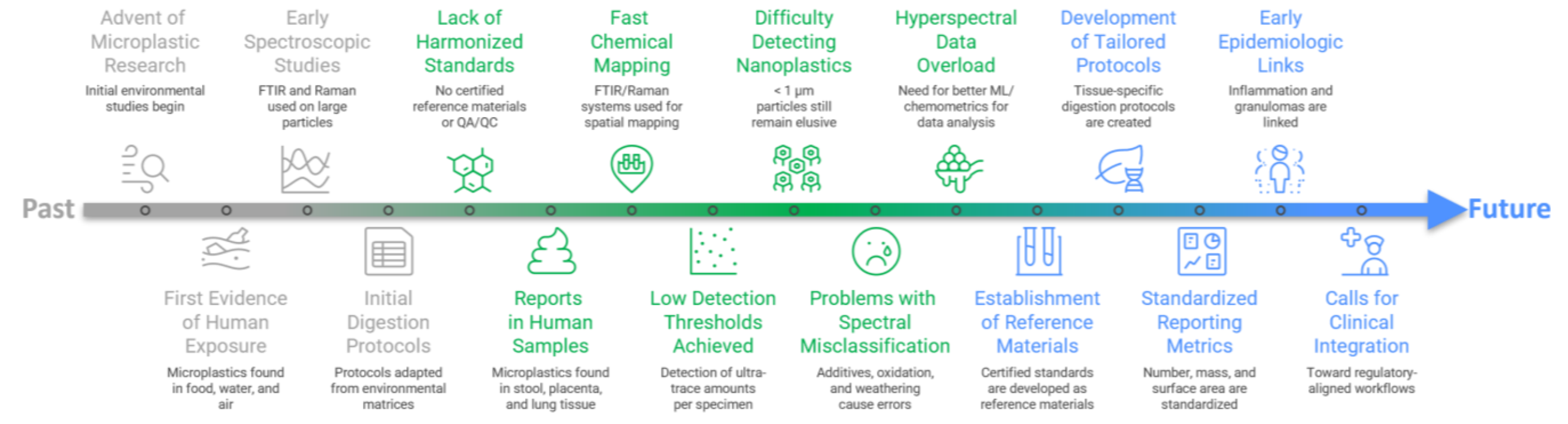

**Figure 1.** Timeline of micro- and nanoplastics detection in human tissues

## Advances in science and technology to meet challenges

**Improved sample pre-processing:** Reliable MNP detection in human matrices depends on removing biological complexity without altering particle properties [13]. Advanced protocols focus on optimized enzymatic digestion, density separation, and microfluidic MNP enrichment for blood, placenta, lung, and stool samples while preserving nanoplastics and minimizing contamination [14]. Isotope-labelled controls, clean-lab conditions, and blanks are essential for recovery quantification and quality assurance. Integration into reproducible pipelines is key for scaling and clinical or toxicological compatibility [13,14]. **Tackling the Nanoscale:** To address the nanoscale regime high-resolution techniques such as O-PTIR and its advancements like photothermal-induced fluorescence and stimulated Raman scattering, as well near-field nano-spectroscopy techniques (AFM-IR, PiFM) are required [7,13]. These tools enable ultra-trace particle characterization at sub-μm resolution with both chemical identification and precise localization in tissue - crucial for early exposure diagnostics and retrospective health monitoring [8,13]. Further, nano-electromechanical FTIR represents a promising advancement that allows picogram-level collection and mass-based spectroscopic characterization [15]. **High-throughput analysis**: This remains a bottleneck, as high resolution limits scan area and speed [13,16]. For particles larger than 10 μm, laser direct imaging solutions have been demonstrated [17], while in the <10 µm and submicron range, promising advances like fast-scanning O-PTIR now enable video-rate imaging with high spatial detail [16]. Implementation in clinical routine will require robust, automated, and contamination-controlled workflows - prioritizing not only resolution, but also reproducibility, throughput, and compatibility with clinical and biobanking pipelines [13,14]. Fluorescence-based techniques - including CLSM, STED, and photothermal-induced fluorescence offer a promising complementary approach for scalable, high-throughput workflows, pending validation against label-free methods [13,18]. **Multimodal and Correlative Imaging:** No single technique can capture the full complexity of MNPs. Future strategies will combine fast scanning methods to survey large areas at lower spatial resolution and flag suspicious particles, with targeted chemical identification at high spatial resolution [13,14]. This selective 'zoom-in' workflow improves efficiency, preserves spatial context, and supports meaningful interpretation of potential biological effects [13,14]. **Machine Learning and Automation:** Machine learning tools can efficiently handle the complexity of multimodal imaging data[12,13]. Pattern recognition, spectral unmixing, and anomaly detection, trained on representative libraries, will improve particle identification and analytical precision [11,12,19]. Automated segmentation and classification support reproducibility, scalability, and future regulatory applications [12]. **Standardization and Translation:** Future progress depends on certified reference materials, harmonized spectral libraries, and interoperable data formats [11,13,20]. Standardized protocols for sampling, digestion, measurement, and interpretation will improve comparability and support cross-study synthesis [14,21]. Aligning with FAIR principles and promoting open science through shared datasets, transparent workflows, and accessible tools will enhance collaboration, reproducibility, and broader use across disciplines [11,14].

### Concluding remarks

The detection of MNPs in across human tissue types and biofluids shifted the field from environmental observations to a pressing biomedical concern. Infrared spectroscopy methods together with mass-spectrometric methods remain essential while emerging laser-based infrared-photonics platforms, based either on direct absorption or photothermal effects, are currently extending sensitivity into the sub-micrometer and nanoscale domain. Advancing proactive and predictive health management requires robust, standardized and contamination-controlled workflows tailored to human specimens, alongside the integration of advanced photonic instrumentation into clinically compatible pipelines. Interoperability between infrared methods, Raman spectroscopy, fluorescence imaging, and mass-based techniques, including co-registration with histopathology, will determine whether biologically meaningful and clinically actionable information can be extracted. Equally critical are machine learning–based spectral interpretation, automated particle identification, and high-throughput Raman and IR imaging to address the data complexity and uncertainty associated with low-level human exposure. If these analytical, computational, and clinical strands are aligned, infrared photonics is ready to deliver label-free, chemically specific microscale and increasingly nanoscale insights into the presence and fate of MNPs in human tissues - thereby strengthening risk assessment, exposure mitigation and future regulatory decision-making in healthcare.

### Acknowledgements

The authors acknowledge financial support by the project microONE, within the COMET Module programme by the Federal Ministry of Economy, Energy and Tourism (BMWET), the Federal Ministry of Innovation, Mobility and Infrastucture (BMIMI), Land Steiermark (Styrian Business Promotion Agency – SFG) and Land Wien (Vienna Business Agency – WAW). The COMET Module programme is executed by the Austrian Research Promotion Agency (FFG). Further support by research subsidies granted by the government of Upper Austria (Grant Nr. FTI 2022 (HIQUAMP): Wi-2021-303205/13-Au) is acknowledged.

# Whole body:
# 14. FTIR-based exhaled breath analysis: Status and future perspectives

**Gabriela Flores Rangel[1], Boris Mizaikoff[1]**

[1] Institute of Analytical and Bioanalytical Chemistry, Ulm University, Albert-Einstein-Allee 11, 89075 Ulm, Germany

E-mail: gabriela.flores-rangel@uni-ulm.de ; boris.mizaikoff@uni-ulm.de

## Status

Breath analysis has become an increasingly important non-invasive diagnostic modality, as it provides direct access to metabolic, inflammatory, and microbiological processes via the analysis of exhaled gases. Human breath contains a complex mixture of inorganic gases, isotopologues, and volatile organic compounds (VOCs), many of which have been identified as clinically relevant biomarkers for respiratory diseases, gastrointestinal disorders, metabolic dysfunctions, and infectious conditions **[1–3]**.

Among the available analytical techniques, Fourier Transform Infrared (FTIR) spectroscopy represents one of the most mature and robust fundamental platforms for breath analysis. By probing the mid-infrared (MIR) fingerprint region, FTIR enables chemically specific and simultaneous detection of multiple breath constituents within a single measurement **[4–6]**. In contrast to narrowband sensing approaches, broadband FTIR spectroscopy allows comprehensive spectral access facilitating multi-analyte quantification, internal referencing, and effective compensation for interfering species such as water vapor and background carbon dioxide.

FTIR-based gas analysis benefits from a long-standing history in industrial process control, environmental monitoring, and laboratory spectroscopy. This extensive technological background translates into high technology readiness levels (TRLs) for breath analysis applications, where reproducibility, long-term stability, robustness during field deployment and quantitative reliability are critical. Established calibration strategies, traceability to reference spectral databases such as HITRAN, and standardized data processing workflows further consolidate FTIR as a reference technology for clinical and translational studies **[7]**.

From an application perspective, FTIR spectroscopy is particularly well suited for scenarios requiring comprehensive breath profiling, longitudinal monitoring, and retrospective data analysis. Full spectral data acquisition preserves the complete chemical information content of breath samples enabling the development and evolution of multivariate and chemometric models that are increasingly relevant to extract clinically relevant insights from complex spectroscopic breath signatures **[8–10]**.

## Current and future challenges

As is the case for any spectroscopic platform transitioning from laboratory environments to applied and clinical use, FTIR-based breath analysis faces technical and application-related challenges that potentially limit their broader deployment **[11]**. One of the main constraints remains the system footprint. Conventional FTIR instruments rely on light source, interferometer and detector assemblies, which may limit miniaturization and portability. In addition, sensitive MIR detector technologies require cooling, which increases the power consumption and system complexity, particularly for mobile or point-of-care/point-of-need implementations.

Sensitivity represents another relevant consideration, especially for the detection of ultra-trace biomarkers present at sub-ppm, ppb or even ppt concentration levels. While FTIR spectroscopy provides broadband spectral coverage and multi-analyte capability, the fundamental operational principle inherently distributes the optical power across a wide spectral range. This may lead to reduced signal-to-noise ratios at selected target wavelengths bands when compared to targeted narrowband approaches using MIR laser technologies particularly in applications focused on single analytes at exceedingly low concentrations **[12,13]**.

Additional challenges arise from breath sampling itself. High humidity levels, variable flow conditions, and pronounced patient-to-patient and within-patient variability complicate quantitative analysis and require robust sampling interfaces and correction strategies **[14,15]**. Hence, even though FTIR spectroscopy allows compensation for interfering species via appropriate spectral modeling and reference-based approaches, these practical aspects continue to require careful system design and optimization facilitating the translation into widespread clinical practice **[16]**.

It is important to note though that the majority of these challenges are related to system engineering and integration aspects rather than fundamental limitations of FTIR spectroscopy. Consequently, ongoing developments in optical design, detector performance, compact interferometer concepts, and advanced data processing are progressively addressing these constraints and defining promising pathways toward application-oriented and targeted usage of FTIR technologies in exhaled breath analysis.

### Advances in science and technology to meet challenges

A range of technological developments have contributed to addressing practical limitations associated with FTIR-based exhaled breath analysis and have expanded its applicability beyond laboratory settings. A major development was the emergence and evolution of miniaturized photon-gas interaction concepts, including specifically hollow waveguides (HWGs) and substrate-integrated hollow waveguides (iHWGs). These structures facilitate extended effective optical absorption path lengths while probing exceedingly small and rapidly exchangeable gas sample volumes, thereby enabling continuous sensing/monitoring of exhaled breath with a time resolution at the level of seconds while preserving the analytical performance [8,17]

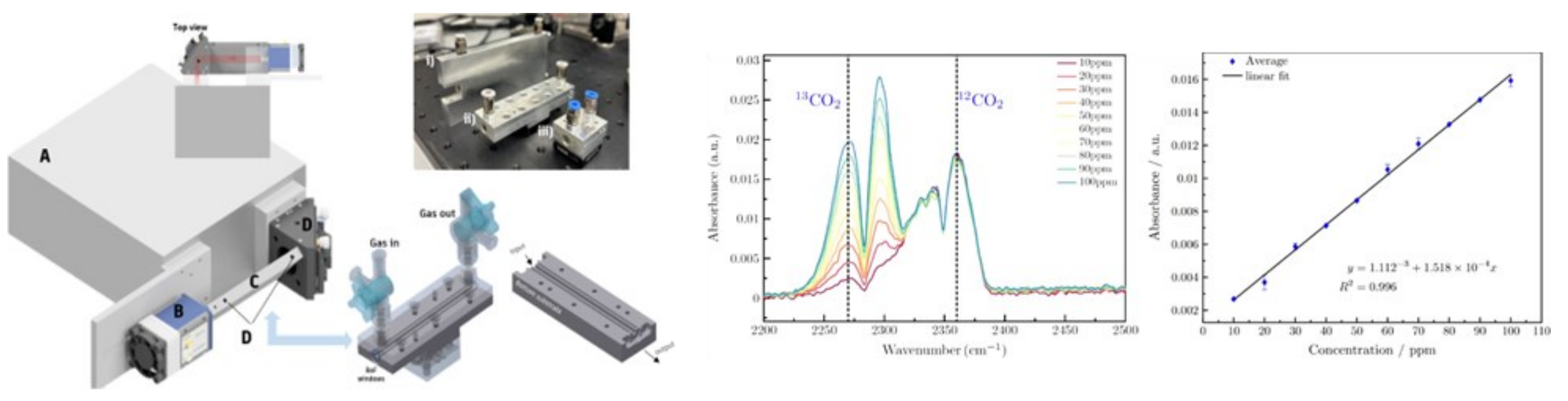

**Figure 1.** Current state of FTIR-based breath analysis using hollow and integrated hollow waveguide gas interaction concepts. Shown are representative experimental implementations interfaced with laboratory-scale FTIR spectrometers, together with selective spectral discrimination and linear quantitative response for $^{13}CO_2/^{12}CO_2$. These results demonstrate how compact iHWGs provide a validated analytical baseline and a modular pathway toward future miniaturized mid-infrared breath-sensing platforms. Adapted from Flores Rangel et al., ACS Sensors, 2025 (CC BY 4.0).

In parallel, advances in MIR fiber and waveguide technologies support more flexible optical layouts and compact system architectures. Advances in silver-halide fibers, chalcogenide fibers, and substrate-integrated hollow waveguides have broadened the available toolbox for MIR light guiding and gas sensing enabling FTIR system designs that better align with application-driven system requirements **[18]**

Detector technologies have also progressed with improvements in sensitivity, response time, and noise characteristics reported for several MIR detector concepts. Together with advances in signal processing and multivariate/chemometric data analysis and mining, these developments contribute to an improved robustness and lower detection limits without requiring fundamental changes of the fundamental FTIR optical architecture **[19]**.

From a clinical and application-oriented perspective, the analytical performance alone is rarely sufficient to ensure the suitability of a breath analysis technology. In practice, parameters such as the response time, robustness at variable sampling conditions during deployment, and low maintenance requirements play a decisive role for routine clinical use. Equally important are standardized measurement procedures and data outputs that can be interpreted in a transparent and clinically meaningful way. In this regard, FTIR-based systems indeed benefit from mature instrumentation, robust calibration strategies, and well-established data analysis frameworks, which support reproducible measurements and facilitate longitudinal and comparative clinical studies.

Within this technological landscape it is noteworthy that laser-based MIR sources including quantum cascade lasers (QCLs), interband cascade lasers (ICLs), and emerging interband cascade LEDs (IC-LEDs) are increasingly explored as complementary components for specific sensing tasks aiming at target species analysis also in exhaled breath matrices **[20]**.

### Concluding remarks

The capability to detect multiple breath constituents simultaneously, to account for interfering species, and to support multivariate and chemometric data analysis strategies renders FTIR techniques well suited for clinical and translational application scenarios where robustness and data interpretability are required.

Within the broader landscape of mid-infrared exhaled breath sensing, broadband FTIR spectroscopy and laser-based approaches address fundamentally different, yet complementary, application scenarios. FTIR systems offer full spectral coverage and robust quantitative performance, which makes them particularly suitable as reference platforms for multi-analyte analysis, method development, and clinical validation studies. Conversely, laser-based sensors trade spectral breadth against selectivity and efficiency enabling compact and energy-efficient solutions tailored to specific biomarkers or use cases. Rather than replacing broadband platforms, these targeted approaches are expected to build upon and coexist with FTIR-based systems contributing to a layered sensing ecosystem that spans from laboratory-grade reference measurements to application-specific, portable or even wearable implementations.

While challenges related to system footprint, sensitivity in specific use cases, and power consumption remain, ongoing developments in advanced gas cell concepts, waveguide technologies, detector performance, and data analysis methods continue to broaden the range of practical FTIR breath analysis applications. In parallel, narrowband mid-infrared light sources such as QCLs, ICLs, and integrated IC-LEDs are being explored as complementary technologies for targeted sensing tasks and future wearable or near-body implementations. From a roadmap perspective, FTIR-based systems provide a stable analytical reference for the development and validation of emerging breath sensing approaches. Together with application-specific laser-based technologies, FTIR contributes to a layered breath analysis ecosystem in which comprehensive spectral platforms and targeted sensing solutions address different application needs.

### Acknowledgements

The authors acknowledge support from the EU Horizon 2022 project M3NIR (Grant Agreement No. 101093008)

# 15. Breath analysis via infrared spectroscopy: Innovations toward clinical implementation

**Mohamed Sy[1], Aamir Farooq[1,*]**

[1] King Abdullah University of Science and Technology (KAUST), Thuwal 23955-6900, Saudi Arabia.

E-Mail: *aamir.farooq@kaust.edu.sa

## Status

Human breath contains a diverse mixture of volatile organic compounds (VOCs) that provide a non-invasive window into physiological and pathological processes [1]. Conventional platforms such as Gas chromatography–mass spectrometry (GC–MS)[2, 3] and selected ion flow tube mass spectrometry (SIFT–MS) [4] remain analytical gold standards, but their size, cost, and operational complexity prevent routine clinical adoption. Portable electronic-nose systems offer faster measurements but lack the specificity and long-term stability needed for diagnostic reliability due to sensor drift and weak discrimination in complex mixtures [5]. Consequently, despite its diagnostic potential, exhaled breath analysis has historically remained constrained to research environments. In recent years, advances in mid-infrared (mid-IR) photonics have begun to shift this landscape toward compact, high-performance breath analyzers. Quantum and interband cascade lasers now provide high-resolution access to the molecular fingerprint region, while multipass and cavity-enhanced architectures deliver sub-ppb sensitivities in portable formats [6, 7]. Figure 1 outlines this emerging workflow, linking non-invasive breath collection to infrared spectral acquisition, machine-learning-based feature extraction, and cloud-enabled health assessment. Complementing these hardware developments, progress in machine learning and signal processing has significantly enhanced the interpretability and robustness of breath spectra. Data-driven models can isolate overlapping signatures, compensate for physiological variability, and detect weak absorbers under realistic clinical conditions [8-10]. They also enable quantitative analysis when reference spectra are incomplete or entirely unknown [9]. Combined, these photonic and computational advancements are shaping a new generation of deployable IR breath analyzers capable of delivering real-time, clinically actionable diagnostics.

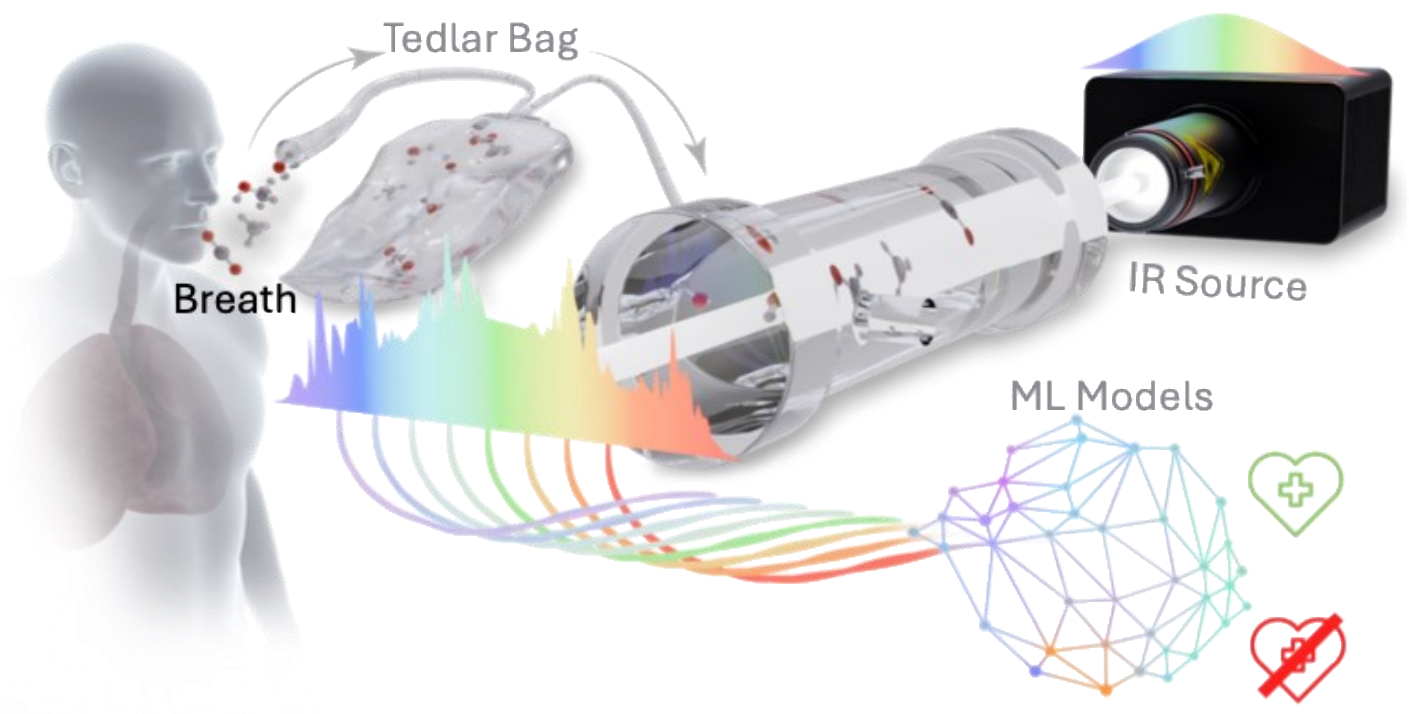

**Figure 1.** Workflow of breath-based IR diagnostics from breath collection to mid-IR analysis and ML-based biomarker extraction.

## Current and future challenges

Despite notable progress, breath-based IR diagnostics still faces challenges across chemistry, instrumentation, computation, and clinical translation [11]. Breath composition is fundamentally complex. Hundreds of VOCs appear at trace levels with overlapping absorption features, and factors such as diet, medication, and environmental exposure introduce variability that can obscure disease-specific patterns.

Water vapor adds further difficulty. As the dominant absorber, it can mask low-concentration biomarkers and requires suppression or correction strategies that remain imperfect [12]. Additional noise sources such as laser drift, detector instability, and temperature or humidity fluctuations also limit reproducibility. Standardization remains limited. Consistent sampling procedures, certified calibration mixtures, and shared spectral datasets are scarce, hindering comparability across sites and slowing clinical and regulatory progress [13]. Machine-learning approaches must also contend with heterogeneous data and the need for clinical interpretability. Limited cohort sizes and inconsistent sampling conditions constrain robust model developments [14-17]. Deployment and clinical integration introduce practical constraints. Devices must be compact, low-power, and reliable for decentralized use, while integration into clinical workflows requires secure data handling and demonstrated diagnostic utility. Regulatory pathways for breath biomarkers are still evolving. These challenges highlight the need for greater reliability, scalability, and standardization to support a transition from laboratory prototypes to clinically deployable systems.

## Advances in science and technology to meet challenges

Overcoming these barriers requires coordinated progress across photonics, computation, manufacturing, and clinical validation. Photonics innovation remains central. Broadband mid-IR laser sources combined with long-path or cavity-enhanced architectures are essential for resolving complex mixtures and achieving ppt–ppb sensitivities [7]. Mid-IR photonic integrated circuits (PICs) offer improved stability, manufacturability, and compactness [18]. Machine learning will play a critical role. Robust spectral-unmixing pipelines, feature-engineering strategies [15], denoising autoencoders [8], and interference-resilient models [16] enhance reliability under realistic conditions. Explainable AI methods [17] will support clinician trust and regulatory review. Standardization is a key enabler. Harmonized sampling procedures, calibration-free sensing approaches, and open spectral databases [19] are required to ensure reproducibility. These needs parallel the foundry-style frameworks that accelerated scalability in integrated photonics. Packaging, portability, and system engineering must also advance. Low-power lasers, microfabricated detectors, and integrated readout electronics will support deployment in clinics and homes. Figure 2 summarizes the hardware, software, and workflow components required for scalable clinical validation and device translation. Coupled with secure mobile-health platforms, these technologies can extend breath diagnostics to continuous monitoring and population-scale health assessment [20].

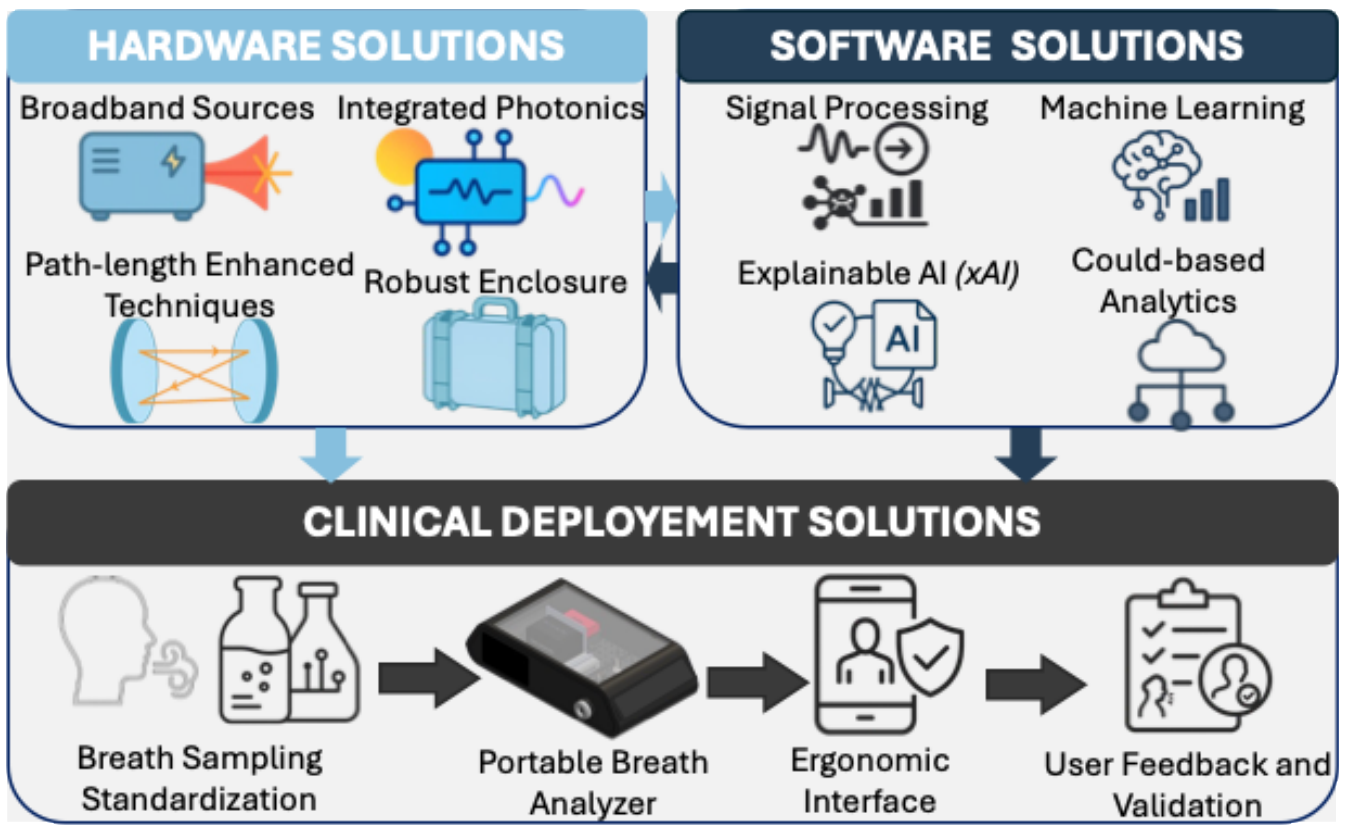

**Figure 2.** Hardware and software solutions enabling clinical deployment and validation of breath-based infrared diagnostics.

## Concluding remarks

Breath-based infrared diagnostics is poised to become a key component of personalized and preventive healthcare. Photonic miniaturization and AI-driven analytics have produced compact prototypes capable of detecting clinically relevant biomarkers in real time. For widespread adoption, the field must continue addressing spectral interference, physiological variability, calibration, and clinical validation while ensuring manufacturability and usability. These needs mirror the coordinated development trajectory seen in integrated photonics roadmaps. With sustained interdisciplinary progress, breath-IR sensing can advance from research instrumentation to scalable clinical tools capable of early disease detection, continuous health monitoring, and widespread non-invasive diagnostics.

**Acknowledgements**

We acknowledge the funding received from King Abdullah University of Science and Technology (KAUST).

# 16. InfraRed-omics as an integrative framework for molecular fingerprinting and medical health phenotyping

**Liudmila Voronina[1,2], Marinus Huber[1,2], Tarek Eissa[1,2], Katharina Dietmann[1,2], Lorenzo Gatto[2], Mihaela Žigman[1,2, *]**

[1]*Ludwig-Maximilians-Universität München (LMU), Garching, Germany*
[2]*Max Planck Institute of Quantum Optics (MPQ), Garching, Germany*

*E-Mail:* *mihaela.zigman@mpq.mpg.de

## Status

Infrared molecular fingerprinting, the comprehensive infrared vibrational spectroscopic profiling of organic molecules within a biological system, can yield valuable insights into physiological function in both health and disease. While genomics and transcriptomics reveal cellular potential, metabolomics focuses on investigating small-molecules, proteomics captures proteins as functional effectors that drive physiology. In contrast, "*InfraRed-omics*" captures the integrated spectroscopic signature of a system as a whole. Although it is not suited for pinpointing individual molecular changes, InfraRed-omics has inherent advantages for profiling complex molecular matrices [1]. When applied to blood as a systemic biofluid, the resulting "*infrared molecular fingerprint*" integrates functional information about the processes in the human body and provides a snapshot of the immediate condition of the system (Fig. 1). As the direct outcomes and effectors of biological functions and phenotypes, these pan-molecular infrared fingerprints inherently reflect the complex interplay between genetic background, a person's demographics, environmental factors and lifestyle influences. Thus, if adequately established, infrared molecular fingerprinting can, in alloy with multivariate analysis and machine learning, predict medically meaningful health outcomes.

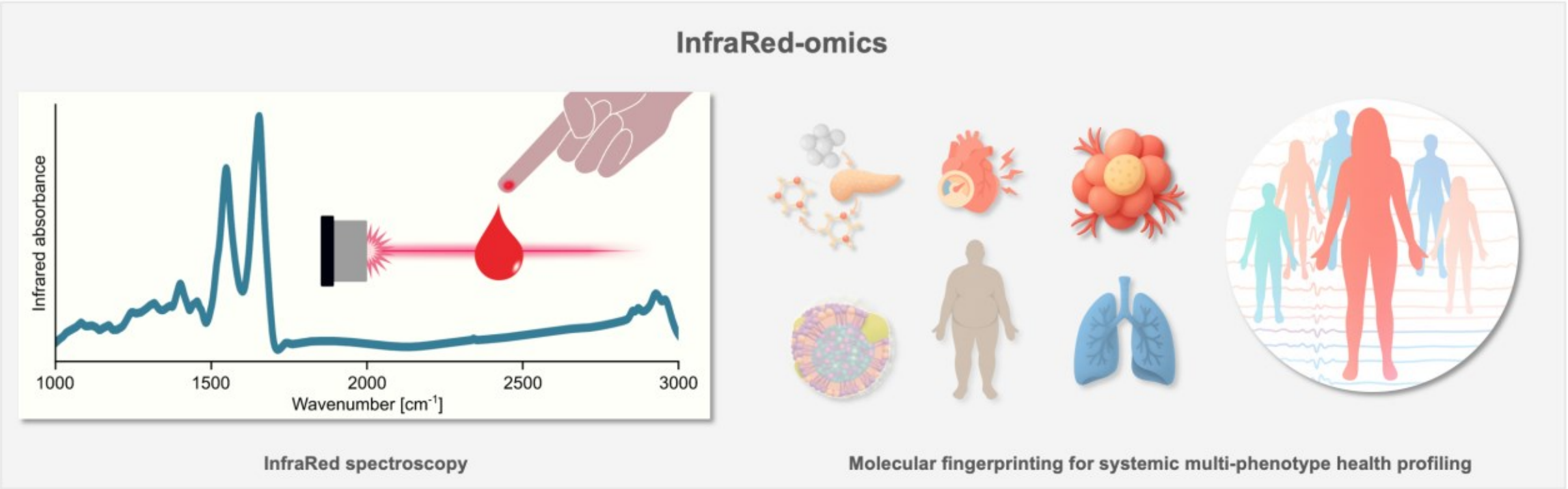

**Figure 1.** The concept of infrared molecular fingerprinting fueling medical health profiling. Average mid-infrared absorbance spectrum of cell-free blood (adapted from Ref. [2]). Demonstrated capacity of infrared molecular fingerprinting to detect a variety of health states, depicting deviations in endocrine-metabolic activity, hypertension, presence of tumor growth, molecular changes reflecting lipid particles, multimorbid metabolic syndrome and functional airway load.

## Current and future challenges

How different are we? The quest to use systemically circulating molecules to characterize human health has been driving extensive scientific efforts. How much of this variation reflects genetics, environment, behavior, diet or even chance? Advances in reproducible infrared spectroscopy workflows [3,4], especially when coupled with machine-learning analytics [5], now allow deeper interrogation of how our systems function and how complex molecular patterns in fluid [3,4,6] and solid tissues [7] reflect health and disease

Most rare and common chronic conditions are not only dynamic and heterogenic, but also multifactorial and develop over heterogeneous trajectories [8-10]. This raises practical questions: How could we effectively define disease, optimize prevention and treatment, and monitor responses and long-term health, given all the biological diversity as well as dynamics? It is ever more seen that many strategies must be tailored - “personalized” or “individualized” - to a person’s genetic, molecular, physiologic, behavioral and exposure profiles [9,17,18]. Modern medicine emphasizes diagnostics and interventions tailored to each person, recognizing that even modest improvements can be clinically meaningful. If the overarching goal is to maintain and restore health across the “*healthspan*” (period of life free from disease), precision approaches often combine highly individualized as well as some more systemic interventions in a coordinated manner. Because these and such strategies are empirical and context dependent, establishment of robust evidence on how to deliver precise preventive and therapeutic care at the level of an individual is fundamentally required. With all that in mind, tackling best ways to compare individuals and health states, identify meaningful subgroups, and design truly personalized interventions remains an immense open challenge.

Within this landscape, conceptual and empirical results of our team suggest that infrared molecular fingerprinting could serve as an enabling technology for modern medicine. The method is condition- and molecule-agnostic yet phenotype-specific, enables high-throughput measurements, requires only minimal pre-analytical sample preparation, low sample volume and low cost. It may support precision health monitoring, identify individuals early in disease etiology who could benefit from specific interventions, effectively stratify populations and identify at-risk individuals even within naturally heterogeneous populations and improving the success rate of clinical trials. All these and such applications, however, require rigorous experimental evaluation and independent validation to establish their true clinical value, medical utility and long-term holistic as well as socioeconomic benefits.

**Advances in science and technology to meet challenges**

**Profiling composite physiological health and disease states:** Quantitative profiling of health begins with establishing well-characterized reference conditions, which serve as comparator states needed to identify deviations from *healthy* physiologies as experimental seed data. In practice, this centers on assembling human biospecimens together with rich medical parameterization, and framing these as datasets to interrogate specific clinical questions. Because human systems change over the course of life and even within an individual’s adult health span, sufficiently powered clinical studies are needed, even at exploratory stages, to draw meaningful conclusions. Any promising findings must then undergo independent testing and clinical validation before they can be considered further.

A central objective is to test which health deviations can be detected through infrared molecular fingerprinting of cell-free blood. In a naturally heterogeneous adult population with more than 5.000 samplings, we found that infrared fingerprints can capture and distinguish commonly occurring health phenotypes [2]. Intriguingly, we were able to reliably single out individuals classified as healthy within this diverse cohort. In a clinical case-control study setting involving over 2.500 participants, we further showed that bulk blood plasma infrared molecular fingerprints contain signatures associated with common cancers [11]. We identified infrared patterns specific to therapy-naïve lung, prostate, breast and bladder cancer, distinguishing them from matched controls, with diagnostic performance tracking the stage of cancer progression [11]. In parallel, infrared molecular fingerprinting was proposed as a tool for rapid triage of brain cancer [12].

Realistic profiling of commonly occurring physiologies necessitates quantifying the *composite* health states, just as most physiological states come with co-occurring “passenger” deviations, comorbidities and risk factors. Multi-molecular, systemic infrared profiling may be particularly suited to address the realm of multimorbidity (coexistence of interrelated health states in an individual), which remains among the major challenges in clinical medicine. Although multimorbidity increases with age, more than half of all people with multimorbidity are younger than 65 years [19]. The appropriate management of such disorders is a key challenge for health systems, dominated by single-disease approaches that are increasingly inappropriate, with diverse multimorbid combinations too frequently underdiagnosed, underestimated, and undertreated. Developing more effective ways to capture these combined physiologically prevailing constellations, that often arise from shared pathway-level perturbations, could thus meaningfully enhance risk stratification and therapeutic strategies.

**Beyond disease detection:** A systems-level perspective challenges the traditional compartmentalization of medicine and underscores the need for integrated therapeutic solutions. We still lack an adequately articulated picture of how dysfunction of individual tissues and organs is reflected in blood-based molecular profiles. Systemic molecular analyses now allow us to examine how organ-level molecular changes integrate at the whole-organism system level. Proteomic measurements, for example, have been shown to link circulating protein abundances and their changes to disease phenotypes - yet the tissues of origin and destinations of many secreted proteins still remain unknown [20]. What, then, is the true gatekeeping of organismal homeostasis and how could we possibly better capture and understand it? Focused efforts where multi-omics approaches are combined with InfraRed-omics can contribute to a more integrated understanding [14]. The added value of infrared fingerprinting lies in its holistic nature: minimal sample pre-processing, short measurement time and library-independent data analytics allow for building of large multi-centric cohorts, that are a prerequisite to multimorbidity profiling.

Medically, it is valuable not only to characterize endpoints but also to capture subclinical, intermediate phenotypes, ideally delivering continuous, quantitative health estimates. We have showcased the identification of actionable at-risk phenotypes, including pre-diabetes and shared pathophysiology of metabolic syndrome pre-state, which represents a risk factor for multiple follow-up diseases. These findings open the possibility of screening for combined conditions [2]. In line with this, we trained a multilabel machine learning classifier capable of simultaneously detecting and distinguishing more than ten clinical phenotype combinations - demonstrating that the approach is sensitive to the diversity of phenotypes and specific to combinations of medically relevant deviations [2]. This level of disease-agnostic profiling is of an advantage, both in its capacity to capture molecularly distinct deviations and in its relative ease and readiness for potential implementation into clinical routines.

**Lessons learned from longitudinal follow-ups and populational screening examinations:** How much of the circulating molecular signal captured by blood-based omics reflects true system biology rather than technical noise? This question becomes especially relevant in longitudinal studies, when the measurements are distributed in time, across multi-centric studies and, most importantly, in the potential scenario of routine clinical use by multiple clinics over course of years. Standardized Fourier-transform infrared (FTIR) spectroscopic fingerprinting of blood plasma and serum across three independent studies indicates that technical variability is modest compared to within-person biological variation over days, weeks, months up to nearly a decade. Therefore, repeated sampling and infrared profiling presents a viable strategy for longitudinal examination of health states [3,5]. This is where the technology could bring in added value, as many other omic and multi-omic approaches remain challenged by robustness, comparability across measurement campaigns, analytical complexity and costs [13].

Longitudinal infrared spectral interrogations demonstrated patterns associated with forthcoming metabolic health deviations and survival outcomes, with predictive performance that purely cross-sectional studies cannot deliver [2]. Taken together, these findings further support the potential of infrared molecular fingerprinting as a medically meaningful modality able to fuel affordable population-level screening with demonstrated potential to complement health checkups.

**Aiding molecular understanding by decomposing chemical complexity:** Considering the use of infrared spectroscopy to gain better understanding on how biological systems function – molecularly and systemically – it is important to acknowledge that there is a whole range of possibilities to chemically decompose such biological matrix and apply infrared spectroscopy to the most informative fractions. This is especially relevant as highly abundant molecules overshadow information of less abundant ones in infrared molecular fingerprints [1]. As proteins present the most abundant molecular contributions to infrared molecular fingerprints, this biomolecular class presents a reasonable target to combine with fractionation techniques such as liquid chromatography. Moreover, infrared profiling is inherently sensitive to post-translational molecular modifications. Spectroscopic characterization of the glycosylation patterns of plasma proteins for example may contribute to further understanding and stratification of physiological health span.

It is also fair to ask whether infrared examinations of bulk circulating molecules provide interpretability and remain useful for any qualitative assessments. Indeed, infrared profiling with regression analyses trained on large-scale continuous clinical chemistry parameters demonstrated their value in effective determination of concentrations of clinical chemistry analytes [2], just as conventional blood tests do - a sustainable strategy fueled with medical explainability.

**Concluding remarks**

There is an assumption that advancing sensitivity of infrared molecular fingerprinting to detect ever smaller changes in composite molecular mixtures will by itself drive progress. Yet, given the strongly overlapping molecular encodings, a sensitivity enhancement of bulk infrared spectroscopic measurements alone may deliver limited gains if not accompanied by rigorously standardized protocols, deeply characterized health and disease physiologies and phenotypes, along with systematic exploration of the chemical underpinnings of infrared molecular fingerprints. New measurement modalities have to therefore provide non-redundant information. Electric field-resolved spectroscopy [15], which enables sensitive and broadband infrared molecular profiling with high acquisition rates, may further extend current capabilities. For example, fraction-by-fraction in-flow infrared spectroscopy could help deconvolve the molecular complexity of biofluids. Furthermore, infrared fingerprinting of biological cells in flow [16], as a label-free flow cytometry technique, has the potential to unlock a new level of information and to provide new strategies for diagnostics, therapy monitoring, and biotechnology. However, these comparatively nascent strategies still require extensive empirical evaluation, standardization and validation.

Translating established workflows for InfraRed-omics from research laboratories to clinical practice holds substantial promise for improving risk stratification, disease diagnosis and prognosis. The demonstrated strengths of FTIR include its robustness and the reproducibility in measuring biological matrices [3], the ability to integrate data across measurement campaigns [2,5], the use of small sample volumes, and short turnaround times. Importantly, preservation of "self-contained" infrared molecular patterns and their ratios across a broad range of molecular classes remains challenging for multi-omics approaches and represents a unique advantage of InfraRed-omics. Evaluating this technology in well-controlled and adequately powered, multicentric clinical studies with rich phenotypic characterization, and integrating results and outcomes across cohorts, may allow infrared molecular fingerprinting - combined with machine learning and complementary technological platforms - to map infrared spectral patterns onto the spectrum of physiologically relevant phenotypes and thereby advance our understanding of how biological systems function, molecularly and systemically.

# 17. Strategies for robust data analysis in infrared molecular fingerprinting of biological samples

**Tarek Eissa[1,2], Marinus Huber[1,2], Joseph Rebel[1], Frank Fleischmann[1,2], Mihaela Žigman[1,2]**

1 Ludwig-Maximilians-Universität München (LMU), Garching, Germany
2 Max Planck Institute of Quantum Optics (MPQ), Garching, Germany

E-mail: tarek.eissa@mpq.mpg.de

## Status

Infrared (IR) spectroscopy has matured from a technique used in analytical chemistry into a viable candidate for high-throughput health screening. Applied to biospecimens, a single spectrum captures a holistic *IR molecular fingerprint* that reflects the sample's overall molecular makeup [1,2]. Combined with multivariate analysis and machine learning (ML), IR molecular fingerprinting can support diverse predictive tasks in clinical applications [2–9]. As with other multi-dimensional spectroscopic and omics approaches, the utility of IR fingerprinting depends on the informative content of the measured molecular signals and on robust study design and data analysis [1,2,10,11]. Here we discuss data-analytic and ML considerations for IR fingerprinting for health screening applications, focusing on liquid-phase mid-IR spectroscopy of crude plasma.

A plasma IR fingerprint reflects the superimposed vibrational signals of its dominant molecular classes—proteins, lipid particles, and water-soluble metabolites—each contributing partially overlapping absorption features (Fig. 1a). As emphasized by analyses of extracellular blood molecules, proteins dominate absolute IR absorbance across most of the spectrum, owing to their high concentrations and diverse functional groups [1,12]. Lipid particles are most characteristic at the ester C=O stretch near 1740 $cm^{-1}$, whereas the more diverse and typically less concentrated metabolites introduce narrower, distinctive peaks [1]. The resulting plasma spectrum is therefore not made up of isolated, interpretable bands but a composite molecular fingerprint whose structure reflects the relative abundances, chemical heterogeneity, and overlapping vibrational modes of many constituents. Importantly, because chemically related and unrelated molecular species often share broad regions of absorption, the IR fingerprint encodes physiologically meaningful variation at the level of the whole molecular landscape. The balance between molecular specificity and mixture-level information is naturally dependent on the biospecimen analyzed and the measurement modality. In gas phase measurements, for instance, IR spectra are dominated by sharper, well-resolved peaks that can facilitate more direct molecular assignment, but at the expense of wide molecular coverage [13].

Transforming IR fingerprints into clinical endpoints requires methods that can link spectral data to diverse output types. Such outputs can include binary disease classifications, multi-class classifications assigning spectra to one of several mutually exclusive endpoints, multi-label classifications where each spectrum can be associated with multiple concurrent endpoints, and regression of continuous health markers—each calling for tailored modeling and evaluation strategies. Existing work already spans this variety of tasks, from IR-based regression models calibrated to quantify clinical parameters such as glucose, lipids, and proteins in biofluids [3,6,7,14], binary and multi-class classification of common cancers [4,15], to multilabel models that jointly predict panels of chronic conditions and multimorbid states from a single IR spectrum [3].

## Current and future challenges

IR fingerprints are sensitive not only to the underlying molecular composition associated with a subject's health state, but also to the conditions under which samples are collected, processed, and measured. Preanalytical variables—including donor fasting status, sample collection procedures, clotting dynamics, freezer-storage conditions—can shift the concentrations of plasma constituents [16,17], thereby altering spectral baselines and band intensities. Similarly, analytical factors—including variation in optical path length, sample temperature during measurement, water and CO2 content, instrument drift, and differences in background subtraction—can all impact the apparent absorbance profile. Distinguishing biological information from artefactual variation requires protocol standardization, quantitative quality metrics, and systematic monitoring of measurement stability. These considerations also highlight the importance of randomization in sample processing and measurements, such that technical drifts and batch effects do not spuriously influence biological groupings of samples in downstream modeling. Nevertheless, longitudinal work on blood-based IR fingerprints demonstrates that person-specific spectra are highly stable over months to years when measured under standardized protocols [11,18].

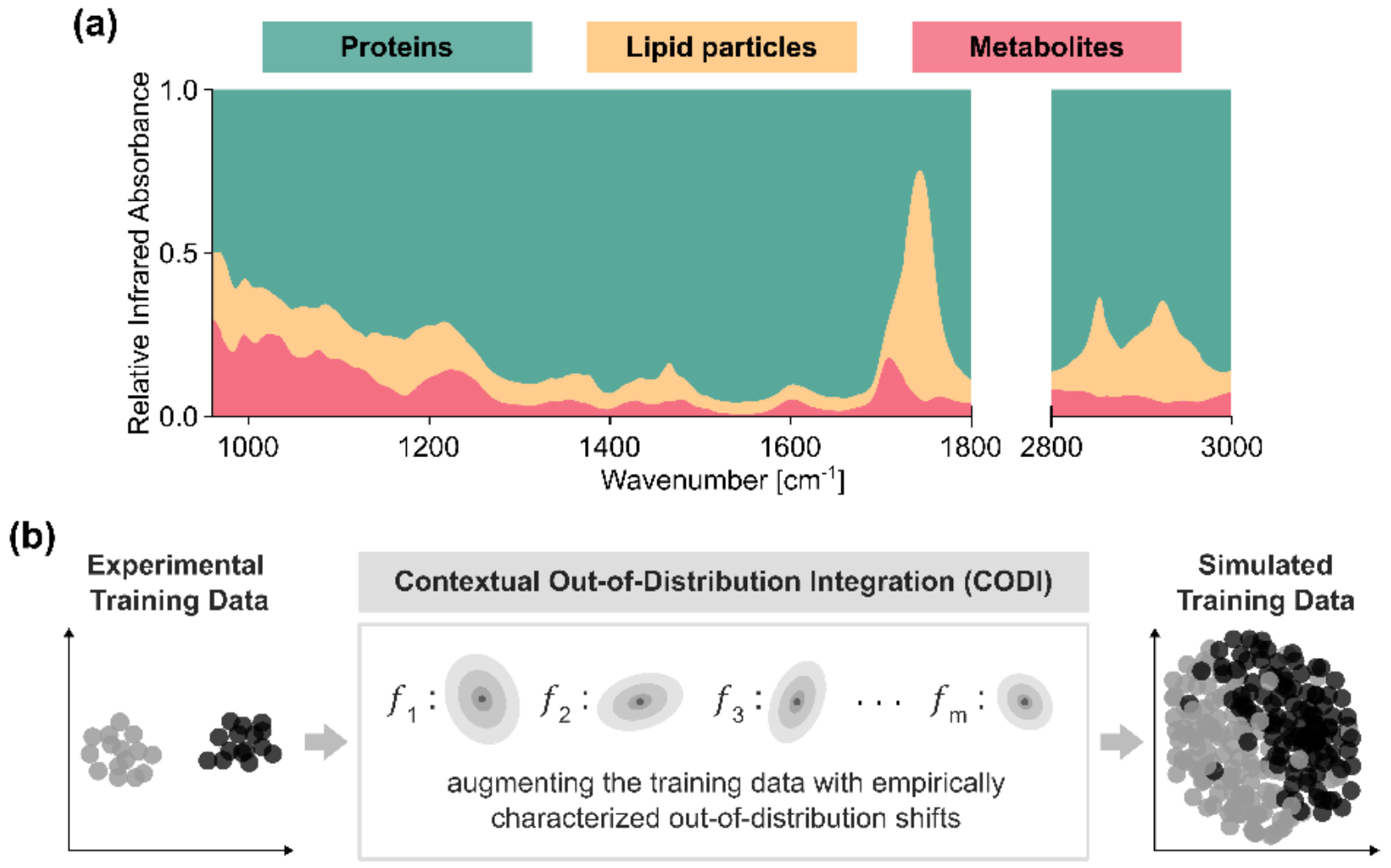

**Figure 1.** (a) Approximate relative contributions of the molecular classes in crude plasma (adapted from Ref. [1]); (b) Schematic of CODI, an approach which augments labelled experimental data with empirically characterized variability to generate simulated data with a more representative distribution (adapted from Ref. [11]).

Health screening applications must also contend with domain shifts and confounding variables driven by differences in populations and recruitment strategies. Differences in study participants' age, gender, comorbidity burden, lifestyle, and disease stage and subtype can all imprint signatures on their plasma IR fingerprints. For instance, if case individuals are on average older than controls, predictive models may learn to recognize such correlated factors rather than the targeted phenotype. Or, a model trained on one ethnic group may not generalize to another. These factors can yield impressive validation performance estimates, yet fail in deployment settings. A robust health screening strategy requires that study populations are reflective of the targeted deployment—covering relevant disease stages/subtypes, appropriate controls, and the variety of common comorbid conditions. This calls for study designs and analysis plans that anticipate domain shift and confounders, for example through stratified sampling or confounder-adjusted analyses.

Another important consideration is the quality of the realistic and obtainable clinical endpoints. IR-based models are only as meaningful as the ground truth on which they are trained. Yet, the "ground truth" can be prone to inaccuracies. Diagnostic categories may be defined using heterogeneous criteria or self-reported information, which can introduce recall bias and mislabeled samples. "Biomarkers" and "biomarker candidates" can suffer from assay variability and short-term biological fluctuation—e.g., when the spectroscopically analyzed sample was donated at a different time than the one used to measure the marker value. This label noise can hinder adequate model training, typically in the presence of severe mislabeling, and also compromise the ability to accurately estimate how well an IR-based model matches the biological phenomenon it is intended to predict. Mitigating this requires attention to endpoint definitions—using standardized diagnostic criteria, benchmark measurements obtained from the same sample donation time, and, where feasible, clinician-adjudicated diagnoses and validation of self-reported values.

Furthermore, reliable predictive performance depends on how developed models are trained and validated. A recurring observation in broader clinical ML studies is that cross-validation within one sample set tends to overestimate model performance, whereas external evaluation often reveals prediction accuracy degradation due to data domain shift [11]. This requires moving beyond simple "train-test" splits and random cross-validation towards validation schemes that explicitly mirror anticipated deployment conditions. In practice, this means separating data into distinct development and evaluation domains—e.g., training on a subset of collection centers, clinical studies, sample donation/measurement time periods, and testing on others—while preventing leakage of participant visits or repeated-measures and sample batches across these splits. Within the development domain, nested data splits can be used for model selection, whereas the held-out external domain is used for an unbiased assessment of model discrimination and failure modes.

Yet another challenge is the conceptual shift from targeting isolated biomarkers to using "molecular fingerprints" for broader health-state phenotyping [19], where the state is inferred from multivariate model outputs. To be usable in regulated clinical practice, there must be clarity about the basic definitions and quality criteria required for the model to function, including specification of the intended clinical purpose, target and reference populations, and clinically meaningful ranges of variation and risk that these models are meant to capture. These considerations intersect with emerging risk-based regulation of ML in medicine, where systems used for diagnosis or health screening are often treated as high-risk and subject to requirements on risk management, data governance, transparency, and post-deployment monitoring [20].

**Advances in science and technology to meet challenges**

Classical analysis of IR spectroscopic data has typically focused on single peaks, bands, or band ratios linked to known functional groups. These univariate descriptors remain useful for assessing where broad chemical signals lie and matching to libraries of individual components to assess signal similarity. However, they capture only a small fraction of the information contained in the IR fingerprint of complex molecular mixtures. In practice, univariate approaches perform well only when the signal of interest is strong relative to sample variability and is dominated by a few high-concentration molecules with unique spectroscopic signatures. When targeting health phenotypes in heterogeneous cohorts, it is more common that a diverse set of molecular species change concentrations together, each contributing signal across the absorption features. ML approaches treat the IR spectrum as a multi-dimensional object, where the dependencies among the whole feature range are used, to better extract molecular data patterns that map to health phenotypes.

For supervised learning tasks of this kind, regularized linear models are reliable workhorses. Ridge and elastic-net regression, penalized logistic regression, and linear support vector machines can handle the strong collinearity and high dimensionality of IR spectra and, when properly tuned, often match the predictive performance of more complex nonlinear models. These learners can extend to multi-output tasks via multinomial formulations or one-vs-rest schemes, enabling prediction of multiple endpoints from a single measurement. More flexible models, including gradient boosting and neural networks, can capture nonlinear interactions and higher-order structure in the relationship between spectra and clinical endpoints, and several of these methods can natively support multi-output predictions. Yet, when the dominant relationship between IR spectra and endpoints is approximately linear, no gain in predictive performance should be expected from the added complexity, and for nonlinear patterns, substantially larger, more heterogeneous datasets are required to learn and generalize the underlying structure to unseen samples.

Beyond model selection, the representativeness of the training data is arguably even more important. As discussed above, differences between training and deployment data distributions rarely arise from a single factor, but from several, making it difficult to collect representative training datasets. Methods that explicitly account for distributional variation in the data are particularly appealing here, where the goal is to train models that remain reliable under domain shift in practice. One approach that exemplifies this strategy is CODI (contextual out-of-distribution integration) [11]. CODI is an approach that uses independently collected calibration measurements to characterize how data can change under realistic sources of variability—e.g., differences in sample handling/storage, measurement drift, or underlying biology—and then introduces these variations *in silico* into the training seed data (Fig. 1b). Rather than assuming perfectly standardized conditions, CODI aids a learning algorithm to rely on molecular features that are both reproducible and related to the endpoint of interest. Simulations informed by the expected scope of data variation thus can enable the anticipation of the true data domain without requiring prohibitively large training datasets.

## Concluding remarks

IR fingerprinting can yield multiplexed readouts of human physiology from single spectra. Realizing this potential requires careful control on several fronts, from clinical definitions and spectroscopic measurements, data analysis and model validation strategies, and regulatory expectations.

# 18. Introduction to IR technologies

**Borislav Hinkov[1,+], and Lukasz Sterczewski[2,*]**

[1] Silicon Austria Labs, Villach, Austria
[2] Wrocław University of Science and Technology, Wrocław, Poland

E-mail: +borislav.hinkov@silicon-austria.com , *lukasz.sterczewski@pwr.edu.pl

### Status

From a technological standpoint, infrared (IR) photonics provides direct access to molecular information with critical importance for healthcare applications. This is because vibrational molecular fingerprint absorptions, e.g., of bodyfluids, tissue, or exhaled gases [1–3] occur in the IR region. Of particular interest is the mid-IR (MIR) band, often defined between ~3–12 µm wavelength, covering fundamental molecular absorption patterns. Access to the MIR range unlocks a wide variety of diagnostic, monitoring, and screening approaches as described in Part I of this roadmap. Interest in the MIR stems primarily from its higher spectral selectivity than the near-IR (NIR) counterpart, as complex molecules are often easier to distinguish when their lower-energy transitions are probed. At the same time, this has practical trade-offs that have to be considered as the task is to turn spectra into information. The advantage of measuring in the NIR is the better performance of available components and typically a larger penetration depth into materials; the disadvantage, however, is a stronger dependence of NIR spectra on temperature due to the occupancy of higher vibrational levels and strongly overlapping overtone bands [4].

Yet, despite decades of scientific maturity, the translation of MIR photonics into scalable healthcare technologies has historically been constrained not by fundamental spectroscopic principles and related limitations, but by the availability, practicality, and integrability of such enabling technologies. Additionally, a strong segmentation of the various technical efforts prevented the development of a "joint vision" for the development of the field.

A defining characteristic of the field is a sharp technological discontinuity that emerges beyond wavelengths of approximately 3 µm. In this spectral region, where molecular specificity is the strongest, suitable materials, sources, detectors, and optical components typically degrade significantly compared to their visible and NIR counterparts. This ">3 µm leap" has long imposed penalties in size, cost, power consumption, and robustness, limiting many MIR systems to laboratory environments. Consequently, healthcare applications have often relied on large-footprint Fourier-transform infrared (FTIR) instruments [5,6] or have shifted towards indirect or surrogate optical techniques (e.g. Raman) [3]. At longer wavelengths, technological maturity is even lower. Consequently, as shown in Fig. 1, one can say that in MIR spectroscopy, challenges scale with wavelength.

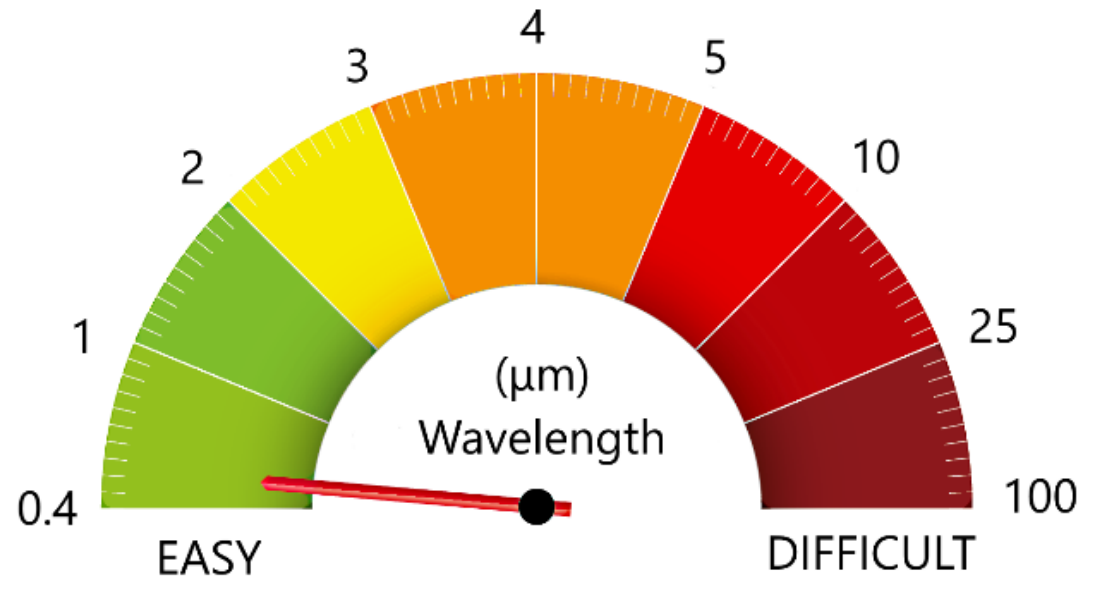

**Figure 1.** A major dilemma of IR technologies: there is a strong decline in device performance and the availability of suitable materials when exceeding ~3 µm wavelength. Please note that this is just a simplified rule of thumb, and details may vary significantly from case to case. Still, it has been preventing people from designing IR instruments on a large scale and triggering the use of alternative technologies such as Raman.

Fortunately, over the past decade several converging developments have altered the MIR landscape. Compact and mechanically robust FTIR architectures, advances in thermal emitters and uncooled detectors, and the maturation of semiconductor laser sources, most notably quantum cascade and interband cascade lasers, and detectors have significantly reduced system complexity and power requirements [7–14]. In parallel, the emergence of MIR photonic integrated circuits (PICs), integrating lasers and detectors with routing waveguides and microfluidic concepts into lab-on-chip platforms, has enabled new strategies for light delivery, interaction, and detection in biologically relevant environments [15–18]. These advances are complemented by improved (monolithic) fabrication processes [19], heterogeneous material integration [20–21], and growing access to foundry-level manufacturing for selected MIR platforms (e.g. InP) [22].

At the system level, MIR photonics technologies now span a broad spectrum of architectural approaches: from broadband spectroscopic instruments and imaging platforms to highly targeted, application-specific sensors optimized for portability, wearability, or point-of-care use. Importantly, no single technology covers all healthcare scenarios. Instead, the status of the field is best described as a diversified but increasingly connected ecosystem of single components (sources, detectors, …), spectrometers, and integration technologies, i.e., early-stage PICs. This ecosystem forms the technological "root system" as shown in Fig. 2 that supports the application "canopy" discussed in Part I and is explored in detail in the chapters that follow here.

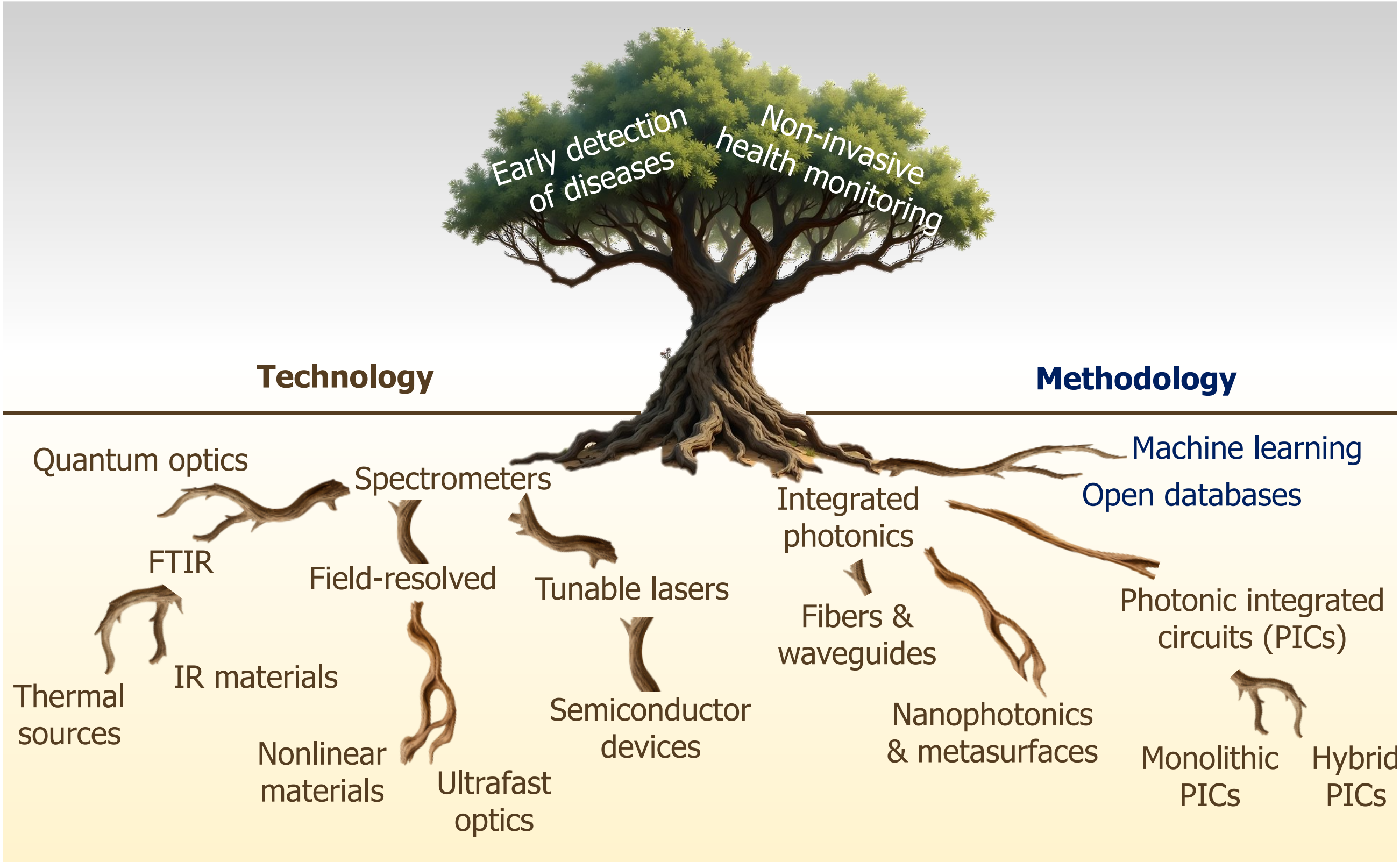

**Figure 2.** Visual interpretation of how societally-relevant applications of IR spectroscopy depend on technology. Virtually parts of the root system are covered in subsequent chapters. This image aims to guide the reader in establishing a hierarchy.

## Current and future challenges

Despite the substantial progress, the widespread deployment of MIR photonics in healthcare remains limited by a set of interrelated challenges that extend beyond conventional performance metrics. These challenges can be grouped into four tightly coupled areas.

*Component-level constraints.*

At the most fundamental level, sources and detectors in the MIR continue to impose some trade-offs between spectral coverage, device efficiency, and cost. Wall-plug efficiency, thermal management, and long-term reliability are particularly critical for portable, wearable, and point-of-care systems. Uncooled detectors and emitters offer attractive routes to scalability, but often at the expense of device performance including device sensitivity or bandwidth, requiring careful system-level optimization. These constraints become more pronounced at longer wavelengths, where material choices and fabrication complexity remain limiting factors. Nevertheless, some progress has been made and solutions for these wavelengths start to become practical.

*Measurement interfaces and reproducibility.*

Healthcare applications demand robust and repeatable interaction between light and complex biological samples. Variations in sample preparation and environmental conditions can dominate measurement uncertainty, often exceeding intrinsic instrument noise. The diversity of optical techniques for measurements in standard transmission or reflection geometry as well as the typically more complex techniques relying on, attenuated total reflection (ATR), photoacoustic or photothermal principles further complicates cross-platform comparison and standardization. Addressing these issues is essential for translating laboratory demonstrators into clinically reliable tools.

*Metrology, standardization, and transferability.*

For MIR photonics to support multi-site studies, longitudinal monitoring, and regulatory approval, measurements must be reproducible, comparable, and stable over time. This requires standardized calibration procedures and reference materials, available spectral libraries, and unified data formats. Without such metrological foundations, even highly sensitive instruments may produce results that are difficult to reproduce or validate across institutions and out-of-lab settings.

*Clinical trust, regulation, and data interpretation.*

Finally, many MIR photonics systems increasingly rely on advanced data processing and machine learning to extract clinically relevant information from multi-dimensional spectral data. In healthcare contexts, these approaches fall under stringent regulatory frameworks and demand transparency, explainability, and robustness. The designation of many medical AI systems as “high-risk” reinforces the need for technologies whose physical measurement principles, data collection and assessment routines, and fundamental performance limitations are well defined and auditable.

Together, these challenges highlight that progress in MIR photonics for healthcare cannot be achieved through isolated component improvements alone but requires coordinated technological advances across the entire photonic systems.

**Advances in science and technology to meet challenges**

Addressing the challenges outlined above is demanding for a set of convergent advances that redefine how MIR photonics technologies are conceived, built, and evaluated for healthcare.

The first major thrust is *platformization through integration*. MIR integrated photonics enables the co-design of sources, waveguides, interaction regions, and detectors within compact and manufacturable material systems. Where fully monolithic concepts are not suitable, heterogeneous integration of dissimilar materials allows to mostly preserve the individual component performance while leveraging scalable fabrication processes. Such platforms reduce alignment complexity, improve robustness, and open pathways toward wafer-scale production. Importantly, integration also enables new measurement geometries, such as evanescent-field sensing and on-chip spectroscopy, that are particularly well suited to aqueous and biological samples.

A second thrust concerns *efficient and application-matched light generation and spectral analysis*. While pursuing universal instruments, current developments emphasize tailoring spectral coverage and resolution, together with modulation schemes to specific clinical questions. Compact FTIR systems benefit from advances in MEMS technology, thermal emitter characteristics, and detector performance, while laser-based approaches exploit tunable and arrayed semiconductor source geometries to enable high brightness, selectivity and sensitivity in the spectrometers. Vertically integrated and surface-emitting source concepts promise further concepts towards miniaturized on-chip spectrometers which promote scalability and support applications outside traditional laboratory settings.

A third, increasingly important thrust is *the co-development of metrology and computation*. Advances in calibration strategies, reference standards, and transfer of learnings enable spectral data acquired on different platforms or at different sites to be meaningfully compared. At the same time, data-driven analysis is being designed in closer alignment with physical measurement constraints, favoring interpretable models, uncertainty quantification, and physics-informed learning. This co-design approach supports regulatory compliance while preserving the selectivity, sensitivity and variability of MIR measurements.

Collectively, these advances shift MIR photonics for healthcare from bespoke instrumentation towards a modular, interoperable technology platform. In this roadmap, they are represented as the interconnected roots of a technological tree, feeding multiple application branches while sharing common foundations.

### Concluding remarks

MIR photonics occupies a unique position at the intersection of molecular specificity, non-invasive measurement, and scalable optical technology. As illustrated throughout this roadmap, its potential impact on proactive and predictive healthcare is substantial, but realization of this potential depends critically on the maturity and coherence of the underlying technologies.

This part of the roadmap frames IR photonics technologies not as isolated components, but as an interdependent root system supporting diverse healthcare applications. Progress beyond proof-of-concept demonstrations requires that sources, detectors, optical interfaces, integration platforms and technologies, and data pipelines evolve in alignment, guided by healthcare-specific constraints rather than laboratory convenience. Importantly, no single technological pathway is expected to dominate across all use cases. Instead, success will emerge from a whole portfolio of complementary solutions optimized for different clinical contexts.

The chapters that follow examine individual elements of this technological ecosystem in detail. Together, they define the scientific technology and engineering state and discuss the advances needed to transform infrared photonics into a trusted, scalable, and widely deployed foundation for future healthcare systems.

# FTIR:
# 19. Compact, high-resolution Fourier spectrometers for molecular sensing from near- to far-infrared

**Jakub Mnich[1], Jarosław Sotor[1], and Łukasz Sterczewski[1,*]**

[1] Laser and Fiber Electronics Group, Faculty of Electronics, Photonics and Microsystems, Wrocław University of Science and Technology, Wyb. Stanislawa Wyspiańskiego 27, 50-370 Wrocław, Poland

*E-mail: lukasz.sterczewski@pwr.edu.pl

## Status and current challenges

Fourier transform spectroscopy (FTS) is a mature experimental technique with widespread adoption in disciplines including material science, pharmacology and life sciences. It draws its strength from apparent simplicity, requiring only a broadband thermal source (typically a heater/globar), an interferometer, and a photodetector. One can say that the principle of operation has not changed much since the XIX-th century **[1]**. On the other hand, despite several decades of development, it is expected that there is still room for innovation. But is this demand for improvements driven by actual needs, or academic curiosity?

It may be easier to answer this question with specific challenges in mind. A recent study by M. Huberus et al. has demonstrated the feasibility of spectroscopically-based cancer diagnostics with binary classification performance in the 78%–89% range using an FTS instrument probing plasma and serum samples **[2]**. Machine learning algorithms have been used to extract relevant absorption features from the spectral range of 3–10.5 µm, typical for most commercial instruments with cooled photodetectors. However, a wealth of biological (and potentially clinically-relevant) information is accessible at even longer wavelengths extending further towards the longwave infrared (up to 15 µm) and far-infrared/THz region (15–100 µm) **[3]**. While this poses a major technological challenge, unlocking new spectral bands is expected to improve the classification performance of algorithms and help detect diseases at earlier stages. This naturally calls for ultra-broadband FTS instruments with lowered cooling requirements. Size, weight, and power consumption (SWaP) play a key role if FTS technology is envisioned to be used on a large scale for point-of-care testing applications to address the need for photonics-assisted health monitoring. While incoherent FTIR cannot match traditional analytical chemistry methods when it comes to limits of detection (LoD), it can still be a useful tool for breath analysis which, unlike solid and liquid samples, calls for enhanced spectral resolution (< 0.1 $cm^{-1}$) **[4]**. Frequency comb-based FTIRs are uniquely well-suited for such gaseous samples.

## Advances in science and technology to meet challenges

Virtually all components of the FTS instrument require intensive engineering. The road to the Fourier spectrometer tailored to health diagnostics is depicted in Fig. 1. We have identified various bottlenecks that limit critical performance metrics like spectral coverage, signal-to-noise ratio, and resolution, all relevant in the context of blood plasma or exhaled breath analysis. Starting from the source, existing solutions utilizing high-power globars (formerly known as "glow bars" – heated metal wires) often cause damage to a biological sample due to excessive heat. Our recent work **[5]** has demonstrated the feasibility of FTS experiments with 1 W of electrical power delivered to the globar (when ~200 mW optical power was collected from the device), and 200 mW of electrical power with ~26 mW illuminating the sample (Fig. 2).

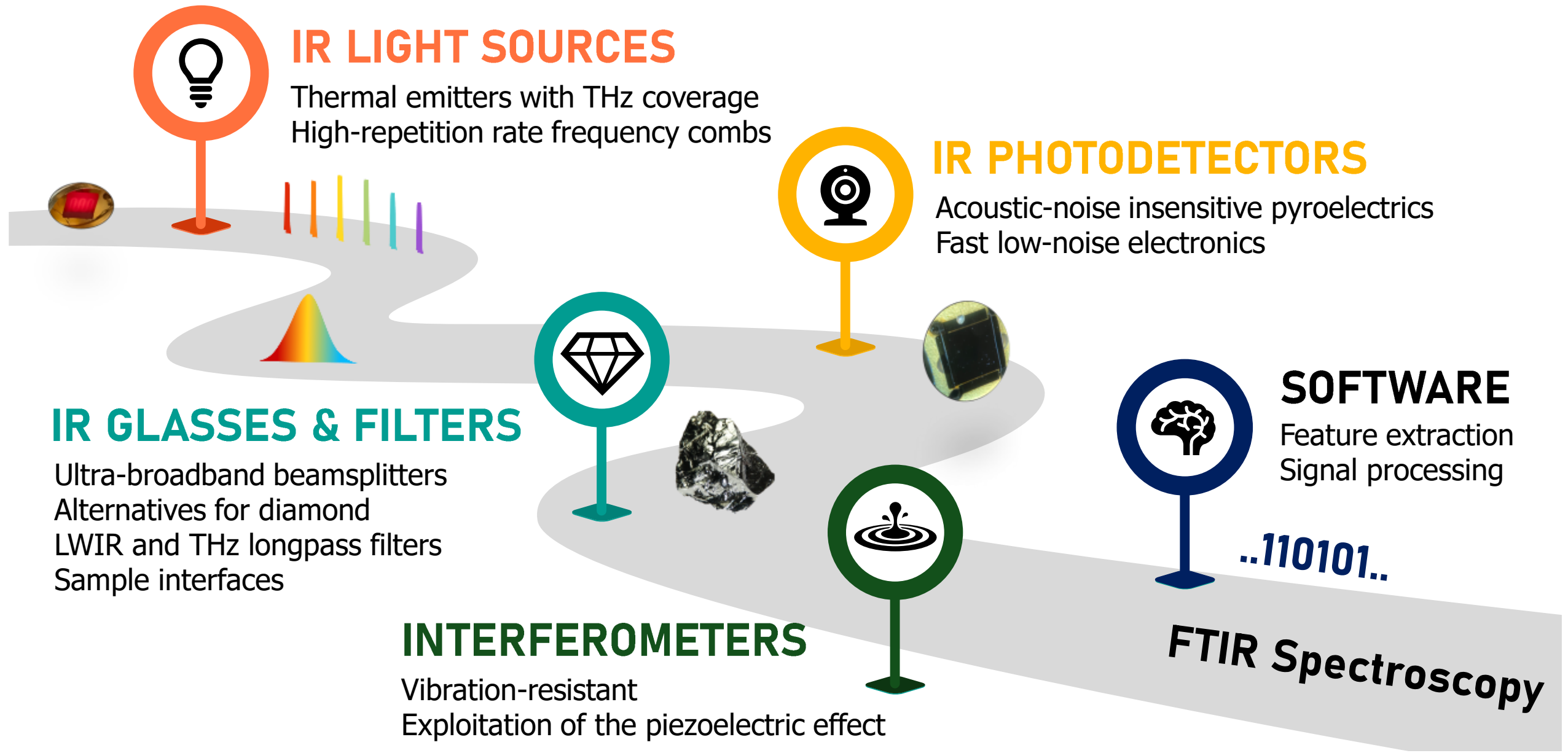

**Figure 1.** Roadmap of critical developments for practical broadband Fourier transform spectroscopy. Parameters optimized for molecular sensing in the gas and liquid state include spectral coverage, resolution, noise, and data analysis capabilities. The spectral coverage is fundamentally limited by infrared (IR) light sources, which often lack the long-wavelength part due to insufficient emissivity at the lowest optical frequencies. For gas sensing, broadband, tunable high-repetition rate frequency combs (also chip-based) are desired. Next, wavelength-flexible IR detectors with a spectrally flat optical response, no microphonic effect, and room-temperature operation are of critical importance. This requirement is accompanied by the need for non-toxic and environmentally-stable IR-transparent glasses, including low- and high-index ones. Low-index materials would be ideal for protective windows, while high-index ones could serve as beam splitters. Next, one needs new interferometer topologies with improved resistance to vibrations and fast scanning capabilities. Finally, software for accurate extraction of spectroscopic features and linking them to open databases would be revolutionary.

In both cases, the long-wavelength part (up to 30 µm) remained useful, excluding wavelengths primarily below 5 µm. Much promise for exhaled breath analysis is also offered by chip-based frequency comb sources **[6]**. These tiny emitters of coherent infrared light with a characteristic spectrum of sharp equidistant lines enable one to surpass the nominal instrumental resolution limitations **[7]** to ultimately reach the MHz range **[8]** – a 1000× more than a typical tabletop instrument with GHz (~0.03 cm$^{-1}$) resolution for 30 cm of interferometer arm delay. This is particularly attractive for complex mixtures of gases (volatile organic compounds), where spectral congestion and overlapping lines can be better resolved at lowered pressures, when absorption linewidths are sub-GHz. A major strength of chip-based comb lasers is that, rather than emitting light over a continuous spectrum, they concentrate optical power around their discrete teeth. This provides many advantages for sensing thicker samples or gaseous analytes, as the resulting noise properties are greatly improved compared to thermal sources, albeit at the expense of spectral coverage in the range of several THz (~100 cm$^{-1}$) – a 100 times lower than for globars. Paradoxically, obtaining broad spectral coverage is often constrained not by the source but by the detector and transmission characteristics of the used optics. In particular, the availability of a high-refractive-index material transparent across multiple spectral regions from the visible to the far-infrared (THz) is rather scarce. While diamond addresses this need (from the ultraviolet up to 100 µm of wavelength) **[9]**, it requires complex fabrication via chemical vapor deposition. Other infrared glasses, in turn, are often toxic or degrade in a humid environment. This clearly motivates the need for research on new materials for beamsplitters and photodetector windows with environmental stability and compatibility. Sample interfaces like prisms or gas cell windows will also benefit from such developments.

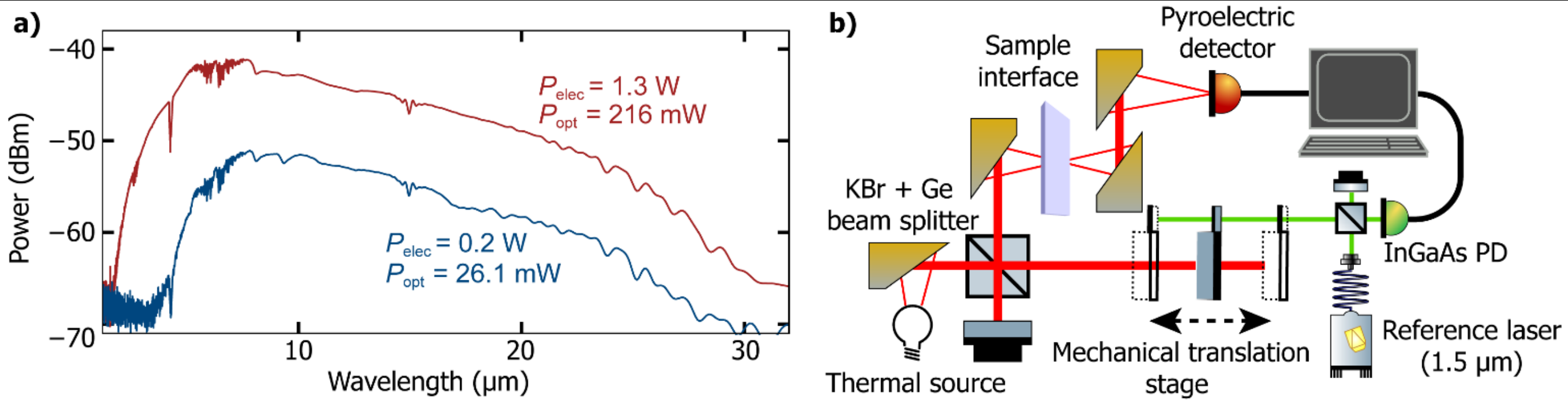

**Figure 2.** (a) Spectrum obtained using a coated potassium bromide (KBr) beamsplitter with an engineered thermal source. (b) Experimental setup of the FTS instrument. Panel (a) is adapted from data published in Ref. **[5].**

If no protective window is an option for the photodetector, as in the case of selected pyroelectric materials like thin-film lithium tantalate (LTO), one can obtain an optical response from the UV (<0.39 µm) to the THz (50 µm) at a static optical configuration and one set of components at room temperature **[10]**. This, however, comes at the expense of the microphonic effect: the detector responds to acoustic noise, which may produce spectral artifacts. Therefore, low-index window materials with a broad transparency range are also of high relevance to protect and environmentally isolate the sensing element.

Another major challenge for any FTS instrument is the delay scanning mechanism. In principle, the interferometer must be vibration-free down to a micrometer range. Simultaneously ensuring simplicity, robustness, and long (several cm) delay required for liquid and gas-phase spectroscopy is nontrivial and often comes with tradeoffs. To date, the most popular linear delay mechanism has been replaced with a rotary star **[11]** or birefringence-based moving wedges **[12]**. Still, the perfect mechanism does not exist. Miniaturization is often difficult for some topologies. Much promise is offered by modern piezoelectric motors and voice coil actuators, which can move at high rates, albeit their weight-handling capabilities are limited.

Once the spectrum is collected, it needs to be interpreted. While it contains much information, it is difficult to analyze directly by humans. In the conventional case, a comparison with a spectral database gives an insight about the sample composition; however, for many samples, no databases exist at all. This motivates the need for standardization and making databases open with the option to contribute. It is a common plague of the infrared community that measured spectra are not deposited in public repositories, which would fuel the development of machine learning **[13]** or, more generally, artificial intelligence **[14]** spectroscopic tools.

### Concluding remarks

To make FTS instruments more suited for large scale health diagnostics, several significant improvements of its components must be made. This requires a joint research and development effort from material scientists, physicists, electrical engineers and programmers. Once room-temperature, portable spectrometers can simultaneously operate over multiple spectral bands from the near-infrared to the THz, a true revolution is expected to begin. This, however, must be accompanied by data standardization and anonymization.

### Acknowledgements

J. M. and Ł. S. acknowledge funding from the European Union (ERC Starting Grant, TeraERC, 101117433). Views and opinions expressed are, however, those of the authors only and do not necessarily reflect those of the European Union or the European Research Council Executive Agency. Neither the European Union nor the granting authority can be held responsible for them.

# 20. Accessible mid-IR spectroscopy for healthcare: Leveraging compact and MEMS-miniaturized FTIR platforms for automated molecular characterization

**Bassam Saadany [1], Matthias Budden [2] and Thomas Gebert [2]**

[1] Si-Ware Systems, Cairo, Egypt & Paris, France
[2] WiredSense GmbH, Hamburg, Germany

E-mail: bassam.saadany@si-ware.com, matthias.budden@wiredsense.com

## Status

Fourier Transform Infrared (FTIR) spectroscopy provides rich molecular fingerprints for chemical and biological materials. However, conventional FTIR instruments are bulky, costly, and unsuitable for field or point-of-care use. The advent of micro-electromechanical systems (MEMS) has enabled a new class of miniaturized FTIR spectrometers, fabricated on silicon, combining optical, mechanical, and electronic functions on a single chip [1–3].

At the same time, efforts toward simplifying access to high-performance MIR spectroscopy have aimed to close the gap between complex laboratory analytics and practical, real-world usability. In the pharmaceutical domain MIR spectroscopy is recognized as a powerful, reagent-free method for identity testing and quantitative analysis. Yet its potential remains underexploited due to high entry costs and the need for application specific reference data as well as specialized expertise in spectral data analysis.

The field appears to be at an inflection point, where miniaturization to MEMS-based spectrometers and the rise of cost-efficient, automated MIR systems form mutually reinforcing paths that enable the widespread use of IR spectroscopy beyond specialized laboratories.

## Current and future challenges

For the wide adoption of FTIR spectroscopy, there is a need for compact, low cost FTIR systems enabling label-free, reagent-free, and real-time analysis of biofluids and tissues. The portability and robustness of FTIR spectrometers should open the door for field-deployable healthcare diagnostics and Point-of-Care systems.

In the pharmaceutical sector, MIR spectroscopy is recognized by the European Pharmacopoeia [4] for identity testing of many raw materials. It offers higher molecular specificity than NIR and supports quantitative analysis. Yet adoption in pharmacies and hospitals remains low due to high instrument acquisition costs lack of user-friendly workflows the need for expert-level interpretation as well as challenges in analyzing heterogeneous, multi-component formulations.

This is notable given the high demand: For example, in Germany alone, nearly 10 million prescription-based compounded medicines were produced in 2024 [5]. These formulations are often tailored for pediatric or geriatric needs or produced in response to supply shortages and require reliable yet accessible quality control. It has been shown that the integration of such [6] FTIR hardware with automated analysis and AI-driven models enables not only routine identity analysis of raw materials for compounded medicines but also quantitative determination of active ingredients even in complex matrices. One example of such a complex analysis is the rapid determination of medically relevant cannabinoids such as THC and CBD in medicinal cannabis flowers and concentrated extracts [7].

Despite rapid progress, several challenges must be addressed to unlock widespread IR adoption:

- Compact, low cost and scalable FTIR spectrometers with lab grade performance are necessary.
- Spectral interpretation for complex biological or pharmaceutical matrices requires advanced chemometrics and robust AI models.
- Calibration stability across humidity, temperature, and sample heterogeneity must be ensured.
- Sensitivity improvements are needed for trace biomarker detection in biofluids.
- Regulatory and clinical validation remain essential for healthcare integration.
- Standardization across instrument classes and sample interfaces such as ATR-, transmission- and reflection-modes is needed for interoperable analytics.

**Advances in science and technology to meet challenges**

This section illustrates how IR spectroscopy can move from laboratory benches into clinical workflows – if instruments are compact, affordable, robust and resolve the mid-IR fingerprint region.

*MEMS-Miniaturized FTIR Spectrometers*

Fully integrated MEMS-based Michelson interferometer fabricated using standard silicon processes was introduced in [8,9]. This established a paradigm shift enabling chip-scale FTIR spectroscopy, eliminating the need for optical alignment, since optical and mechanical components are monolithically aligned during fabrication, thus enabling mass production for portable applications.

In such MEMS FTIR systems, the moving mirror of a Michelson interferometer is implemented using a silicon MEMS platform with electrostatic comb-drive actuation. The displacement—typically up to ±250 μm—creates an interferogram that yields a broadband IR spectrum. This monolithic silicon architecture has proven compactness, robustness, and calibration stability [8–10]. Alternative MEMS architectures are being advanced by Fraunhofer IPMS [10], IMEC [11], and others, including translatory actuators and MEMS grating-based interferometers. Yet the silicon monolithic interferometer remains among the few commercially validated MEMS FTIR platforms with long-term calibration stability and reproducibility [9,13].

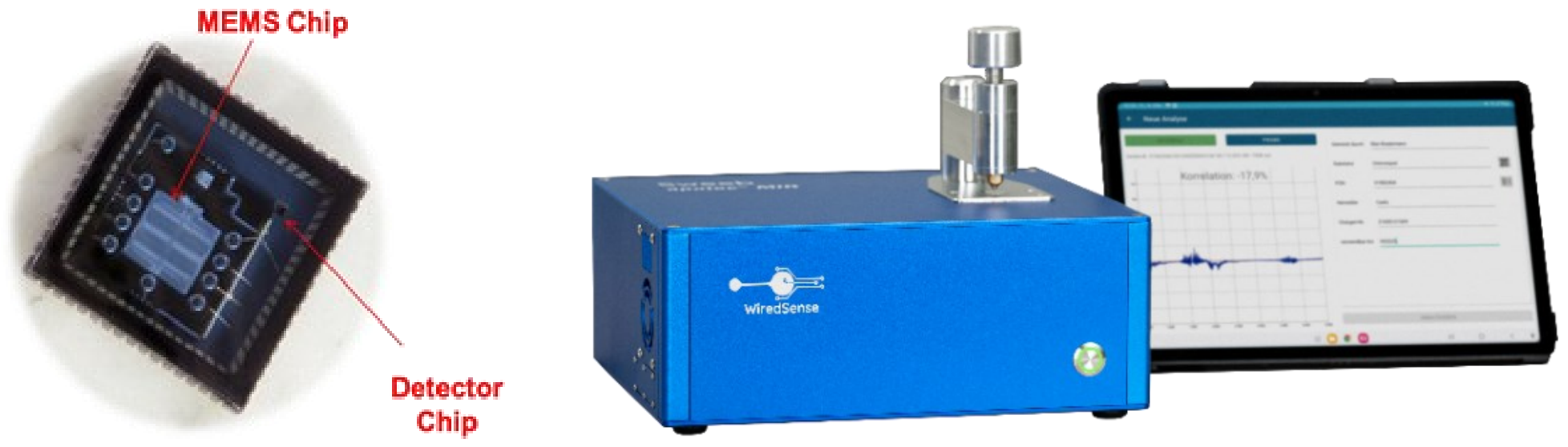

a) MEMS FTIR core module b) Compact FTIR in ATR mode 1.3-25 μm

**Figure 1:** a) Photograph of MEMS FTIR spectrometer core (Si-Ware), showing the integrated MEMS interferometer, photo detector chip and coupling micro optics in a small package. The device implements a full Michelson interferometer on silicon with electrostatic actuation [8,9]. b) Photograph of the compact FTIR System (Sweeb from WiredSense), equipped with a diamond attenuated total reflection (ATR) sample interface that enables easy access to high-performance spectra across the whole MIR fingerprint range (~1.3 μm–25μm wavelength) and supports automatic cloud based spectral analysis [6]

*Accessible FTIR Based on Advanced Pyroelectric Detection Technology*

In parallel, high-performance, compact, and cost-efficient MIR spectrometers were built around high-sensitivity, fast uncooled pyroelectric detectors with high linearity [14]. These advances in detector technology, that have been shown as viable alternatives to helium-cooled bolometers for THz absorption measurements [15] enable FTIR instruments that are significantly more affordable, more robust, and easier to operate than traditional lab-scale systems yet maintain high analytical performance in the long wavelength MIR "fingerprint" region.

A central enabler of a low-threshold MIR analytical system is a cloud-based spectral analysis platform, offering automatic preprocessing, chemometrics, and AI-based quantification and reporting. This combination directly addresses barriers that have historically limited MIR adoption in decentralized settings.

*Biomedical, Healthcare and Pharmaceutical Applications*

A recent study [16] demonstrated COVID-19 detection from saliva and nasopharyngeal swabs using a portable MEMS FTIR spectrometer, identifying spectral biomarkers associated with viral infection. The measurement and analysis steps combining two spectral regions: MIR (1.75–4.0 μm) and NIR (1.3–2.6 μm) are shown in Fig. 2, where MIR region with dry samples achieved an accuracy of 89% (±4.2% at 95% confidence interval), with 83% sensitivity, 95% specificity, and an AUC of 0.93. Work on screening for infectious diseases has shown high diagnostic accuracy for malaria detection using MEMS FTIR platform, and extended to other healthcare applications such as fibrosis diagnostics, renal disorder screening, and spectral histopathology [17,18]. Another example is the hair analysis using MEMS FTIR [19], a consumer-facing spectroscopic device used to analyze hair composition and damage in real time.

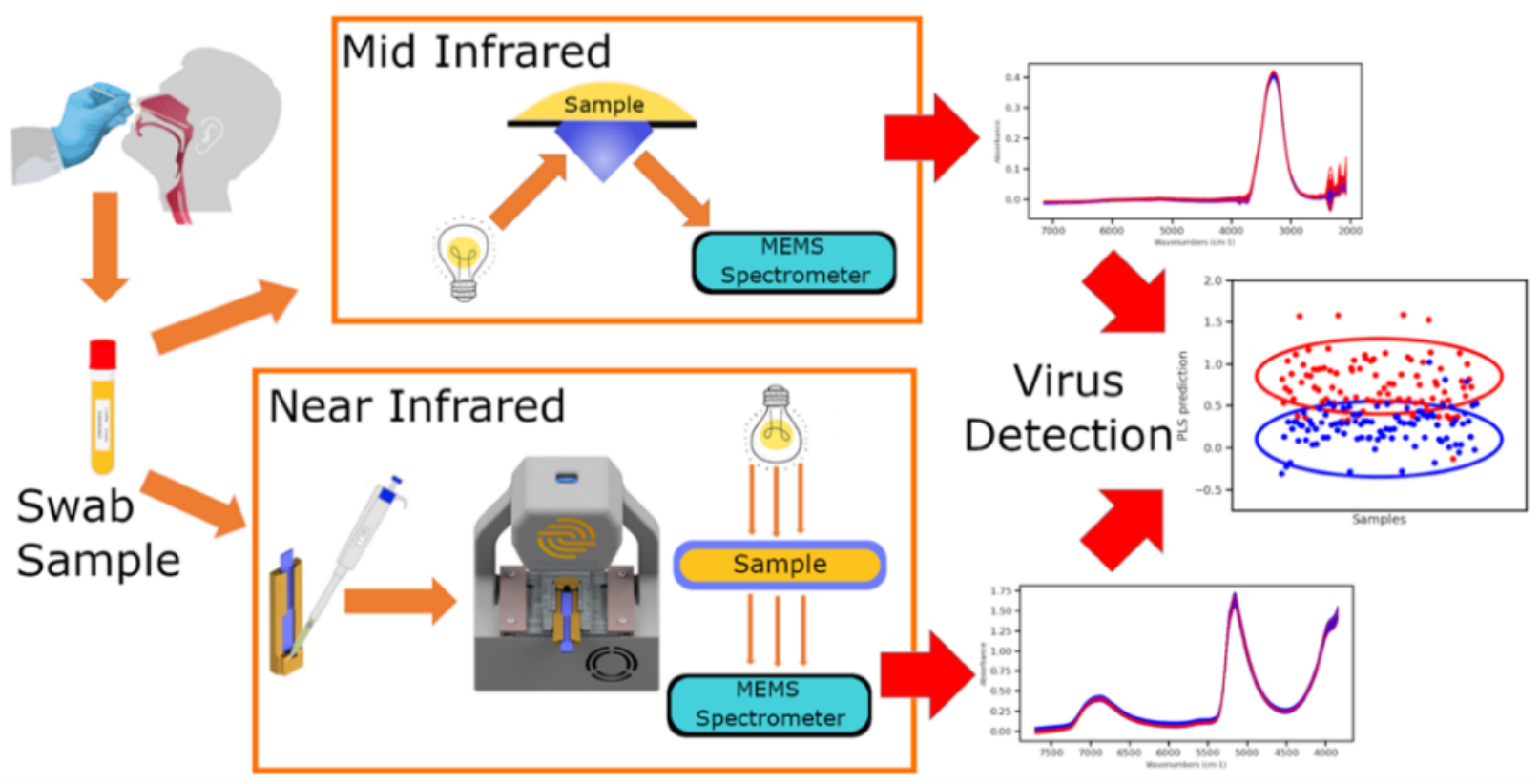

**Figure 2** Measurement and Analysis steps for COVID-19 detection, samples are scanned with MEMS FTIR in both MIR ATR configuration and NIR transmission and machine learning [16].

## Concluding Remarks

IR spectroscopy stands at the threshold of widespread deployment across healthcare and pharmaceutical practice. The scientific foundations are solid, with more than 14,000 FTIR-related publications indexed in 2024 alone. By uniting miniaturization with accessible, cloud-driven MIR spectroscopy, the field gains both physical portability and analytical simplicity.

Looking forward, the convergence of compact and ultra-compact MEMS FTIR, AI, and edge computing can enable continuous, connected biochemical sensing. Future systems may be embedded into wearables, point-of-care devices, and telemedicine infrastructure. We envision IR spectroscopy as a cornerstone of future decentralized diagnostics, digitalized pharmacy workflows, and personalized healthcare – transforming infrared photonics into an everyday analytical tool.

# 21. Thermal emitters and pyroelectric detectors for FTIR towards spectroscopic twin and point of care applications

**Johannes Kunsch[1], Marco Schossig[2] and Shankar Baliga[3]**

[1] Laser Components Germany GmbH, Olching,, Germany
[2] Infrasolid GmbH, Dresden, Germany
[3] Laser Components Detector Group, Chandler, USA

E-mail: j.kunsch@lasercomponents.com

### Status

Major applications for FTIR like point of care analysis of saliva at routine dentist visit or whole blood analysis by a community nurse are still untapped and one reason is that most recent FTIR platforms lack on scalability. Classic globar-based FTIR has been an established benchmark technology experiencing moderate growth over the decades. Since 2024, there have been notable developments: German pharmacies have begun installing affordable FTIR instruments for material identification. These devices utilize a thin $LiTaO_3$ pyroelectric crystal [1] rather than the popular DLaTGS [2]. This material choice has been supported by a review paper [3] performed comparably against a helium cooled bolometer in a Tunable Diode Laser Absorption Spectroscopy experiment at 64.7 µm [4] and shows time constants well below 100 µs [5]. High performance pyroelectrics with thin membranes by atomic lift-off have been reported recently [6], but this approach does raise RoHS discussions because of lead content. Additionally, a concept towards a "people's FTIR" was introduced [7, 8].

A study involving thousands of patients demonstrated that infrared fingerprinting with commercial FTIR systems (4 $cm^{-1}$ resolution, cryogenically cooled semiconductor detector) can predict metabolic syndromes up to 6.5 years before onset [9]. Recent research in lung cancer diagnostics using similar cohorts indicates that classic FTIR remains comparable to fs-laser-based IR spectroscopy [10]. These findings support the development of a spectroscopic twin approach, offering a unified, high-throughput method with strong potential for point-of-care medical diagnostics.

Thermal detectors and thermal emitters are combined within this chapter since both need black layers resulting even in overlapping research experience of one of the authors [11, 12]. And in fact, at the moment the absorbers of thin LTO detectors and miniaturized emitters as used in [7] are of a similar NiCr nanorod type. Metamaterial based approaches have been reported [13] as well as laser blackened surfaces [14], but so far not with ultra broad band characteristic. Until now, massive globars do dominate since they can be stably operated at up to 1300°C in air, which is almost double the temperature of miniaturized emitters at a lifetime of 100.000 hours. A figure of merit that characterizes efficiency of an FTIR has been introduced in a QFTIR paper [15]

### Current and future challenges

At first, it must be distinguished among spectroscopic twin for clinics and a workhorse version for population screening that should be as close as possible to the clinics version in terms of medical output. However, the commercial equipment that has been used so far is not globally scalable for point of care (POC) diagnostics since it is overall still too bulky in terms of size, weight, and power consumption, uses a cooled detector and is non-transparent for the user. The latter becomes a problem since the European Union Artificial Intelligence Act deems the use of AI in medical diagnostics as high risk [16] and the best strategy is to focus on simplicity and transparency from beginning.

The initial challenge is to convince the community, that a redesign of FTIR details, i.e. a technology with TRL 10 level, becomes mandatory for large scale medical. One crucial technical argument is displayed in Fig. 1. It highlights the fact, that the vast majority of commercial uncooled instruments does use a pyroelectric detector in voltage mode, i.e. a slow design resulting in the need for individual correction functions [17].

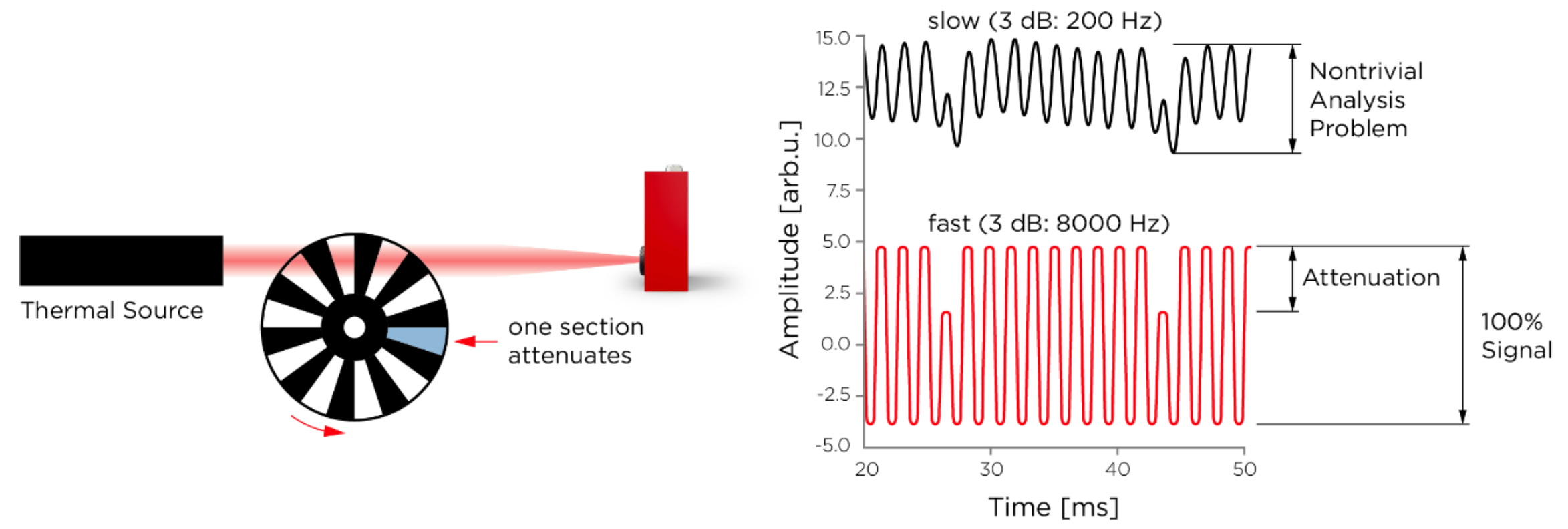

**Figure 1. Left:** Schematic experimental setup of the "Signal Waveform Challenge": A chopper blade with reduced transmission in one section in front of a thermal emitter. **Right:** Corresponding signals of pyroelectric receivers to this experiment. The grey line (MPY-01, WiredSense) needs correction functions [17] while the red line (PR-No1 IR, Laser Components) has "What you see is what you get" quality.

Cooled photodiodes easily overcome this limitation and are therefore used routinely. However, they have disadvantages such as smaller active area, limited spectral range with individual and temperature dependent spectral cut-off behaviour, and they act as an additional heat source. Recent key papers [9,10,18] stopped at 930 $cm^{-1}$. Uncooled detection has the potential to catch up by wider spectral range [19, 20]. Even an extension down to 400 $cm^{-1}$ has been used at quantitative vaccine investigations [20] which makes sense to use in general since the spectroscopic twin is a method open for adaptation even of future findings into old data. The FTIR platform introduced to pharmacies does support a wide spectral range since it uses a diamond ATR and avoids sample tourism. Noise, drift and overall comparability of health parameters require further investigation and optimization.

Another challenge is reducing the power consumption of sources into single digit number Watt range by reducing size while still having an isothermal hot spot with high emissivity, adequate temperature, mechanical stability, long lifespan, and low drift and noise all of which warrant further investigation.

## Advances in science and technology to meet challenges

Efficiency improvements at IR sources do require fundamental material improvements towards high temperature at ambient air including substrates that hold thin layers. Hot miniature sources at inert atmosphere could be a workhorse type compromise, but this requires advanced infrared packaging methods. A cheap method in polishing diamond would be very helpful.

The performance of thin LiTaO3 detectors must meet established "bread and butter performance" of DLaTGS detectors. This is trivial for frequencies up to a few hundred Hz, but a real challenge in the single digit kHz range [1, 3] that is vital for FTIR applications. Besides material improvements one strategy is to further improve electronics. Balanced detection is a proven method for boosting receiver performance because it effectively reduces noise, particularly common-mode noise, and increases signal fidelity. By comparing the outputs of two matched detectors and subtracting the signals, balanced detection cancels out shared noise sources while preserving the true signal, resulting in improved measurement accuracy. Therefore, this technique should be applied to thin LiTaO3 detectors as well, since matched detectors can be manufactured (see Fig. 2).

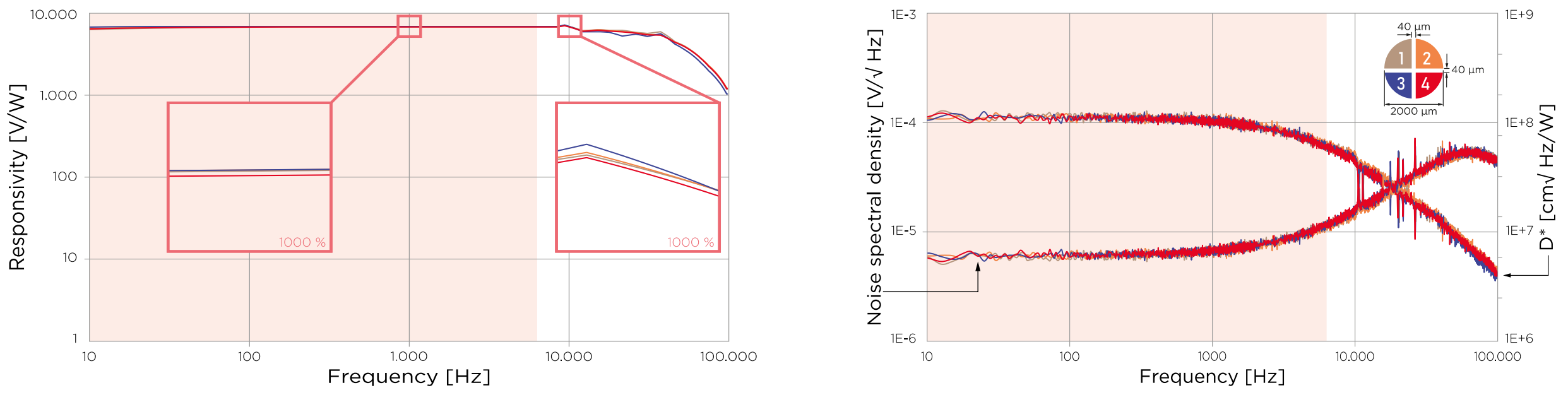

**Figure 2. Left:** Individual responsivity vs. frequency charts of a four-channel pyroelectric receiver (PR No3 IR, Laser Components) including zoom. **Right:** Corresponding individual noise vs. frequency and performance vs. frequency charts. The channel color code and the detector geometry are displayed and described in the upper right. **Summary:** The feasibility of producing four matched thin LiTaO3 detectors has been demonstrated; however, only two are needed for balanced detection, making the setup simpler. The reliable frequency range for balanced detection, indicated by the colored background in the graphs, meets the requirements of FTIR measurements.

## Concluding remarks

Usefulness of FTIR based on uncooled detection has clearly been proven more than a decade ago already [19] with very little practical impact so far. Therefore, it is time to use and amplify recent momentum and start with classic concepts for population screening by workhorses. Using existing thermal miniature sources or operating globars at reduced power and using uncooled fast pyroelectric detectors are the fastest way to bring the spectroscopic twin into real life at POC. The next 1-2 years are needed to create a start version of "Medicopeia" (in analogy to Pharmacopeia) and define and create the instrument platform including procedure manual where medical data creators and medical companies can jump on. The idea is, that the spectroscopic twin will promote adoption of mid- and long wave infrared in a manner similar to how telecommunications advanced the use of near infrared. This application is data-driven and collection of data for whole blood, saliva and urine needs to be initiated. Assuming a cost of 1 million Euro per disease per biofluid this leads easily to projects that are out of bootstrapping budgets and will likely require synchronized multinational efforts, multidisciplinary approaches and dedicated permanent application hubs. In an optimistic scenario, widespread implementation could begin within five years.

## Acknowledgements

We acknowledge the courtesy of WiredSense for measurements of Fig. 1 and Fig. 2.

# 22. Beam splitters for ultra-broadband FTIR spectroscopy

**Timothy Olsen[1,*], Christopher Harrower[1], and Łukasz Sterczewski[1]**

[1]Omega Optical LLC, Delta Campus, 21 Omega Drive, Brattleboro, VT 05301, USA
[2]Laser and Fiber Electronics Group, Faculty of Electronics, Photonics and Microsystems, Wrocław University of Science and Technology, Wyb. Stanislawa Wyspiańskiego 27, 50-370 Wrocław, Poland

*E-mail: tolsen@omega-optical.com

### Status and current challenges

Fourier transform infrared (FTIR) spectroscopy relies on an elegant and powerful concept of multi-wavelength interference on a photodetector. At the heart of the technique is the beam splitter (BS) – an optical element that splits the optical beam into two arms and then recombines them with a variable delay. This setup, known as the interferometer, is one of the most powerful concepts in optics and currently enables us to search for new physics and gravitational waves in the celebrated LIGO experiment **[1]**.

Whereas many materials in a bulk form serve as efficient BS, the optical coverage requirements for modern, ultra-broadband FTIR spectroscopy pose a significant technological challenge. No material is absorption-free; therefore dips in the transmission characteristics are inevitable. Additionally, relatively few materials have a sufficiently high refractive index $n$ to provide an equal beam splitting ratio. A notable exception is germanium, whose $n \approx 4$ in the mid-infrared and reflects ~50% of s-polarized light at 45° incidence. Unfortunately, it suffers from pronounced absorption above 10 μm of wavelength and a drastically degraded performance above 15 μm, which makes typical plane-parallel Ge plates of limited use in FTIR. Other important materials of choice are potassium bromide (KBr), zinc selenide (ZnSe), calcium flouride ($CaF_2$), and cesium iodide (CsI), each with distinct transmission ranges for certain applications. There is an important disadvantage of bulk materials for beam splitting, which is shown in Fig.1. In short, the lack of a built-in compensating element causes the interferometer arms to be unbalanced. Different optical paths seen by the beam results in a degraded interferometric efficiency and phase errors (Fig. 1a).

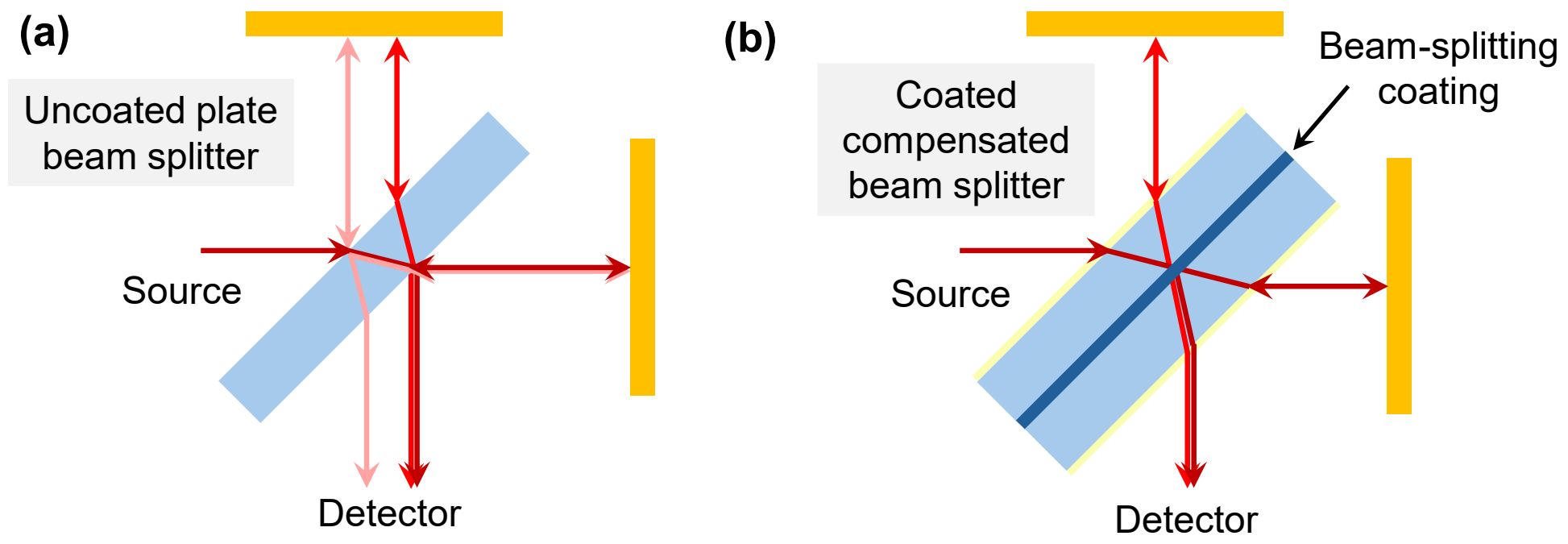

**Figure 1.** Two commonly used beam splitter topologies for broadband FTIR spectroscopy. (a) Raw, uncoated plate beam splitter made of raw high-index material like ZnSe, germanium, or silicon. The lack of a built-in compensator results in different optical paths seen by the beams in the two arms. The up-going beam passes through the beam splitter three times, while the other one passes through only once. This distorts the interferogram and reduces fringe contrast. (b) A proper, compensated beam splitter with additional protective coatings on the outer surfaces is symmetrical. In each arm, the beam travels through the beam splitter 4 times.

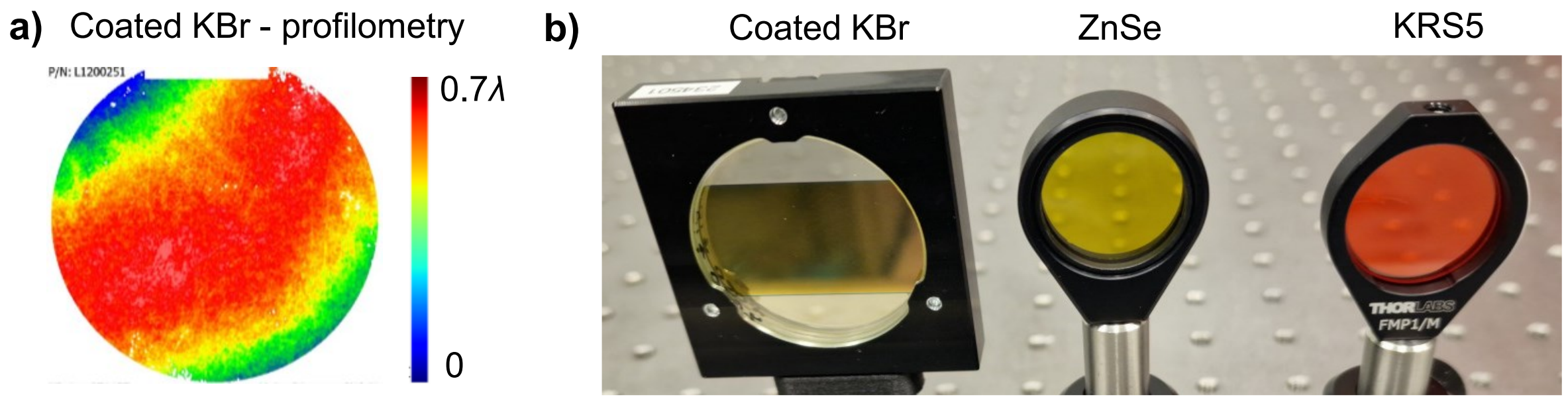

**Figure 2.** (a) Beam splitter made of two stacked and coated KBr plates after assembly. To ensure good interferometric performance, the tilt is reduced below the test wavelength of $\lambda$=.632 µm, confirmed with a phase shifting interferometer. (b) Photography of three commonly used beam splitters for broadband FTIR spectroscopy. Picture taken by Mr. Jakub Mnich at Wroclaw University of Science and Technology, Poland.

A practical solution to this problem is the BS design shown in **Fig. 3b**. A stack of two identical plates with a beam-splitting coating interface between them provides identical paths, which minimizes phase errors and wavefront distortion. The latter is of paramount importance when incoherent sources, such as globars **[2]**, are used, as even the tiniest tilt of the wavefront drastically reduces interferogram amplitude – the signal measured by the photodetector in FTIR. Such BSs are often made of materials with exceptionally wide transmission ranges like KBr that have hygroscopic properties. Notably, it is the broadband BS coating on the KBr substrate that makes the interference possible. To ensure the BS's resistance to ambient conditions, special coatings are deposited on its surfaces. It is non-trivial to ensure that this additional optical interface does not introduce additional losses and typically requires a proprietary composition of thin films developed over many years.

In this topology, the assembly process is also critical as it is very easy to distort the surface figure of the optic during the assembly process. Carefully designed mounting techniques are used to assure the required flatness for good interference at the beamsplitter surface of the optic. **Fig. 2a** shows the surface flatness of measurement of an example KBr beam splitter manufactured by Omega LLC, USA, taken with a phase shifting interferometer. The highest deviation of thickness does not exceed 0.7$\lambda$ at the test wavelength of $\lambda$ =0.632 µm. In **Fig. 2b,** a coated compensated KBr beam splitter (Omega LLC) is shown together with two other popular plate beam splitters made of ZnSe (working up to ~20 µm) and KRS5 (toxic $TiBr_{42}I_{58}$, up to 40 µm). Unfortunately, the performance of raw plane-parallel plates is typically sub-optimal for interferometric applications, which require additional manufacturing steps.

### Practical ultra-broadband beam splitter with commercial availability – performance characteristics

Whereas so far we have not discussed diamond BS material **[3,4]** which is close in index to ZnSe – its price and sub-optimal BS ratio (~70/30) often exclude it from many applications. This material is typically grown in chemical vapor deposition (CVD) chambers and provides an unparalleled transmission window from the ultraviolet (0.2 µm) to the THz range (100 µm / 3 THz). Unfortunately, compared to an engineered KBr beam splitter, it introduces severe dips in the spectrum between 4–6 µm of wavelength due to multi-phonon absorption **[5]**. This is why many applications prefer the characteristics offered by KBr.

An example transmittance/reflectance ($T(\lambda)/R(\lambda)$) curve of the compensated KBr beam splitter is shown in **Fig. 3**. While the curves end at 25 µm, it has been confirmed experimentally that the BS remains useful up to 30 µm **[6]**. From these characteristics, it is possible to retrieve the interferometric modulation efficiency **[7]** defined as $\eta = 4R(\lambda)T(\lambda)$. This value defines the obtainable fringe contrast for a given beam splitting ratio. Although it is obviously optimal with $\eta = 1$ for a 50/50 ratio, even at 70/30 or 80/20, this value reaches favorable 0.84 and 0.64, respectively. This is why even at the longest wavelengths ($\lambda$>15 μm), the modulation efficiency (shown in Fig. 4) is still high for the KBr beam splitter, as calculated using data from Fig. 3.

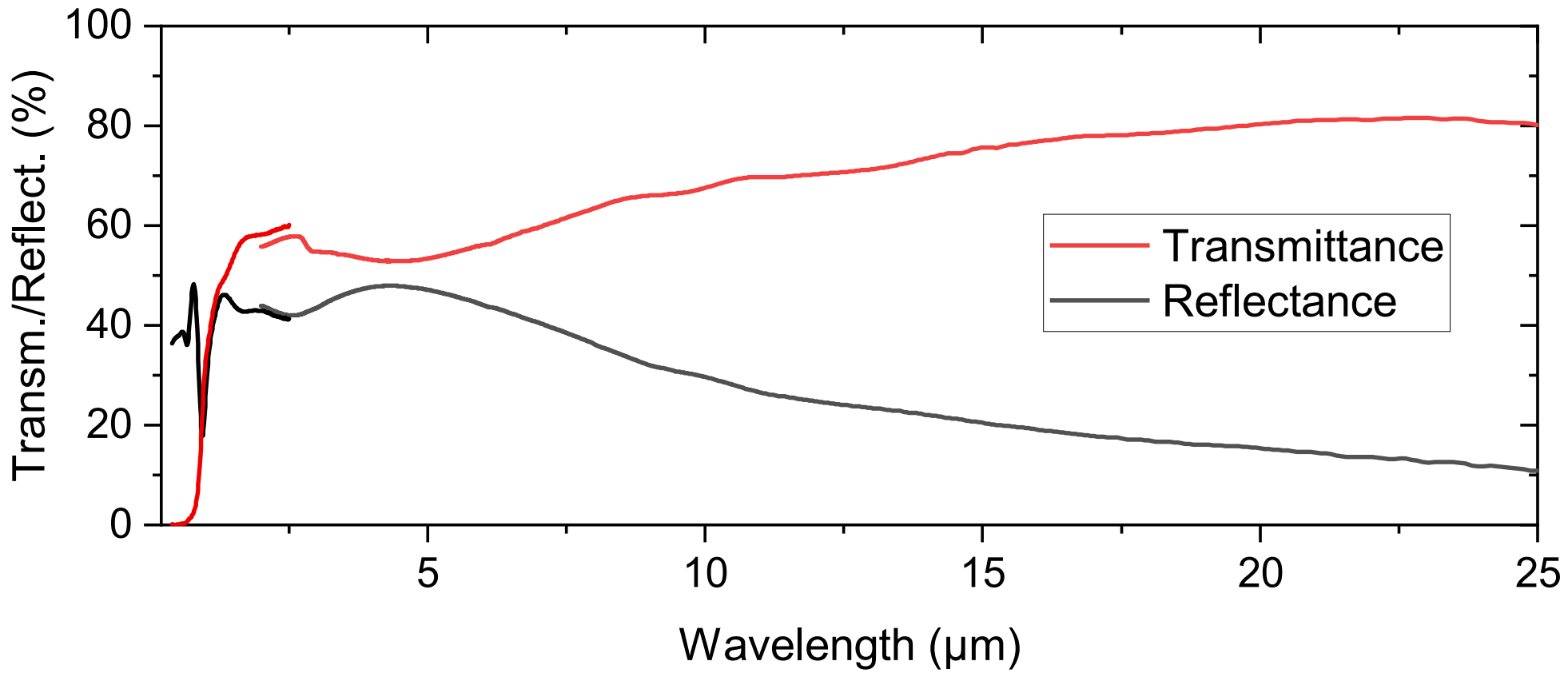

**Figure 3.** Transmittance/reflectance curve of the compensated KBr beam splitter. At ~2 μm one can see overlapping measurements from two different instruments, which disagree in the single percent range. Due to the BS coating, the transmission reaches practical (>40%) values above 1 μm.

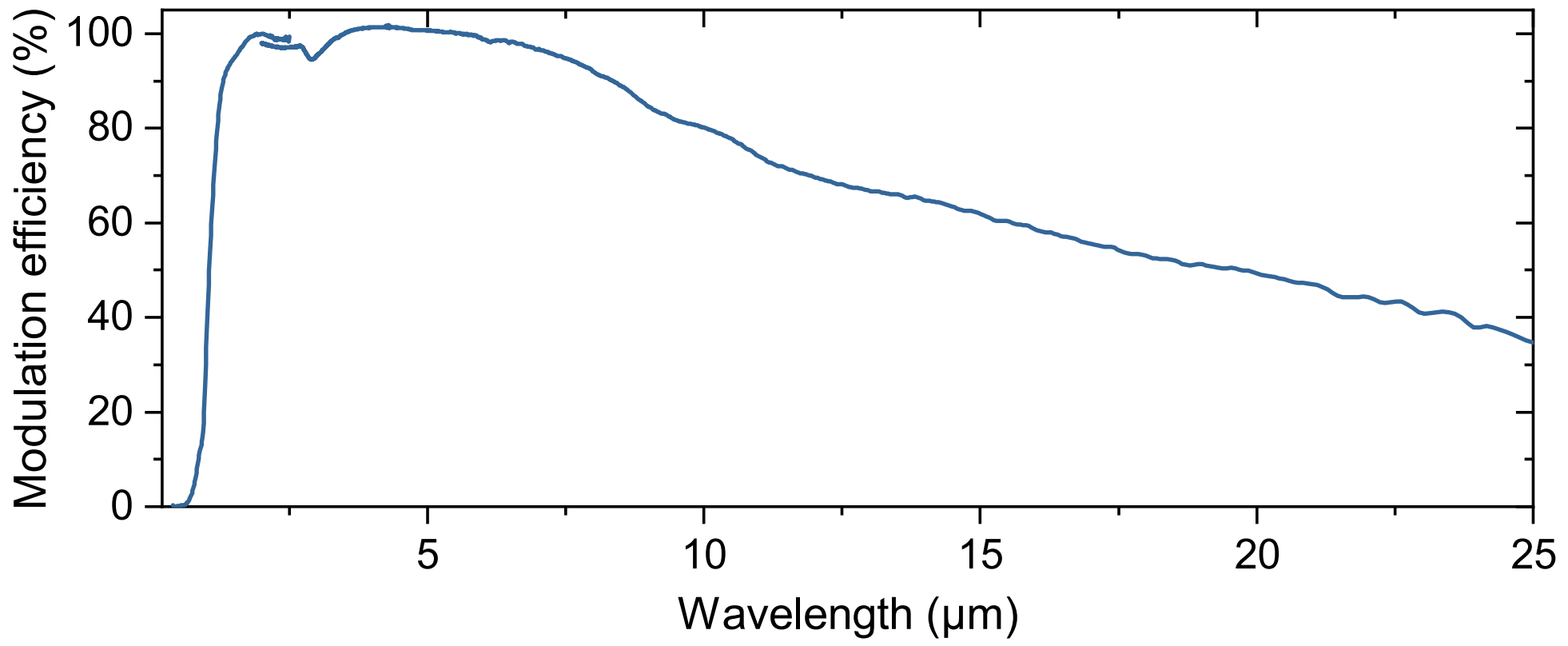

**Figure 4.** Modulation efficiency calculated as $\eta = 4R(\lambda)T(\lambda)$ using data from **Fig. 3**. Optimal performance is obtained to almost 7 μm; however a drop to 50% (−3 dB) occurs at ~17 μm with a slow decay rate.

### Concluding remarks and outlook

Beam splitters represent an essential component of Fourier spectrometers. Despite the maturity of this spectroscopic technique, the availability of ultra-broadband high-performance beam splitting optics is scarce. Fortunately, this is now changing with the growing market demand for custom Fourier spectrometers. There is also much interest in covering the THz range (0.1–10 THz) **[8]**, which calls for extending the beam splitter portfolio. For instance, float-zone silicon is known to yield enough modulation efficiency for the THz range. Omega Optical is currently developing these new BSs.

### Disclosures

T. O. and C. H. are employees of Omega Optical LLC offering coated beam splitters.

### Acknowledgements

L. A. S. acknowledges funding from the European Union (ERC Starting Grant, TeraERC, 101117433). Views and opinions expressed are, however, those of the authors only and do not necessarily reflect those of the European Union or the European Research Council Executive Agency. Neither the European Union nor the granting authority can be held responsible for them.

# 23. Quantum infrared spectroscopy with undetected photons

**Ivan Zorin[1], Chiara Lindner[2], Shigeki Takeuchi[3], Sven Ramelow[4,5] and Paul Gattinger[1]**

[1] Research Center for Non-Destructive Testing, Science Park 2, Altenberger Str. 69, 4040 Linz, Austria
[2] Fraunhofer Institute for Physical Measurement Techniques IPM, Georges-Köhler-Allee 301, 79110 Freiburg, Germany
[3] Department of Electronic Science and Engineering, Kyoto University, Kyotodaigakukatsura, Nishikyo-ku, Kyoto 615-8510, Japan
[4] Ferdinand-Braun-Institut (FBH), Gustav-Kirchhoff-Str. 4, 12489 Berlin, Germany
[5] Humboldt-Universität zu Berlin, Institut für Physik, Newtonstraße 15, 12489 Berlin, Germany

E-mail: ivan.zorin@recendt.at

## Status

Quantum optics provides additional degrees of freedom for generation and manipulation of light and thereby enables advanced measurement techniques that rely on the fundamental quantum nature of electromagnetic radiation. Metrology with undetected photons [1], which exploits quantum interference of entangled photons in a nonlinear interferometer, has emerged as a very promising approach to mid-infrared (mid-IR) spectroscopy. The techniques united under this umbrella can be used to access mid-IR information, without direct mid-IR sources and detectors thus circumventing the need for classical IR devices, and they operate at ultra-low power levels (pico- to nanowatts) unattainable for traditional instruments such as Fourier transform infrared spectrometers (FTIR). Sensing with undetected photons, particularly its use in mid-infrared (mid-IR) measurements, is a very recent development, as evidenced by the fact that most scientific publications and reports emerged in the last five years. Research in this area remains mostly driven by groups working on fundamental quantum optics; progress towards practical implementations and benchmarking for routine IR spectroscopy (including optimization and miniaturization) is in early stage [2–4] and expected to be further advanced. These methods exhibit strong potential for established IR biomedical applications, as they offer room-temperature operation, reduced costs, simpler maintenance, high stability, and high signal-to-noise for ultra-low incident power levels.

The paradigm of IR spectroscopy with undetected photons is based on nonlinear quantum interferometry [5] and the principle of induced coherence without induced emission [6]. Entangled photon pair sources can be engineered such that one photon's frequency is in a spectral region that can be efficiently detected (e.g., the visible or near-IR), while its partner is in the wavelength range relevant to the measurement of interest. The method was first introduced and demonstrated in quantum imaging by Lemos et al. [7], led by Nobel laureate Anton Zeilinger, and soon after applied to spectroscopy in the mid-IR [8]. The basic arrangement of a nonlinear quantum interferometer, commonly referred to as an SU(1,1) interferometer, along with the basic operational principles relevant for IR spectroscopy are shown in Fig. 1. In the basic scheme, a pump photon decays into a pair of correlated signal and idler photons via spontaneous parametric down-conversion (SPDC) in two sequentially arranged nonlinear crystals. Due to the photon path indistinguishability for overlapped beams, interference can be simultaneously observed in both domains. The interference pattern and its visibility depend on the amplitudes and phases of all interacting photons. An alteration in the idler path modifies the observed interference. Thus, information about a sample with a certain IR transmission amplitude $|\mathbf{T}|$ and dispersion $\boldsymbol{\gamma}$ can be retrieved in the spectral domain attainable by Si-detectors. Since the method is fundamentally interferometric, sensing can be implemented using either a quantum FTIR (QFTIR) detection scheme [2,3,9,10] or visible grating spectrometers [11], as the Fellgett advantage is effectively absent in this regime. A third possible modality is to use the spectrally dependent, spatial interference fringes captured with a 2D CCD camera [12].

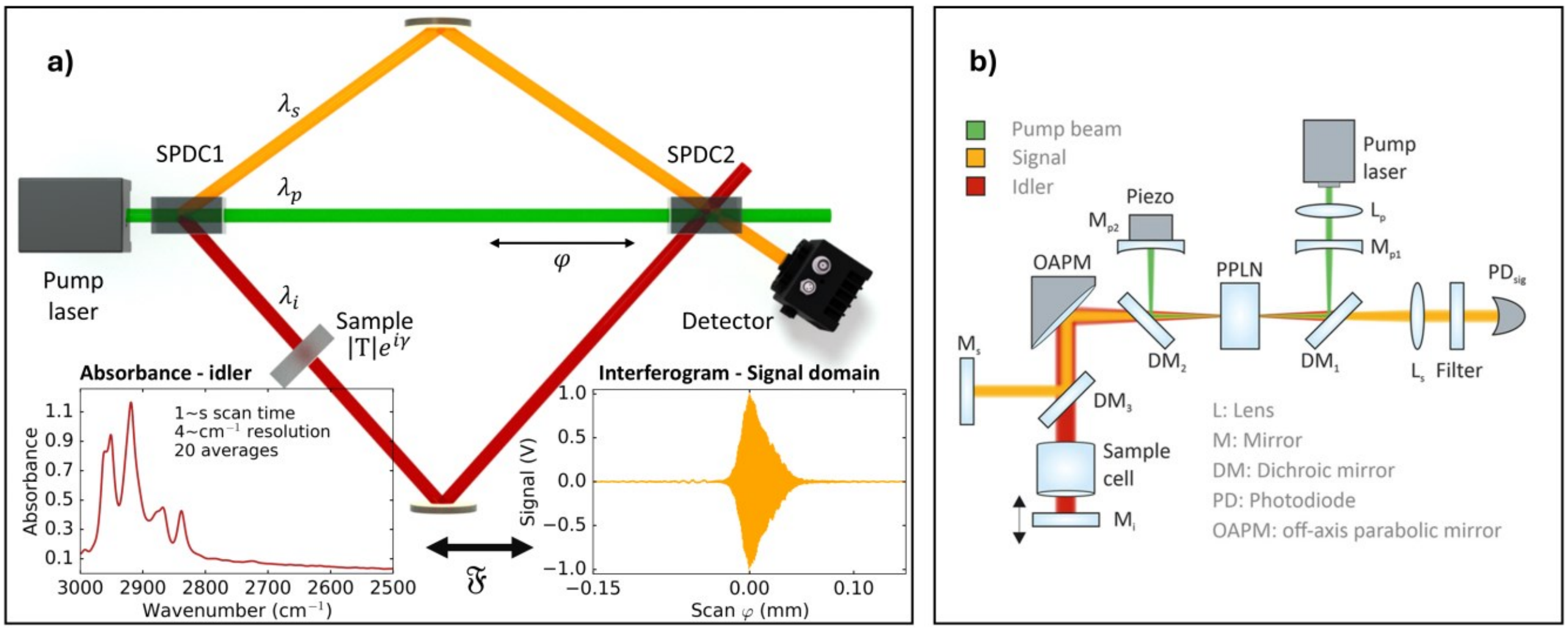

**Figure 1. (a)** Scheme of a nonlinear interferometer for routine mid-IR QFTIR spectroscopy. Broadband signal $\lambda_s$ and idler $\lambda_i$ correlated photon pairs (near-IR and mid-IR) are generated in two nonlinear crystals pumped by a visible laser $\lambda_p$. The photon paths are aligned to ensure indistinguishability; interference can be observed in both domains. A sample placed in the idler arm partially alters the indistinguishability of the detected $\lambda_s$ photons at the second crystal and therefore modifies the interference pattern. Thus, the mid-IR samples' properties can be retrieved by detection of the $\lambda_s$ photons using e.g. Si-technology. **(b)** Schematic setup of the pump-enhanced nonlinear interferometer for high-sensitivity QFTIR (see details in [20], adapted from Lindner et al., APL Photonics 8, 051301, 2023; licensed under a Creative Commons Attribution (CC BY) license).

The practical implementation of a system for mid-IR spectroscopy is more complex than the schematic shown in Fig. 1, with the differences hidden in detail. For example, the two-crystal configuration is implemented with a double pass through a single crystal; the photon paths can be made collinear using type-0 quasi-phase matching; and the spectral bandwidths can be engineered to be rather broadband. Such an advanced nonlinear interferometer for sensitivity-enhanced quantum-FTIR spectroscopy in shown in Fig. 1 (b).

## Current and future challenges

The main challenges hindering broader adoption of this technology include extending the spectral range and improving the signal-to-noise ratio. Significant engineering efforts are required to develop instruments suitable for routine spectroscopic scenarios. Research groups in quantum sensing with undetected photons acknowledge these limitations and work to enhance the spectral brightness of SPDC emission and broaden its bandwidth. Furthermore, the technique is advancing towards applications in chemical imaging and microscopy [13,14].

## Advances in science and technology to meet challenges

The spectral range of quantum mid-IR spectroscopy can be extended by employing alternative nonlinear materials, such as non-ferroelectric oxide crystals with longer wavelength cut-offs [15,16]. Another promising approach involves advanced crystal engineering, for example, using multiple poling designs for the ultra-broadband QFTIR [17], depicted in Fig. 2. As the nonlinear interferometers operate in the shot-noise dominant regime [2,3], the signal-to-noise ratio can be essentially increased by increasing the photon rate. This can be done using cavity-enhanced schemes shown in Fig. 1(b)[18]; in the high time-resolution regime the approach can benefit from pulsed pump schemes[19].

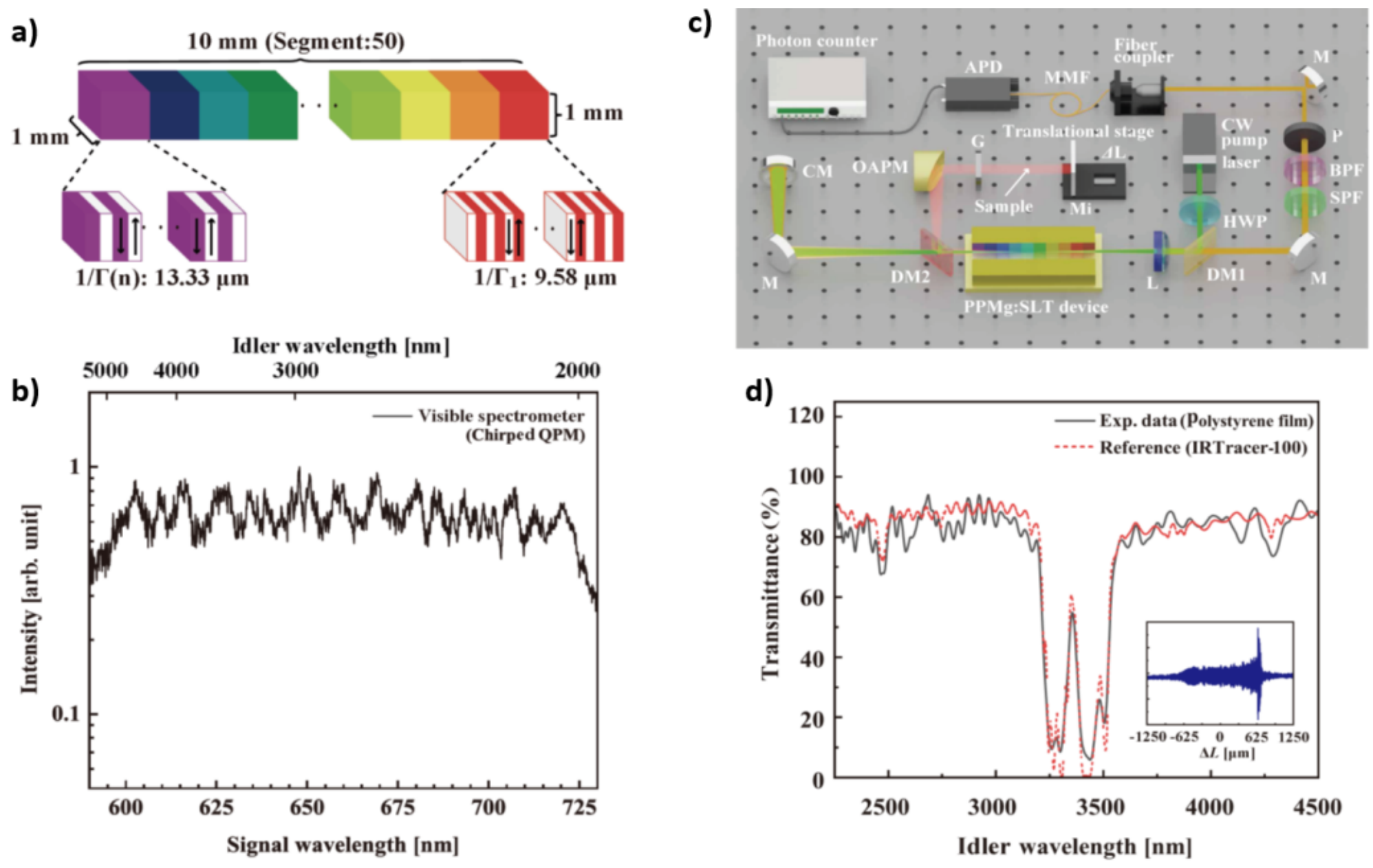

**Figure 2.** Ultra-broadband quantum infrared spectroscopy adapted from Ref. [19]. (a) Schematic view of a chirped quasi-phase-matched (QPM) device made of periodically polled Magnesium-doped stoichiometric Lithium Tantalate (PPMg-SLT). The designed wavelengths of visible and MIR photon pairs are 595 to 725 nm and 2000 to 5000 nm; (b) Experimental signal (visible) spectrum for visible and MIR photon pairs generated from the chirped QPM device; (c) Schematic view of the ultra-broadband QFTIR (details in Ref. [19]); (d) Transmittance spectrum of polystyrene film taken by the QFTIR (Exp. data) and conventional FTIR.

## Concluding remarks

Quantum IR spectroscopy with undetected photons exhibits strong potential for spectroscopic and medical applications due to its unique advantages such as shot-noise limited operation and ultra-low probing powers. These techniques are still in their early development phase and require further research and engineering to enable deployment in field conditions and practical use by spectroscopists and medical professionals; however, attempts for adoption are already in place. Moreover, IR sensing with undetected photons offers promising opportunities for scaling and miniaturization alike classical instruments [20].

## Acknowledgements

IZ and PG acknowledge funding by research subsidies granted by the government of Upper Austria: Quick (Wi-2022-597365/18-Au). CL acknowledges funding by Fraunhofer-Gesellschaft (Lighthouse Project QUILT). SR acknowledges funding by BMFTR (SIM-QPla 13N15944, QEED 13N16384) and the Einstein Foundation Berlin (EJF-2021-681). ST acknowledges support from MEXT Q-LEAP (JPMXS0118067634) and JST ERATO (JPMJER2402).

# Coherent Light Sources:
# 24. A perspective on vertically emitting quantum cascade lasers with low power dissipation

**David Stark[1], Réka-Eszter Vass[1], Killian Keller[1], Alessio Cargioli[1], Mattias Beck[1], Jérôme Faist[1]**

[1] Department of Physics, Institute for Quantum Electronics, ETH Zurich, 8093, Zurich

E-mail: starkd@ethz.ch , jfaist@ethz.ch

## Status

Quantum Cascade Lasers (QCLs) [1] based on the mature InP material platform, delivering high-performance emission across 3.4–16 µm range, are excellent candidates for sensing of multiple targets in complex biomolecular or gas mixtures. Within this material platform, single-mode distributed-feedback (DFB) QCLs have emerged as the standard targeting narrow features (~< 6 $cm^{-1}$), while DFB QCL arrays [2] provide a route to cover broader spectral windows via numerous discrete wavelengths (~100–200 $cm^{-1}$). Yet translating these sources into scalable and portable applications is challenging due to the power and cost requirements of a single wavelength. These constraints stem from the intersubband gain of QCLs, which entails both typical threshold current densities in the $kA{\cdot}cm^{-2}$ range and in-plane, edge-emitting devices. The resulting multi-watt dissipation and elevated testing and packaging costs hinder widespread adoption.

To circumvent these limitations imposed by the intersubband gain, two complementary research directions have emerged: (i) device miniaturization to reduce power dissipation (see review [3]), and (ii) surface-emitting QCLs utilizing outcoupling elements — conceptually belonging to the broader Horizontal-Cavity Surface-Emitting Laser (HCSEL) family [4] — to enable wafer-level testing and scalable array operation (see review [5] and representative primary works [6,7]). More recently, these two research directions have been unified to a third direction (III): the Quantum Cascade Surface Emitting Laser (QCSEL) architecture [8, 9], combining miniaturization and surface emission (Figure 1). Ultimately, the vision is to provide the mid-infrared counterpart to the Vertical Cavity Surface Emitting Laser (VCSEL), unlocking portable low-power multi-wavelength sensing applications at scale. In the following, we compare these three research directions and then highlight the key challenges that outline future developments.

## Current and future challenges

Table 1 compares a selection of state-of-the-art publications across the three directions, chosen to represent: (I) edge-emitters combining lowest continuous-wave (CW) power dissipation with maximum efficiency; (II) vertical emitters demonstrating the highest reported efficiencies across different device architectures; and (III) recent results for the emerging QCSEL architecture. We analyze these architectures and their specific challenges considering the target specifications for portable spectroscopy: room-temperature CW operation and single-mode emission delivering milliwatt-level optical power within a sub-Watt thermal budget.

To effectively reduce power dissipation via device miniaturization, the challenge lies in maintaining low waveguide and mirror losses to avoid compromising the threshold current density and wall-plug efficiency. Cheng et al. [10] achieved this by combining a short 0.5-mm buried-heterostructure Fabry-Pérot cavity with coatings on both facets and epi-down mounting for optimal heat extraction. This approach enabled a power dissipation below 0.43 W, with single-mode emission provided by the wide free spectral range of the short cavity. In contrast, Briggs et al. [11] targeted deterministic spectral control and simplified packaging by employing a 1-mm lateral-DFB ridge mounted epi-up with only a back-facet coating. This configuration resulted in a maximum power dissipation of 1.6 W and an efficiency of about 1.1%.

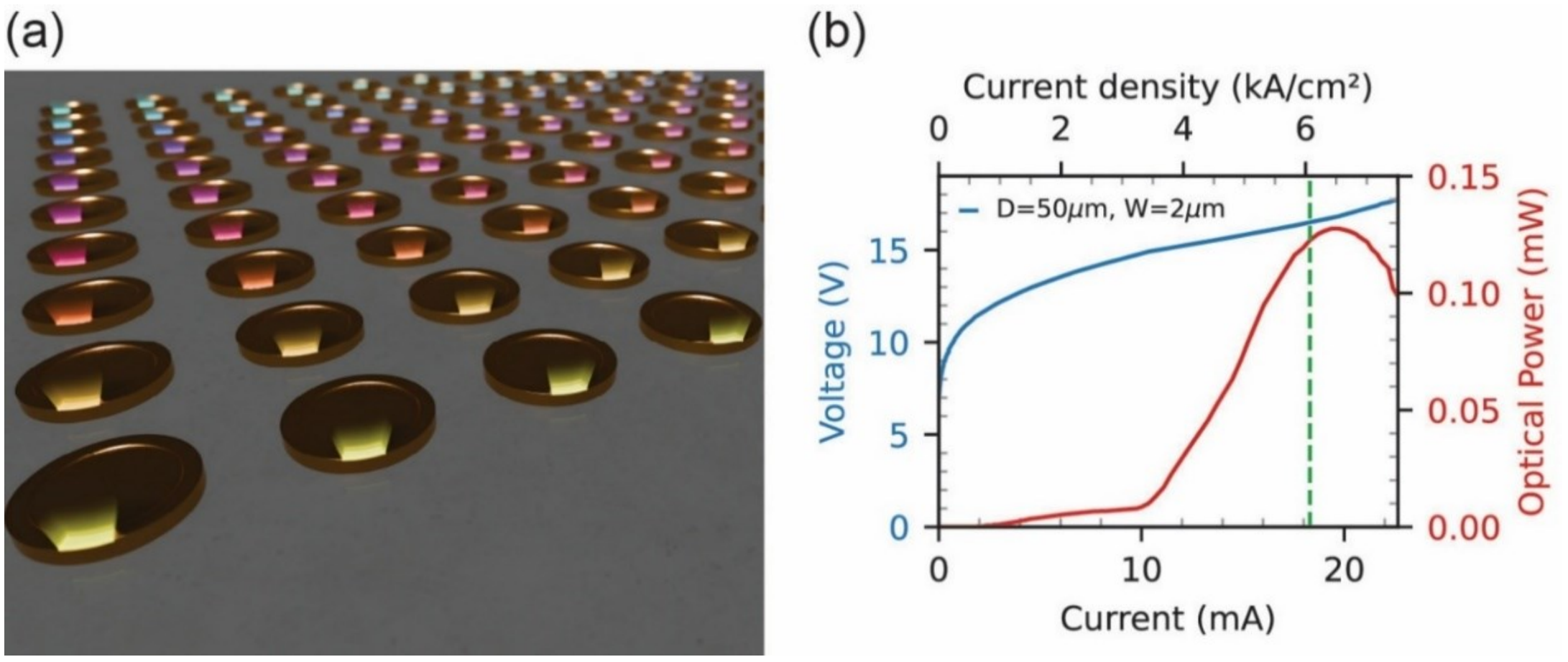

**Figure 1. (a)** Conceptual visualization of the Quantum Cascade Surface Emitting Laser (QCSEL) platform, illustrating the potential for dense, multi-wavelength arrays on a single chip. **(b)** Representative Light-Current-Voltage characteristics of a miniaturized ring QCSEL, adapted from [9]. The device exhibits a threshold current approaching 10 mA and a maximum pulsed efficiency of 0.038% (indicated with the green dashed line).

To enable vertically emitting QCLs—whether through the surface or the substrate—the challenge is to integrate an efficient outcoupling element. The most common approach is to integrate a second-order grating near the waveguide, providing not only vertical outcoupling but also the feedback for single-mode selection. Mujagić et al. [12] employed this strategy using a ring-ridge cavity with a second-order surface grating, demonstrating pulsed single-mode operation with an efficiency of about 1.3% in an epi-up configuration. To address the thermal limitations and bi-directional emission of dielectric gratings, Yao et al. [6] implemented a linear ridge with a buried second-order grating and an overlying metal layer. This scheme is naturally suited for epi-down mounting and enabled CW operation with ∼1.3% efficiency.

Common to the vertical emitters discussed above is a large active area (>$10^{-4}$ $cm^2$), which inevitably results in multi-watt power dissipation. To overcome this, recent QCSEL demonstrations [8, 9] achieved extreme cavity miniaturization by leveraging buried-heterostructures, deep dry-etching, and wafer-level coatings. In this architecture, mode-control is targeted by the short cavity perimeters, while surface emission is provided by short, off-resonant second-order grating sections. This approach has successfully yielded CW operation with sub-Watt power dissipation for both linear and ring geometries. However, the critical challenge is evident in the efficiency, which remains limited to below 0.1% in pulsed mode. This discrepancy highlights that while geometric scalability has been solved, the next phase of research must rigorously address the extraction efficiency, beam profile, mode selection, and thermal management.

## Advances in science and technology to meet challenges

To transform the QCSEL from a feasibility demonstration to the source of choice for low-power sensing applications, future developments must prioritize boosting the efficiency beyond the 1% level set by low-power edge-emitters. First, the optical extraction efficiency must be engineered using insights from far-field measurements. Strategies to enhance outcoupling can be adapted from established vertical emitters: extending the grating section length, employing high-contrast surface or metallic gratings, or integrating a horizontal metallic mirror to enforce unidirectional emission. Simultaneously, minimizing parasitic mirror and waveguide losses remains crucial. We anticipate that by further suppressing these losses, ring cavities with diameters below 20 μm will become feasible, promising threshold currents of just a few milliamperes.

Second, robust mode control must be demonstrated. Future designs can either adopt the vertical emitter approach—harnessing a second-order grating along the full cavity perimeter—or exploit the geometric mode selection of ultra-short perimeters. Specifically, for microring cavities, we expect that deterministic single-mode selection via perimeter tuning is now within reach.

**Table 1.** Representative mid-infrared QCL devices operating near 20°C used to benchmark (I) low-dissipation edge emitters, (II) vertical emitters, and (III) QCSELs. Unless noted, devices emit in the 4.4–4.8 μm range.

| Category | Ref. | Cavity | Driving | Mounting | Area ($10^{-5}$ cm$^2$) | $P_{out}$ (mW) | $P_{diss}$(W) | WPE (%) |
|---|---|---|---|---|---|---|---|---|
| (I) Low-dissipation edge emitters | [10] | Linear BH (Fabry-Perot) | CW | Epi-down | 1.6 | 11 | 0.43 | 2.49 |
| | [11] | Linear Ridge DFB | CW | Epi-up | 4.0 | 17 | 1.61 | 1.06 |
| (II) Vertical emitters | [12] | Ring Ridge DFB | Pulsed | Epi-up | 12.3 | 126 | 12.9 | 1.34 [a] |
| | [6] | Linear Ridge DFB | CW | Epi-down | 31.8 | 94 | 7.1 | 1.32 |
| (III) QCSELs | [8] | Linear BH FP | CW | Epi-up | 1.1 | 0.025 | 0.59 | 0.004 [a] |
| | | | Pulsed | | 1.5 | 0.94 | 1.38 | 0.073 |
| | [9] | Ring BH | CW | | 0.9 | 0.003 | 0.37 | 0.0008 |
| | | | Pulsed | | 0.3 | 0.13 | 0.33 | 0.038 |

[a] Emission wavelength near 8 μm.

To optimize these parameters and validate theoretical efficiency limits, the combination of small footprint and vertical emission enables the automated characterization of large numbers of devices to efficiently map the parameter space. Beyond the optical cavity, future developments should integrate dedicated active regions optimized for narrow gain bandwidth and high internal efficiency at target wavelengths. Lastly, although sub-Watt dissipation levels have been proven feasible, thermal extraction holds significant room for improvement through the implementation of double-channel geometries and established epi-down packaging.

### Concluding remarks

The availability of battery-operated, low-power, and low-cost semiconductor mid-infrared lasers holds the potential to revolutionize proactive healthcare. By enabling the scalable and dense integration of discrete wavelengths, these sources will unlock low-cost, multi-species sensing, paving the way for next-generation breath analyzers, continuous biomarker monitoring, and wearables.

With its ultra-compact footprint and surface-emitting capabilities, the Quantum Cascade Surface Emitting Laser (QCSEL) architecture offers a compelling pathway to realize this vision. In this chapter, we have positioned this emerging architecture with respect to state-of-the-art low-power edge emitters and large-area vertical emitters, identifying key challenges, benchmarking performance, and outlining strategies to bridge the current efficiency gap. By overcoming the remaining hurdles, the QCSEL aims to replicate the transformative impact of the near-infrared VCSEL—democratizing laser-based mid-infrared spectroscopy and transforming high-end diagnostics into ubiquitous components of everyday life.

### Acknowledgements

The authors gratefully acknowledge the financial support from Innosuisse—Swiss Innovation Agency (Innovation Projects Grant No 115.134.1).

# 25. Cascaded laser sources for spectroscopic applications in the MIR

**Robert Weih[1*], Josephine Nauschütz[1], Julian Scheuermann[1], Jordan Fordyce[1], Johannes Koeth[1]**
[1] nanoplus Advanced Photonics Gerbrunn GmbH, Oberer Kirschberg 4, 97218, Gerbrunn, Germany

E-mail: *robert.weih@nanoplus.com

## Status

In 1971 the scheme of cascading multiple intraband transitions in series was proposed [1] and two decades later experimentally demonstrated [2] opening the scientific field of quantum cascade lasers (QCLs). Based on the cascading scheme a single injected electron can generate multiple photons enabling quantum efficiencies > 1. Moreover, the emission wavelength is not mainly defined by the semiconductor bandgap but rather by a layer thickness within a superlattice like structure. A wide spectral range reaching from 2.6 µm up to more than 250 µm (cryogenic temperature operation) was covered so far. Furthermore, multi-watt output power has been demonstrated for Fabry-Pérot and single mode devices operating in continuous wave (cw) mode at room temperature [3,4]. Commercially available devices are available roughly in the range between 3.2 and 14.0 µm. Another device making use of the cascading scheme is the interband cascade laser (ICL) which was first proposed in 1995 [5]. Unlike in the QCLs an interband transition in a type-II quantum well is utilized. The broken bandgap alignment of the two binaries GaSb and InAs within the 6.1 Å semiconductor family enables tunnelling between conduction- and valence band states at a semi-metallic interface. In 2008 continuous wave operation at room temperature was achieved at a wavelength of 3.75 µm [6] and soon after record low threshold current densities below 100 A/cm$^2$ could be demonstrated [7]. Since then, several design optimizations have led to continuous performance improvements paving the way towards practical usage in several applications [8–10]. Meanwhile, ICLs that operate in cw mode at room temperature and above are commercially available in a wide spectral range from 2.6 to 6.5 µm. Nowadays, ICLs and QCLs are key building blocks for numerous applications in the MIR spectral region, ranging from industrial process control to environmental and medical sensing [11,12]. They can be electrically driven and are very reliable and robust. Nevertheless, several applications would benefit from compact sources that can cover a broad spectral range rather than a small tuning range around a designated wavelength.

## Current and future challenges

Even though the first realization of ICL and QCL devices was 3 decades ago the development is still ongoing, and continuous improvements are being reported. For ICLs, significant design improvements have recently been found as for example the mitigating of valence intersubband absorption within the active quantum wells [9], the implementation of hybrid cladding layers [13] or the incorporation of InAsP layers in the active region that enabled lasing at wavelengths as long as 13µm [14]. Apart from these improvements, ICLs were realized on Si [15] opening new possibilities for monolithic integration of MIR emitters with silicon photonics. Moreover, further improvements in the design of active stage design are ongoing hand in hand with the development of advanced simulation tools [16]. As for QCLs, device improvements for classical Fabry -Pérot or DFB lasers are being developed. Miniaturized single-mode ring QCLs ($\lambda$ = 4.8 µm) with a room temperature continuous wave threshold power of only 267 mW have recently been demonstrated [17]. Furthermore, a ring laser architecture was developed with a tuning range of 33 cm$^{-1}$ [18] that extends the tuning range of a classical DFB QCL. This might even be improved by employing heterogeneous cascaded gain material which is a unique advantage of QCLs resulting from the intraband transition nature. Even though this principle was already employed in 2006 [19] it was developed further within the last decades and is meanwhile being used in industrial environments to provide a laser source for applications that require a wide spectral coverage within the MIR. In Fig. 1 measured spectra covering the wavelength range from 7.99 µm to 9.08 µm from a monolithic DFB QCL array are shown. The array consists of 8 lasers processed on a QCL wafer that employs a heterogeneous cascaded active region based on 3 different stage designs. By altering the number of emitters, the spectral position of each emitter, or even the bandwidth coverage of the array, various medical applications that address broad spectral features, such as for example glucose or protein monitoring, can be tackled with such devices. Beyond classical QCL sources, advanced devices such as frequency combs have been demonstrated [20] and proven to be a s suitable light source for real time broadband spectroscopic applications.

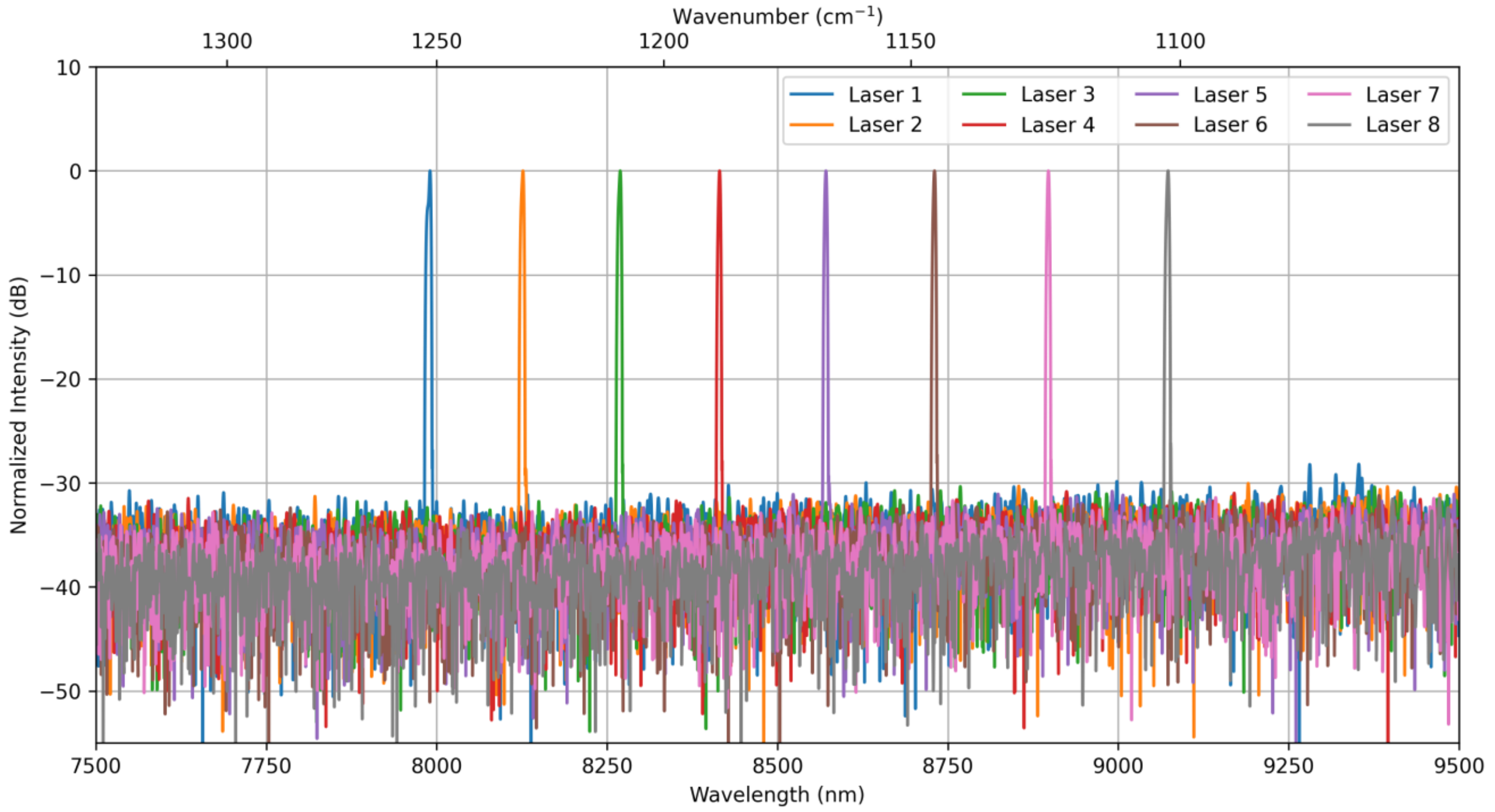

**Figure 1**. Emission spectra of an industrial grade monolithic QCL DFB array covering a spectral range of 150 cm$^{-1}$. The lasers are driven in pulsed mode (400 kHz, 150 ns) at room temperature, and every emission channel reaches an average output power of > 10 mW and a SMSR of 30 dB.

## Advances in science and technology to meet challenges

Scientifically there is still room for performance improvement regarding the efficiency and spectral coverage both for ICLs and QCLs. Advanced simulation capabilities need be used in combination with mature epitaxial growth and processing technology to address key characteristics such as wall plug efficiency, overall power consumption, output power and tuneability. Furthermore, ICLs can serve as a basis for surface emitting devices such as VCSELs whereas for QCLs additional outcoupling structures are necessary due to their dipole matrix orientation. To address the demand for compact and sensitive sensors further developments are required regarding monolithic integration of passive waveguides and detectors or hybrid integration with MIR compatible photonic platforms

## Concluding remarks

Already today ICLs and QCLs are mature mid-infrared laser sources with unique properties, each with their own advantages. ICLs stand out with very low operation power of only a few tens of mW in the 2.6–6.5 µm region, making them ideal sources for spectroscopic applications that rely on low power consumption. Further research will enable them to extend this wavelength range to wavelengths beyond 7 µm. QCLs can already address this long wavelength range with cw performance up to >17 µm and are therefore ideally suitable for LWIR applications. Furthermore, their capability to realize broad gain bandwidth within a single epitaxial stack stands out and can serve as a basis for applications that require large spectral bandwidth.

# 26. Background-free measurement – Field-resolved infrared spectroscopy

**Ka Fai Mak[1] and Alexander Weigel[2,3,4]**

[1] School of Optical and Electronic Information, Huazhong University of Science and Technology, Wuhan, China
[2] Center for Molecular Fingerprinting, Budapest, Hungary
[3] Max Planck Institute of Quantum Optics, Garching near Munich, Germany
[4] Faculty of Physics, Ludwig Maximilian University, Munich, Germany

E-mail: kafai.mak@hust.edu.cn

## Status

Field-resolved spectroscopy (FRS) of infrared light based on femtosecond lasers has recently emerged as an alternative to standard intensity-based detection techniques [1,2]. In this type of spectroscopy, an ultrashort infrared pulse, lasting only few tens of femtoseconds, impulsively excites resonance molecular vibrations in the sample. The resulting oscillating dipoles emit radiation as free induction decay (FID) at the resonance frequencies, continuing to last even after the passing of the ultrashort excitation pulse (Fig. 1). Such electric-field response can be measured directly in the time domain using electro-optic sampling (EOS). Originally developed for time-domain THz spectroscopy, EOS's (and thereby FRS's) square-root dependence on the electric field provides an intrinsic advantage in detection dynamic range over traditional techniques that scale with intensity, with a maximum signal-to-noise ratio of up to 17 orders of magnitude demonstrated [3]. Another strength of FRS is the capability to measure strongly attenuated and weak infrared signals [2], which allowed recently to study even vacuum fluctuations of the electric field [4].

Since the electric field is to be measured, phase stability is required for the femtosecond excitation mid-infrared pulses. Intrapulse difference-frequency-generation (IDFG) is a parametric frequency conversion technique that can generate broadband mid-infrared femtosecond pulses with intrinsic carrier-envelope-phase (CEP) stability. When driven with traditional 1-µm pump sources, such conversion suffers from low efficiency. Consequently, new 2-µm femtosecond lasers were developed to pump non-oxide crystals with higher nonlinearities [5,6] and at a lower quantum defect. Such lasers can be broadly categorized as rare-earth (e.g. Ho- and Th-doped) lasers that can sustain high average power, or transition-metal-doped II-VI lasers (such as Cr:ZnS/Cr:ZnSe lasers) that can directly produce few-cycle pulses with octave spanning spectra [6,7].

Compared to state-of-the-art FTIR spectroscopy, FRS has already demonstrated more than an order of magnitude improvement in detection sensitivity[2]. Further improvement, and when applied to health state monitoring, could substantially enhance early screening of medical conditions [8].

## Current and future challenges

### *Mid-IR source power and bandwidth*

Cr:ZnS/Cr:ZnSe lasers, similar to Ti:sapphire lasers, inherently suffer from a high quantum defect and large heat deposition. This limits the achievable average power at both 2-µm and in the longer-wave mid-infrared. This, in turn, limits the repetition rate at which spectroscopic measurements can be taken if the pulse energy is to be maintained. Ho- and Th-doped lasers, although supporting high average power, continue to be limited by their narrower emission bandwidth, restricting the corresponding spectral coverage of the generated mid-infrared pulse.

Furthermore, the efficiency of converting to the mid-infrared is still in the low tens of percent in the best cases, achieved using crystals such as zinc germanium phosphide (ZGP)[6], whose transmission cut off are at

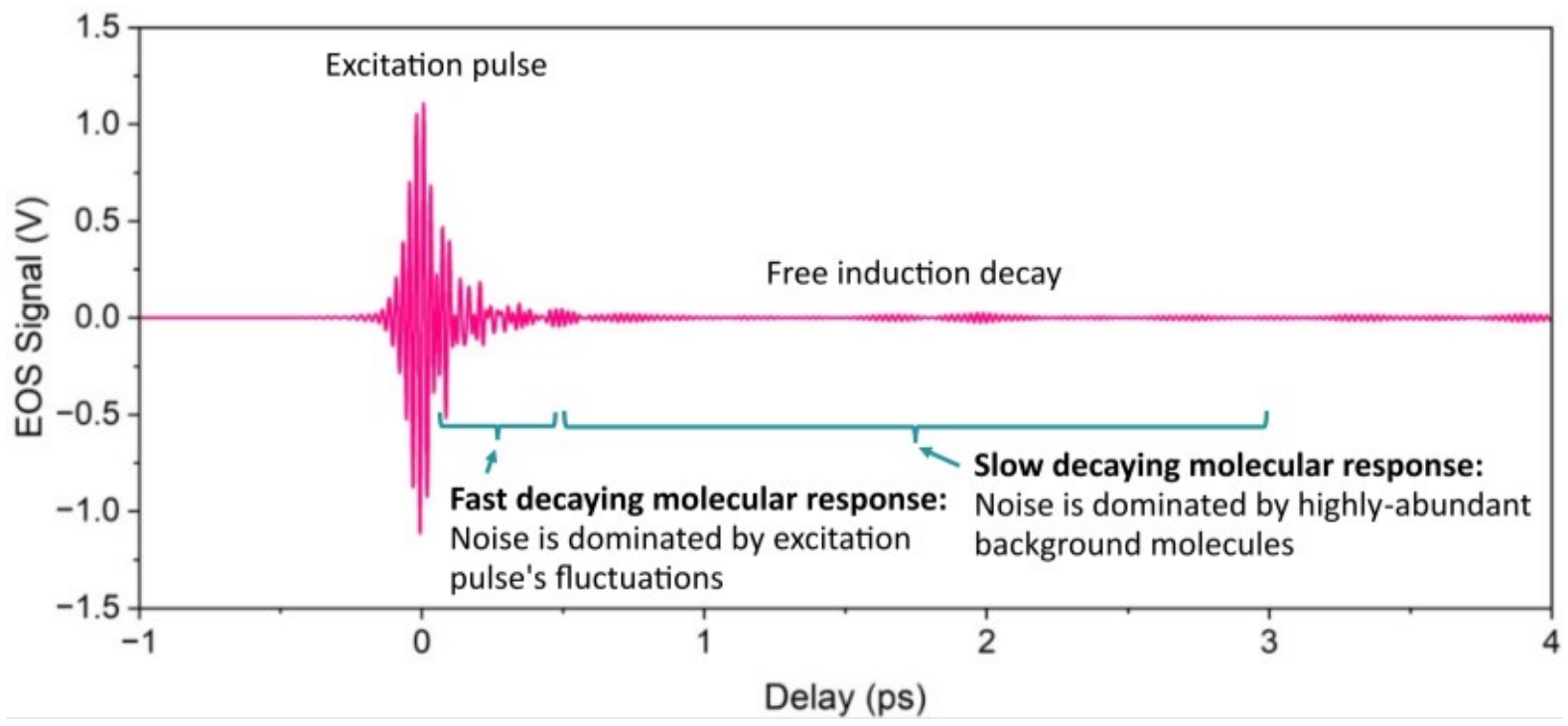

**Figure 1.** EOS measurement indicative of the electric-field response when a femtosecond excitation pulse passes through a mixture of molecules with resonant absorption features.

12 µm. Those crystals that offer broader transmission and phase-matching bandwidth, such as gallium selenide (GaSe), are less efficient [3].

*Detection technique*

With FRS and other time domain techniques, the relative time between a signal and a gating pulse needs to be scanned, often using a slow mechanical delay line. This could result in the laser parameters having already drifted before a scan is completed. Techniques that offer faster scanning speed is required to mitigate unavoidable drifts.

Secondly, even though FRS can time-filter FID with long decay times, the FIDs from larger biomolecules such as proteins in aqueous solution diphase rapidly in time. Thus, significant portions of their signal strengths overlap with the tail of the excitation pulse, hindering their temporal isolation. Furthermore, weaker signals from molecules at low concentrations are masked by the stronger signal from more abundant but clinically less relevant components such as water and albumin.

*Cost*

Technical issues aside, current spectroscopy demonstrations are mostly conducted under laboratory settings with expensive setups. The 2-µm femtosecond lasers that drive the frequency conversion are themselves costly, since they need to provide high power, high beam quality and, in many cases, CEP stabilization. The cost could potentially be reduced by optically pumping these short-wave infrared drivers with laser diodes instead of another fiber laser[9]. However, with diode lasers, there is currently a trade-off between high beam quality and high output power, with single mode diodes typically emitting only at hundreds of milliwatt.

**Advances in science and technology to meet challenges**

*Mid-infrared source*

Ho- and Th-doped lasers are promising frontends for generating 2-µm ultrashort pulses at high average power. Their deployment will require sustained development of such rare-earth-doped femtosecond lasers and, at the same time, nonlinear pulse compression technologies in the 2-µm wavelength range, such as with efficient and misalignment tolerant multi-pass cells [10].

To overcome the low conversion efficiencies of IDFG, newer drivers and frequency conversion methods are being developed. For example, mode-locked Fe:ZnS lasers have recently been demonstrated to directly generate femtosecond pulses in the 4-µm range [11]. CEP-stabilization could in principle be achieved as well. In future, such output can drive the much more efficient process of self-phase-modulation. This would allow an order of magnitude higher average power for the broadband long-wave mid-infrared output.

*Detection technique*

To raise the acquisition speed for delay scanning, a versatile approach is the use of two laser oscillators with detuned frequencies for scanning[12]. Recently, this has been applied to FRS, enabling to record full infrared waveforms at kHz rates with attosecond-precision reproducibility [13].

To unmask the signal from low concentration molecules, the common signal from abundant molecules can be removed interferometrically [14]. Furthermore, the weak FID signal could be selectively amplified, while the excitation pulse is suppressed[15].

*Cost*

A major cost component in Cr:ZnS/Cr:ZnSe lasers is their optical pump source. Recent advances in semiconductor lasers, specifically with photonic-crystal surface-emitting laser (PCSEL) technology that can emit high power light while retaining single-mode brightness [16], could lead to a more affordable replacement for the multistage fiber-lasers currently in use.

More generally, a promising way to lower cost for the entire spectroscopy instrument is to transfer as much of the components—from the laser source, the frequency conversion stage to even the detection stage—to on-chip platforms. Not only would this enable mass production and lower the costs, the footprint of devices could also be drastically reduced, enabling new mobile or even wearable applications. Very recent demonstration of on-chip Ti:Sapphire laser [17], on-chip mode-locked erbium femtosecond laser [18], a Cr:ZnS chirped pulse waveguide amplifier [19] and other on-chip nonlinear frequency conversion demonstrations, [20] set the stage for a bright future for chip-based broadband mid-infrared light source.

**Concluding remarks**

Field-resolved infrared spectroscopy using electro-optic sampling has evolved into an alternative to intensity-based measurements for high sensitivity, high-dynamic-range detection. It benefits from recently developed laser-based superoctave-spanning mid-infrared sources and ultra-rapid scanning approaches. FRS has the intrinsic capability for background-free measurements by time-windowing, which can be complemented by interferometric extinction of the excitation pulse after sample excitation. There is still substantial fundamental and engineering research to be done to further improve the techniques and reduce the system costs. Newer platforms such as on-chip photonics could lead to cheaper, miniaturized solution for applications beyond the laboratory setting.

# Integrated Photonics:
# 27. Hybrid mid-Infrared PICs for predictive health: Bridging research, innovation, and industrialization

**Ryszard Piramidowicz[1,2,3] and Stanisław Stopiński[1,2,3]**

[1] Warsaw University of Technology, Institute of Microelectronics and Optoelectronics, Warsaw, Poland
[2] VIGO Photonics, Ożarów Mazowiecki, Poland
[3] LightHouse, Lublin, Poland

E-mail: ryszard.piramidowicz@pw.edu.pl

## Status

Mid-infrared (mid-IR) light offers a unique feature of interacting with molecular vibrations, thus providing a possibility of directly detecting the presence and monitoring the quantity of specific gases and liquids with exceptional precision and unambiguity [1-3]. Developing accurate photonic sensors operating in the mid-IR addresses the need for professional diagnostic systems and consumer-oriented digital health devices. The latter are key for predictive monitoring – breath analysis, metabolic profiling, and real-time fluid diagnostics [4] – as they provide large, continuous data streams essential for AI- and ML-based predictive algorithms.

Translating the promise of mid-IR optics into compact, affordable, and scalable devices requires photonic integration, combining sources, detectors, and waveguides on a single chip. Current diagnostic systems rely on discrete components, resulting in bulky, power-hungry devices incompatible with consumer applications. Therefore, developing integration technologies optimized for mid-IR operation is essential. This need is further reinforced by market forecasts and strategic roadmaps for integrated photonics, which consistently highlight mid-IR photonic integration as one of the key directions for future technological progress [5].

Despite significant advances in the fabrication of discrete mid-IR components – including light sources, detectors, and passive waveguiding circuits – the primary bottleneck remains the integration technology required to translate these elements into highly compact and functional photonic systems. Monolithic mid-IR integration technologies, while offering clear long-term advantages, are still in the early stages of development, as they require the simultaneous realization of high-performance passive and active components within a single material platform. Consequently, hybrid and heterogeneous integration approaches represent a more viable solution in the short to medium term.

To date, a variety of such schemes have been demonstrated, including flip-chip bonding and butt-coupling interfaces between quantum cascade lasers (QCLs) and integrated waveguides [6, 7], integration of laser bars with passive waveguide circuits [8], heterogeneous integration of QCLs and interband cascade lasers (ICLs) on silicon using evanescent coupling to passive waveguides [9, 10], as well as adhesive bonding of InAsSb PIN photodiodes onto silicon photonic circuits employing grating-coupling interfaces [11], or wafer-level bonding approaches enabling the integration of bolometric detectors directly with germanium-on-insulator waveguides [12].

## Current and future challenges

The vision of proactive and predictive healthcare relies on continuous access to biochemical data through wearable and IoT-enabled devices that operate seamlessly in daily life. From smartwatches that track glucose or hydration to smart toilets that analyze urine, and breath sensors that detect VOCs, the demand for real-time, non-invasive biochemical sensing is rapidly growing.

Delivering such functionality in compact, low-cost, and energy-efficient form factors demands breakthroughs in photonic integration, heterogeneous materials, and on-chip light sources and detectors operating in the mid-infrared (2–15 μm) spectral range. This, in turn, requires addressing a wide range of technological challenges and enablers that will shape the future of MIR photonics, particularly in areas such as materials integration, power efficiency, thermal management, and scalable manufacturing.

New paradigm of designing - in conventional PICs, passive waveguides are designed to tightly confine the mode within the core. For gas and liquid sensing, however, the optical mode must extend beyond the core to ensure efficient light–medium interaction. This feature must be combined with low propagation loss and the ability to design waveguide bends to increase interaction length. Therefore, mid-IR sensing waveguides must be designed in an entirely new way, balancing confinement, loss, and interaction efficiency.

New material platforms - current photonic integration technologies, developed mainly for telecom and datacom applications in the C+L band, are not suited to mid-IR sensing. Thus, new material platforms for waveguiding components must be developed and optimized. The most promising include germanium-on-silicon (Ge-on-Si), silicon-germanium (SiGe), and aluminum gallium arsenide (GaAs/AlGaAs) [6-8, 13-14]. In addition, novel light sources – such as quantum or interband cascade lasers – and detectors, including type-II superlattice (T2SL) devices, need to be tailored for PIC integration, which introduces further material-related challenges [15, 16].

In Fig. 1, a photograph of a multi-channel mid-infrared integrated transmitter is presented. The device comprises three quantum cascade lasers operating at approximately 4.8 μm and a superlattice antimonide detector, all integrated with a passive photonic circuit implemented in Ge-on-Si technology [7]. The device is designed to serve as a multi-wavelength light source for various applications, including gas-sensing.

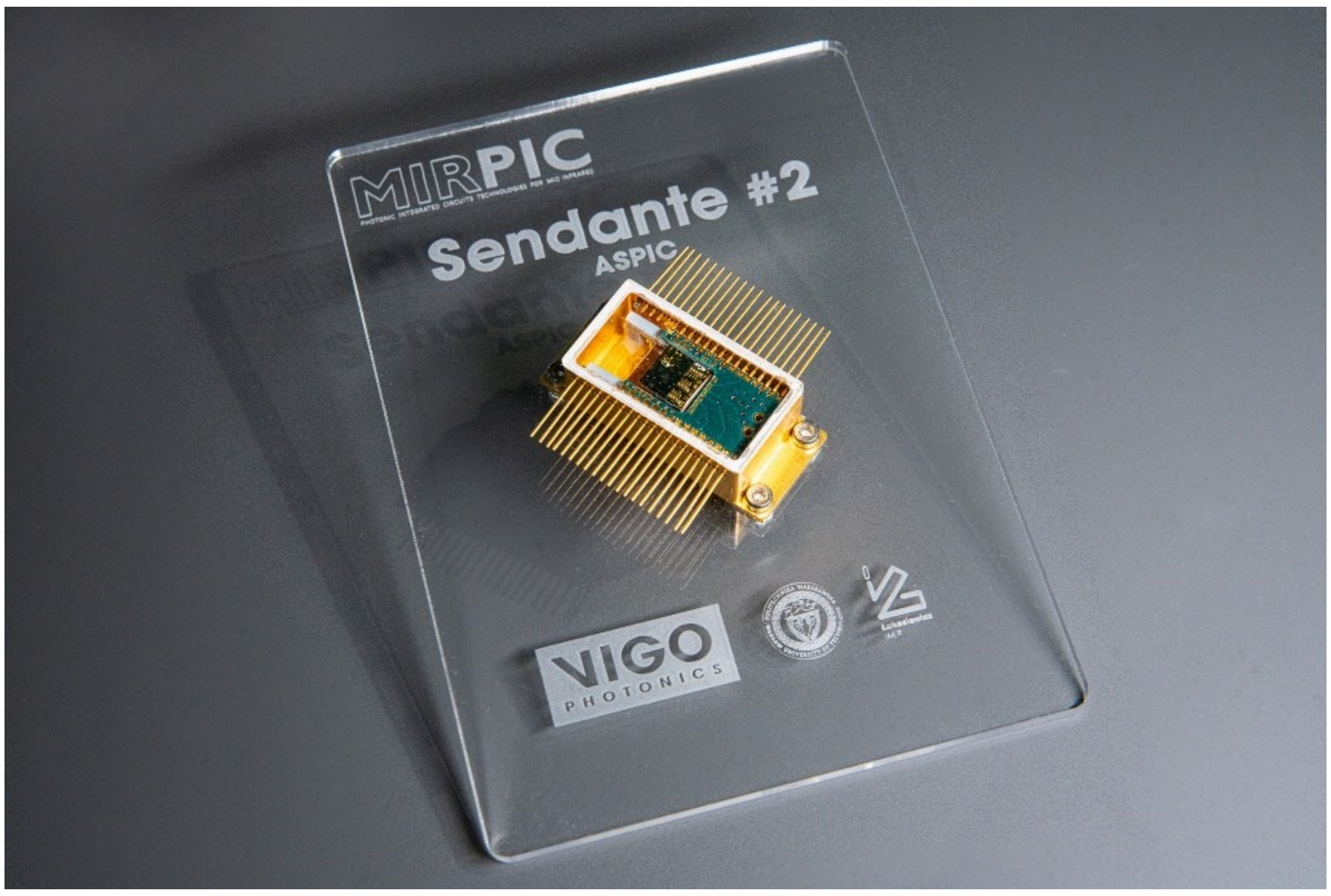

**Figure 1.** Photograph of the multi-channel mid-infrared integrated transmitter.

New approach to integration technology - developing reliable and scalable hybrid or heterogeneous integration remains a major challenge in mid-IR photonics. Addressing this challenge requires an efficient, low-loss optical interface between active devices and passive waveguides, with insertion losses preferably below 1.0 dB, consistent with values achievable in state-of-the-art near-infrared photonic integration platforms. In the case of relatively straightforward butt-coupling schemes, the high refractive index of most semiconductor materials necessitates the development of dedicated anti-reflection (AR) coatings compatible with mid-infrared wavelengths. In contrast, evanescent-field coupling – while potentially offering significantly lower coupling losses – requires the implementation of more advanced heterogeneous integration techniques, such as heteroepitaxy, die- or wafer-level bonding, or micro-transfer printing.

Assembly and packaging – both are critical for any photonic platform; however, mid-IR predictive health applications add additional complexity. Electrical interfacing is coupled with demanding thermal management due to the low efficiency of quantum cascade lasers, while the optical interface must remain exposed to the analyte, affecting long-term durability and reliability. An essential aspect of assembly and packaging technology development is the implementation of appropriate electrical driving schemes. In particular, quantum cascade lasers require high drive currents and voltages, typically on the order of 0.5–2.0 A and 10–20 V, respectively. This results in substantial power dissipation and potential inter-channel crosstalk, especially in multi-wavelength sensor architectures, and imposes stringent requirements on thermal management and stabilization, which are critical for maintaining precise wavelength alignment and stable optical coupling in the presence of thermally induced expansion. Furthermore, certain measurement and modulation techniques require high-speed electrical interfaces with bandwidths of 1.0 GHz and beyond, introducing an additional level of complexity for device packaging and interconnect design.

Manufacturability and scalability - the key challenge lies in fabricating active devices, such as lasers and detectors, which rely on intricate quantum-well and superlattice structures that require precise epitaxial growth (MOCVD or MBE) – a slow and difficult-to-scale process. Moreover, scaling integration technologies like micro-transfer printing to high-volume manufacturing demands sub-micrometer alignment precision and accurate optical interface formation.

**Enabling Technologies for mid-IR Photonics**

Building on the success of PICs at telecom wavelengths requires advancements in mid-IR technologies, which are still confined to laboratory-scale devices. Several promising pathways can accelerate the development of mid-IR PICs for predictive health monitoring, addressing the key challenges outlined above.

Micro-transfer printing (MTP) has already proven effective for heterogeneous III-V/SOI integration and can enable the scalable production of mid-IR PICs with multiple active devices – light sources and detectors - on a single chip [15, 16]. MTP enables parallel micro-assembly of active coupons onto a passive PIC carrier. However, adapting this technology to mid-IR requires extensive R&D, as quantum/interband cascade lasers and superlattice photodetectors feature far more complex epitaxial stacks than their NIR counterparts.

Advanced sensing waveguides have also progressed rapidly. Low-loss structures with confinement factors exceeding 100% have been demonstrated using suspended geometries in SiN and $Ta_2O_5$ [17, 18]. Mastering the scalable fabrication of suspended and plasmonic waveguides, which offer strong light–matter interaction at the metal–dielectric interface, is a key enabling step for high-performance mid-IR sensing [19].

A third critical enabler is interband cascade laser (ICL) technology [10, 20]. Conventional quantum cascade lasers require high voltage and current, resulting in significant heat generation and reduced efficiency. ICLs, in contrast, combine broadband emission with significantly reduced power consumption, making them promising replacements, particularly for the shorter mid-IR range (up to 6–7 μm).

**Concluding remarks**

Mid-infrared integrated photonics holds the potential to transform the predictive health sector in a way similar to how near-infrared technologies transformed the telecom and datacom markets. However, significant challenges remain to be addressed by the R&D community, from defining suitable material platforms to developing efficient integration technologies and scalable manufacturing processes, and ultimately identifying the ultimate killer application. Overcoming these challenges requires the continuous advancement of enabling technologies, such as micro-transfer printing, sensing waveguides, and ICLs, which can pave the way for the broad market adoption of highly compact mid-IR-based devices and systems.

**Acknowledgements**

This work received support from the National Center for Research and Development through projects MIRPIC (TECHMATSTRATEG-III/0026/2019-00), HyperPIC (FENG.02.10-IP.01-0005/23, IPCEI ME/CT, and from the EU Horizon Europe under GA #101213727, Chips Joint Undertaking.

# 28. GaSb-based integrated lasers emitting at 2–3 µm: addressing the power-efficiency requirements for miniaturized devices

**Mircea Guina and Jukka Viheriälä**

Optoelectronics Research Centre, Tampere University, Tampere, Finland

E-mail: mircea.guina@tuni.fi

## Status

Spectroscopic techniques operating in the 2–3.5 µm wavelength range enable to access the fundamental vibrational modes of molecular bonds, particularly C–H, N–H, and O–H stretching vibrations. Thus, this spectral region is relevant for non-invasive diagnostics, as many volatile organic compounds can be detected through breath analysis [1]. Furthermore, it is relevant for detecting metabolites in biofluids such as glucose, lactate, urea, creatinine, fatty acids, and lipids [2]. However, large-scale deployment of spectroscopic techniques in these applications, using portable or wearable systems has remained elusive, primarily due to the lack of compact, power-efficient, and cost-effective light sources offering broad wavelength coverage. To this end, GaSb-based laser diodes incorporating "type-I" quantum wells (QWs) are beneficial, albeit a severe decrease in efficiency above 2.8 µm [3]. Compared with GaSb-based ICLs, they employ less complex fabrication and have much lower power consumption, typically operating at ~1 V and 100's mA drive currents. Furthermore, both types of laser diode technologies exhibit much lower diode voltage and simplified designs compared to QCLs based on InP materials and operating at longer MID-IR wavelengths.

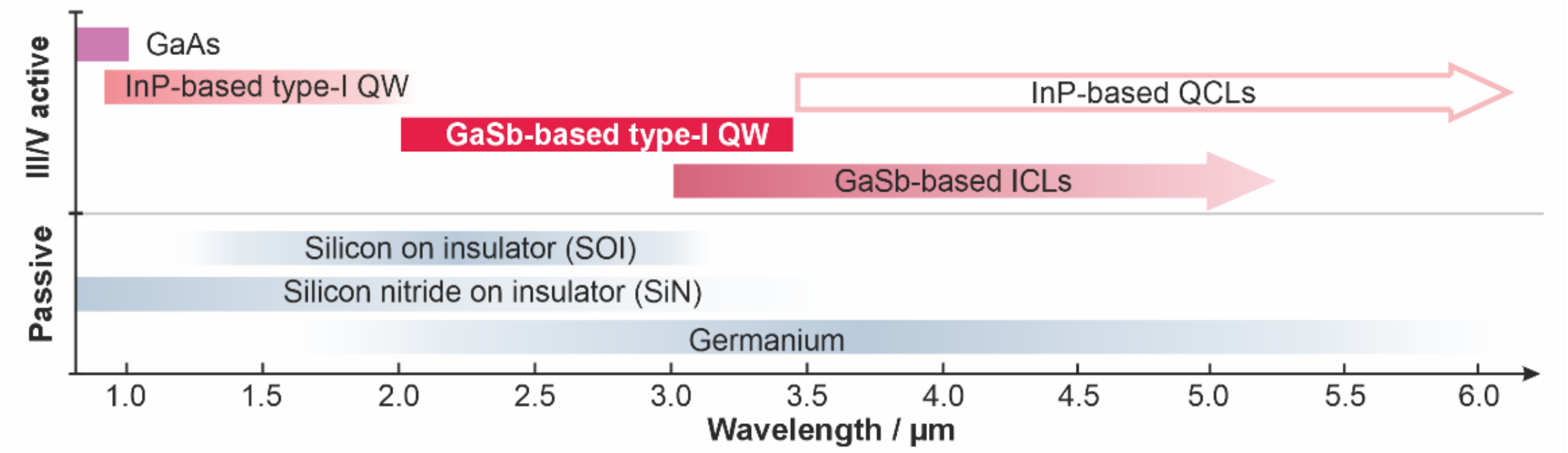

**Figure 1.** Wavelength coverage of III-V optoelectronics systems and major photonic integration platforms.

From the point of view of technology readiness level, GaSb-based semiconductor laser diodes operating in the 2–3 µm spectral window are available commercially and have been deployed in sensing applications, e.g. TDLAS has been used for detection of $H_2O$, $CO_2$, $CH_4$, $NH_3$, HF, and HCl [4]. However, the adoption of GaSb laser technology remains confined to niche and relatively low-volume applications. The limited uptake is primarily due to restricted wavelength tunability for monolithic laser diodes, high cost, the complexity of the external-cavity configurations used to achieve broad wavelength tuning, and output power constraints at wavelengths approaching or exceeding ~3 µm due to Auger recombination and thermal leakage [5].

### Current and future challenges

In general, GaSb-based optoelectronic technology is less available then InP and GaAs platforms, lacking a strong market pull. This has resulted in higher unit costs, limited vendor diversity, and constrained industrial scalability, preventing the uptake in high-volume industrial sensing solutions. To this end, advanced laser functionality and system-level practicality are increasingly pursued through hybrid-PIC technologies, following concepts well established in InP/silicon-photonics (SiPh) platforms [6]. However, extending conventional sub-micron silicon-on-insulator (SOI) technology into the 2–3 μm wavelength range is limited by strong absorption in the $SiO_2$ cladding above ∼2.5 μm, leading to propagation losses exceeding 10 dB/cm, whereas Ge-on-Si waveguides typically exhibit higher losses on the order of ∼3 dB/cm.

From a generic development perspective, accelerating the deployment of hybrid GaSb/SiPh platforms for the 2–3.5 μm spectral region requires deployment of a comprehensive approach, spanning the full technology stack, from gain chip optimization to demonstration of advanced PIC functionality. Key focus areas include: (i) validation of predictive simulation frameworks for GaSb gain chip design; (ii) advancement of heterostructure design and fabrication enabling broader wavelength coverage, improved efficiency beyond 3 μm, and advanced waveguide configurations; (iii) development of design tools for co-simulation of complex GaSb/SiPh architectures; (iv) advancing the integration technologies, including flip-chip and micro-transfer bonding, and (v) demonstration and validation of PIC functionalities in application-relevant use cases.

### Advances in science and technology to meet challenges

Recent advances in GaSb/SiPh-based lasers have addressed the absorption losses of SiPh waveguides, the need for customized GaSb gain-chips with improved coupling, the design of PICs with application-specific functionality, and the integration process. Thus, using micron-scale SOI waveguides with reduced modal overlap in the $SiO_2$ cladding, propagation losses as low as ∼0.56 dB/cm at∼2.7 μm have been achieved [7]. Alternative $Si_3N_4$ waveguides exhibit losses of ∼0.15 dB/cm at 2 μm and ∼0.7 dB/cm near 2.7 μm, beyond which $SiO_2$-related absorption increases rapidly, yet decreasing again in the 3–3.5 μm region [8].

Regarding the development of GaSb gain chips optimized for PIC integration, self-consistent modelling frameworks validated against experiments indicate that optimized heterostructures with asymmetric QWs can deliver ultra-broad gain bandwidths of ∼300 nm, substantially increasing the wavelength versatility attained with a single chip [9]. In terms of integration, GaSb/SiPh hybrid PIC technology has evolved rapidly from laboratory-type "*end-fire*" coupling schemes used for proof-of-concept demonstrations to more scalable approaches. *Flip-chip* integration has enabled the first fully functional hybrid GaSb/SOI PIC lasers [10], validating the integration of wafer-scale passive circuitry combined with high-performance GaSb gain chips. Most recently, *micro-transfer printing* (*μ-TP*) of GaSb coupons onto SOI has been demonstrated [11], clearing up a critical step towards parallel multichip integration, with improved yield and reduced assembly cost.

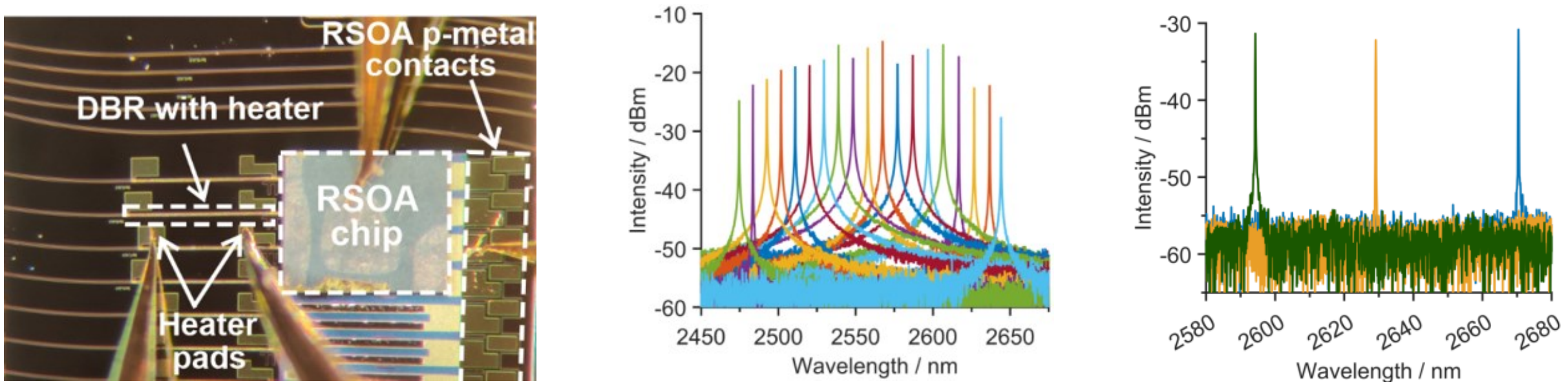

**Figure 2.** (**left**) A hybrid PIC laser incorporating a GaSb reflective amplifier (RSOA) flip-chip bonded to a SOI circuit [10]; (**middle**) tuning spectra of a PIC-based GaSb/SiN laser exploiting Vernier effect based on ring-resonators [8]; (**left**) spectra of a multi-wavelength PIC-based GaSb/SiN laser exploiting distributed gratings [12].

Building on these foundations, there is a growing trend to advance the functionality of GaSb-based integrated lasers and address the needs of specific applications. For example, using the GaSb/$Si_3N_4$ platform, integrated lasers with a wavelength tuning exceeding 170 nm around 2.6–2.7 µm have been demonstrated [8]. Moreover, exploiting a cascaded configuration of on-chip distributed $Si_3N_4$ gratings enabled demonstration of multi-wavelength operation using a single GaSb RSOA [12] (see spectra in Fig. 2, right picture). In parallel, a 2 µm distributed Bragg reflector GaSb/$Si_3N_4$ lasers with narrow linewidths down to the few-kHz has been demonstrated [13]; this approach has been also used for wavelength extension to ~2.7–2.8 µm, demonstrating narrow linewidth room-temperature operation with power levels approaching 20 mW level (results to be published). In terms of proof-of-concept application developments, we would note increased collaborative efforts linking the technology developers to system integrators, targeting both next generation breath-analyzers (www.raven-sensors.eu) and detection of biomarkers (www.photonmed.eu) exploiting spectroscopic techniques in the 2–3 µm region.

Amongst other recent developments in the field, we note the demonstration of a hybrid laser incorporating a micro-ring resonator integrating three different gain chips, that enabled wavelength tuning bands of 22 nm, 22 nm, and 17 nm centered at 1.95 µm, 2.11 µm, and 2.37 µm, respectively [14]. Concerning the longer-term development path, a monolithic GaSb PIC platform integrating both active and passive components would be highly attractive, and preliminary conceptual steps in this direction have been taken [15]. Monolithic integration would not only reduce the complexity inherent to hybrid approaches, but would also enable a new class of functionality, such as mode-locked and frequency-comb lasers in the 2–3 µm range, which remain largely absent from the current mid-infrared components library. However, this requires tackling substantially higher technological barriers, calling for major advances in fabrication to demonstrate low-loss passive GaSb components, such as splitters, couplers, and ring-resonators. As an intermediate step, a hetero-epitaxy approach has been used to demonstrate a GaSb laser operating at 2.3 µm, with the active layer grown on a patterned Si platform and aligned with passive SiN waveguides [16].

**Concluding remarks**

GaSb-based hybrid PIC technology provides a compelling compact and scalable platform, where performance, functionality, and low-power operation can be addressed simultaneously. Advances in broadband gain-chip design, scalable hybrid integration, and PIC-enabled wavelength control already outlined a clear route toward compact, tunable laser sources covering the 2–3 µm range and beyond. A key advantage is the use of *type-I* QWs, which operate at low voltages (~1 V) and moderate drive currents (tens to hundreds of mA), making this technology particularly well suited for portable and wearable spectroscopic systems. Looking forward, future work should address higher levels of system complexity, including on-chip integration of detectors, sensing waveguides, and application-specific photonic circuits. In this context, monolithic integration approaches are expected to attract increasing attention, as they promise further reductions in footprint, and packaging complexity enabling next generation mid-infrared sensing engines.

**Acknowledgements**

We acknowledge the contribution of numerous PhD students and postdocs, especially Nouman Zia, Samu-Pekka Ojanen, Heidi Tuorila, Joonas Hilska, Markus Peil, and Ifte Khairul Alam Bhuiyan. Our recent work concerning the development of GaSb optoelectronics technology has been financially support by Horizon 2020 Framework Programme project *NETLAS* (860807), HORIZON EUROPE Framework Programme project *PhotonMed* (101139777-2), and project *RAVEN* (101135787). General support from the Research Council of Finland via the *PREIN* Flagship program (320168) is also acknowledged.

# 29. Monolithic mid-infrared photonic integrated circuits for compact medical devices

**Borislav Hinkov[1] and Felix Jaeschke[1,2]**

[1]Integrated Photonics Technologies Unit, Silicon Austria Labs, Villach, Austria
[2]Optoelectronic Research Center, Tampere University, Tampere, Finland

E-mail: borislav.hinkov@silicon-austria.com

## Status

The analysis of the optical properties of molecules has revolutionized our understanding of their role in health-related diagnostics, disease monitoring, and medical therapy. A key enabler for the detection of molecular species is optical techniques in the mid-infrared (MIR) spectral range, allowing for high selectivity and sensitivity measurements and structural-property analysis [1,2]. In practice, those measurements are traditionally performed using Fourier-transform infrared (FTIR) spectrometers. While they are powerful broadband tools covering MIR-THz wavelengths (>2.5–25 μm) and are among the standard equipment of analytical chemistry and bio-medical labs, they are bulky and often require time-consuming and expensive offline analytics, significantly limiting their performance in everyday applications.

First steps towards a paradigm change for much more compact, versatile, and portable medical spectroscopic systems were taken with the inventions of the quantum cascade laser (QCL) and interband cascade laser (ICL) by Faist et al. (1994) and Yang et al. (1995), respectively. The same technological foundation enabled the QC detector (QCD, Hofstetter et al., 2002) and IC infrared photodetector (ICIP, Yang et al., 2010). However, chip-scale integration into full photonic integrated circuits (PICs) with complementary electronics is still facing major challenges. Such MIR photonic integrated circuits (PICs) are a class of emerging devices that promise to demonstrate the full potential of MIR technology, e.g., for healthcare applications [3]. Apart from heterogeneous or hybrid integration (e.g., Si-photonics + III-V technology [4]), monolithic sensors are highly promising as scalable, miniaturized in-situ sensors. In that respect, scalability is a much-needed feature for high-volume production at competitive prices.

Early adoptions of monolithic MIR PICs combine QCLs and QCDs with plasmonic waveguides in a simple linear geometry [5], usable for in-situ chemical-reaction monitoring of secondary-structure changes in proteins [3]. In particular, plasmonic waveguides open the route towards simple integration strategies [3,5,6-9], are compatible with QCL- (i.e. InP-)based fabrication protocols and enable high levels of flexibility in implementing additional features including: (i) octave-spanning mode-guiding capabilities [6] for future combination with broadband MIR frequency combs [10], (ii) surface-passivation and -functionalization for enhanced and molecule-specific sensing [7], and (iii) on-chip mode-routing capabilities [8,9] complementary to existing beam-steering approaches in the MIR [11].

In parallel, alternative monolithic MIR PICs have recently been demonstrated, implementing (epitaxial) dielectric waveguides [12, 13], with lower mode-guiding losses and even better on-chip mode directing capabilities, at the cost of a more complex technological realization and limited direct sensing capabilities.

## Current and future challenges

Four main challenges need to be overcome for MIR PICs to enable increased functional densities and capabilities, allowing large-scale deployment of MIR PICs in, e.g., medical wearables:

*(i) Material restrictions*

Although materials used for visible and near-infrared PICs are already established (e.g. PDMS for microfluidics), their optical characteristics – including the complex-valued refractive index – are completely different in the MIR, often rendering them unsuitable in MIR PICs. Issues with direct integration compatibility of Si/Ge waveguides with III/V material platforms such as InP, GaSb or InAs is another limitation [4]. And also, novel and promising IR materials such as thin-film lithium niobate (e.g. for high-speed modulators) is spectrally limited to below ~4.9 µm with high losses at longer wavelengths [14].

(ii) *Lack of modularity*

In contrast to NIR PICs, where modular on-chip integration is available for active components (e.g. lasers, detectors, active modulators) and passive elements (e.g. waveguides, splitters and attenuators with minimum limitations), enabling fully custom-tailored PICs with application-specific geometries [15], modular platforms do not exist yet in the MIR. However, first generations of complex, functional MIR PICs have been demonstrated already, following a modular concept [8, 9, 16].

(iii) *Functionality and functional density*

For improved miniaturization characteristics, MIR PICs will require increased functional densities. Typical limitations currently arise from electrical crosstalk and thermally induced performance degradation due to the high density of active and passive components. Especially the low wall-plug efficiency (still below 30% at room temperature (RT) in continuous wave (CW) operation), resulting in high thermal dissipation of QCLs (often multiple watts) is a critical issue [17]. In addition, for further increasing the functionality of current MIR PICs [3,9,16], critical component integrations have still to be demonstrated such as on-chip integrated amplitude and phase modulators or “beam forming” elements such as on-chip polarizers.

(iv) *Combination with integrated electronics*

While co-packaged optics (CPO), integrating optical components directly with suitable integrated electronics, are gaining momentum in optical communication in data centers, MIR PICs currently completely lack capabilities beyond lab-scale current/voltage supply and extraction. Application-specific integrated electronic circuits (ASICs), being standard in NIR PICs, are still largely missing in MIR integrated photonics, due to a current focus on improving the *optical* characteristics of the PICs. Special requirements to be covered include much higher bias current and voltage (typically beyond 10 V and on the order of 100s of milliamps for QCLs), with severe thermal design limitations for a PIC.

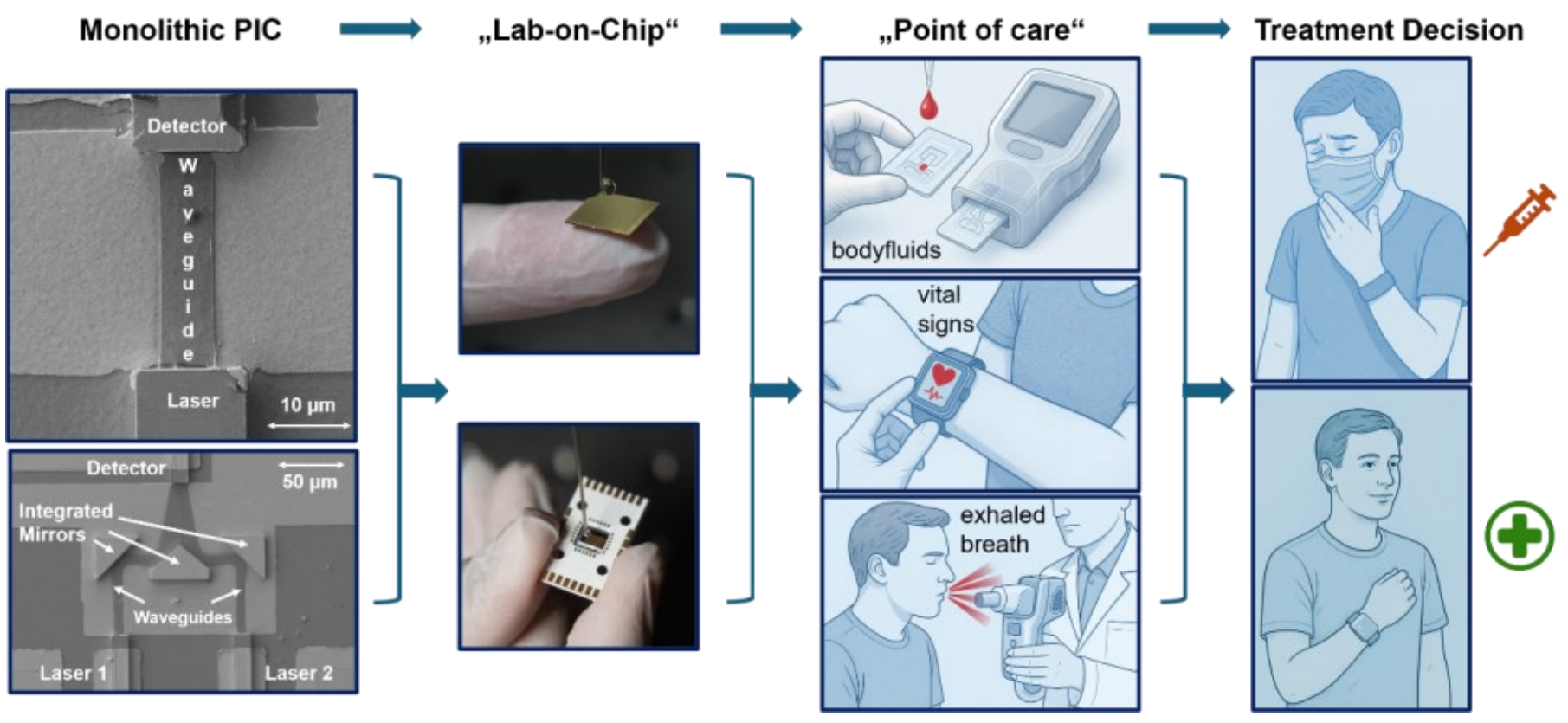

**Figure 1.** Illustration of the full future (monolithic) MIR PIC based chain of „point of care“ medical analysis and treatment decision. Left column is from Ref. [3,9]

## Advances in science and technology to meet challenges

While still being much less mature than their NIR counterparts, MIR PICs have already demonstrated relevant functionality in real-world applications including real-time in-situ protein confirmation monitoring in $D_2O$ [3] and water-contamination analysis in solvents [16].

Existing *material restrictions* are already being tackled by, e.g., the exploration of similar MIR materials as their existing NIR counterparts. One example is the use of polyethylene, with very similar properties as HMDS in the NIR [8]. Another crucial step in providing more suitable materials for MIR PICs comes from expanding the possible combinations with plasmonic concepts beyond recent pioneering work [3,5,16]. For example, the plasmonic-based surface-enhanced infrared absorption (SEIRA) using metasurfaces has demonstrated to be a powerful tool [18]. It still requires only relatively simple and scalable fabrication protocols but remains to be further utilized in MIR PICs.

The lack of *modularity* of current InP- or GaSb-based MIR PIC platforms beyond monolithic integration motivates novel research on concepts of hybrid and heterogeneous integration [19,20]. These works complement monolithic approaches, where individual, high-performance active or passive elements are already available in other material systems. Initial attempts have been made to integrate SiGe-waveguides or TFLN-based modulators into III-V MIR platforms [4,14].

Next steps in *functional density increase* require further improving the individual PIC components, while simultaneously carefully optimizing the overall PIC design. A particularly important problem is the thermal dissipation of MIR active components [17], which eventually needs to be mitigated together with minimizing on-chip electrical crosstalk effects. One possibility is to operate the PIC components well below their maximum operation conditions, thus significantly lowering parasitic thermal effects. On-chip electrical shielding of components is another important aspect to be further explored in future technological improvements, as well as advanced packaging allowing access to optical and electrical interfaces, e.g. through optical fibers and ASICs. Ultimately, consumer-grade MIR PIC devices will require co-integrated electronics, capable of operating under the necessary high-power requirements in a „plug and play“ manner.

It can be assumed that those ASIC developments can strongly benefit from previous work in NIR PICs, together with the transition in datacom towards CPO. The resulting fully integrated PIC systems will require industrial-grade housing and packaging of the combined MIR PIC and ASIC based on reliable and standardized industrial processes.

**Concluding remarks**

In summary we want to highlight the tremendous potential of portable devices using MIR PICs for medical applications, acknowledging the amount of progress needed in available materials, modularity through synergy of different material platforms, increased functional density and integration with electronics.

From a practical standpoint, currently available MIR PICs are affected by low-cost efficiency, which impacts their potential market penetration and the ability to overcome established (MIR) technologies such as the powerful lab-scale FTIR spectroscopy. As demonstrated in the case of NIR PICs, market penetration also heavily relies on wide technological availability to academia, industry including fab-less companies, requiring standardized design rules and guidelines summarized in process design kits (PDKs) and used in foundry services through multi project wafer (MPW) PIC fabrication runs.

The final step for the use of MIR PICs in the medical field will require, beyond technological maturity, affordable pricing and large-scale PIC availability to gain the acceptance of the medical community, i.e. first and foremost medical doctors and practitioners. For this, standardized measurement routines, possibly adopted from existing medical protocols where available, leading to medical device certification, are key.

**Acknowledgements**

The authors want to thank Gottfried Strasser and Benedikt Schwarz for expert technical advice and support and Mauro David, Georg Marschick, Elena Arigliani, Dominik Kukola, Xaver Gsodam and Niklas Brandacher for technical contributions to the PIC developments. Further, we want to acknowledge financial support from the EU Horizon Europe Marie-Curie doctoral co-fund project "CRYSTALLINE" under grant number 101126571. We further acknowledge received funding from the EU Horizon 2020 Framework Program for the project "cFLOW" under grant number #828893. BH acknowledges funding by the Austrian Science Fund FWF for the project "LIQI-sense" (M2485-N34).

# 30. Advances in infrared waveguides and fiber optics for medical applications

**Polina Fomina[1], Alexander Novikov[2], Viacheslav Artyushenko[2,3], Boris Mizaikoff[1]**

[1] *Institute of Analytical and Bioanalytical Chemistry, Ulm University, Albert-Einstein-Allee 11, 89075 Ulm, Germany*
[2] *art photonics GmbH, Rudower Chaussee 46, 12489 Berlin, Germany.*
[3] *Art Fiber Systems, Lisa-Meitner-Str. 9, 89081 Ulm, Germany*

E-mail: polina.fomina@uni-ulm.de ; an@artphotonics.de ; slava@afs.art ; boris.mizaikoff@uni-ulm.de

### Status

Among the various mid-infrared (MIR) fiber technologies, there are 3 types most widely used for transmission in 2–20 µm: IR-glass fibers, polycrystalline silver-halide fibers (PIR), and hollow waveguides (HWG). These IR fibers are used for assemblies of laser cable for IR-lasers, pyrometry endoscopes and spectroscopy probes required for reaction monitoring in industry, for medical diagnostics and environment monitoring [1-7].

The most common IR-glass fibers are drawn from fluoride and chalcogenide glasses. Fluoride fibers offer low attenuation in MIR: Zirconium fluoride $ZrF_4$ covers 0.3 to 4.5 µm, and Indium Fluoride $InF_3$ extends up to 5.5 µm. Chalcogenide infrared fibers (also known as CIR) are usually made of $As_2S_3$ or $As_2Se_3$ composition and possess good transmission in the 1.1 to 6.5 µm and 2–11 µm range, respectively, very high refractive index and high numerical aperture (NA) – compared to fluoride fibers. PIR fibers are produced by extrusion from Silver Halide (AgClBr) crystals and offer exceptional transmission in the MIR range (3–17 µm) with several other advantages – they are non-hygroscopic, non-toxic and non-brittle. PIR fibers have a large NA and are extensively used in mid-infrared FTIR spectroscopy applications for real-time reaction monitoring *in-line*. HWGs have a hollow core with a reflective inner coating to guide the light. HWGs are perfect to transmit a low divergent IR-light in Mid IR-spectrum from 3 to 17 µm. HWGs do not suffer from the Fresnel reflection losses experienced by solid-core IR fibers; however, they are more sensitive to beam-coupling conditions and bending than PIR fibers [8].

MIR optical fibers enable flexible light delivery in the molecular “fingerprint region,” supporting chemically specific, point-of-care diagnostics in real-time [1, 9]. Fiber-based ATR spectroscopy can differentiate tumor from healthy tissue, while fiber-coupled Quantum Cascade Laser (QCL) systems have demonstrated non-invasive biochemical sensing such as glucose detection and skin analysis [10]. QCL-based ATR spectroscopy has further shown promise for detecting early osteoarthritic changes in articular cartilage [11]. Development of MIR spectroscopy directly on a fiber tip opens the door to ultracompact *in vivo* probes [12]. Moreover, Silver-halide fibers have been used for temperature-controlled laser surgery, where fiber-optic pyrometry provides real-time thermal feedback during $CO_2$ laser operation [7].

In addition to optical fibers, conventional waveguide systems are also used in mid-infrared spectroscopy. However, only a limited number of materials exhibit sufficient transparency in the MIR spectral range [13]. One of the most widely employed approaches is ATR spectroscopy, in which the IR beam undergoes internal reflection inside the waveguide crystal, generating an evanescent field at the interface. This evanescent wave penetrates a short distance into the sample, enabling spectral acquisition [14].

ATR waveguides are typically fabricated from materials such as zinc selenide (ZnSe), zinc sulfide (ZnS), germanium (Ge), diamond, silicon (Si), and gallium arsenide (GaAs), among others. The choice of waveguide material is critical and should be guided by the intended application, particularly in biomedical analysis where samples are commonly measured *ex vivo* (e.g., body fluids, tumor cells, or excised tissue placed directly onto the ATR crystal) [14].

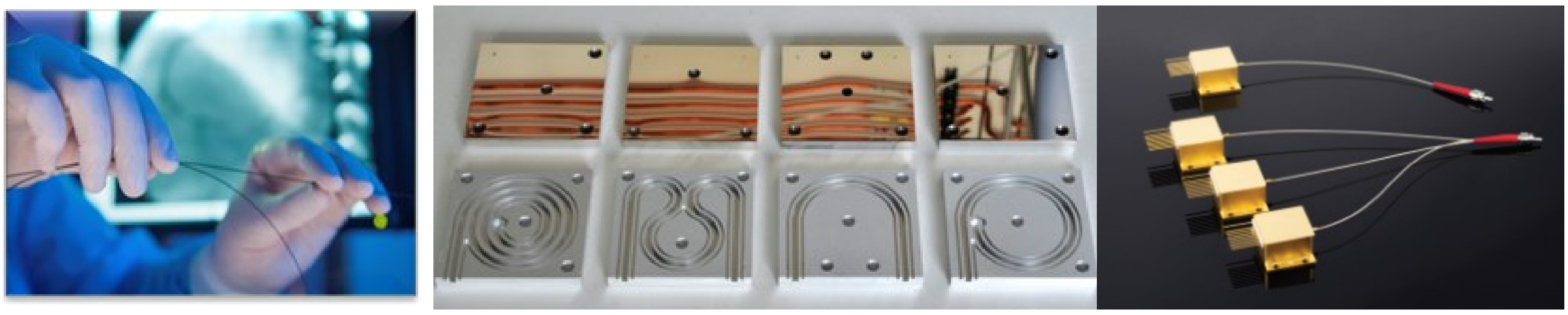

**Figure 1.** Examples of optical fibers and hollow-core waveguides used for mid-IR spectroscopic applications. The images illustrate (from left to right) optical fibers (image courtesy art photonics GmbH), substrate-integrated hollow waveguides (iHWGs) propagating radiation in the mid-infrared spectral band [13] (image courtesy Institute of Analytical and Bioanalytical Chemistry, Ulm University), and assembled fiber-coupled QCL modules (image courtesy art photonics GmbH).

### Current and future challenges

MIR fibers face several performance limitations. High refractive indices of MIR materials such as chalcogenide glass ($n \approx 2.4$–$2.7$) and AgClBr ($n \approx 2.1$–$2.2$) cause Fresnel reflection losses exceeding 20% for two fiber end faces, which significantly reduces coupling efficiency and delivered optical power. In practice, maximum utilized length of IR fibers is usually limited to several meters not only because of relatively high attenuation losses, but also due to less sensitive detectors and less powerful light sources available in the MIR region.

Several challenges are associated with ATR waveguides, as well. Certain materials, such as ZnSe and GaAs, are mechanically fragile, which complicates their use with hard or abrasive samples. In addition, many ATR crystals are chemically sensitive and can degrade under harsh conditions, including exposure to low pH environments. Another limitation is the toxicity of many commonly used waveguide materials. This restricts their direct contact with biological tissue and body fluids, necessitating sample extraction for *ex vivo* analysis and preventing true *in situ* or *in vivo* measurements [15].

### Advances in science and technology to meet challenges

Several technological developments now address these limitations. Suppression of Fresnel losses by microstructured anti-reflection end-face treatments [16], improves coupling efficiency and power throughput. Improvements in crystal purification and fiber extrusion processes reduce scattering losses in PIR fibers and boost their transmission. Advances in powerful light sources (e.g. light-emitting diodes (LEDs), QCLs, supercontinuum) and high-sensitivity detectors (e.g. mercury cadmium telluride detectors (MCT), photoacoustic-based) improve overall system performance, increasing spectra acquisition speed and signal-to-noise (SNR) which also allows the use of much longer fibers [17]. Upconversion spectrometers, which convert MIR signals into the NIR, further mitigate fiber-loss constraints by enabling detection with low-noise NIR detectors [18].

Silicon waveguides are increasingly gaining attention for biomedical applications. Commercial silicon internal reflection elements (IREs) are typically fabricated from thick silicon substrates with precisely polished edges. Silicon is attractive due to its availability, low cost, chemical inertness, and non-toxicity. It also meets key optical requirements for ATR spectroscopy: it possesses a high refractive index, is sufficiently transparent in the MIR region for accurate spectral measurements, and can be mechanically processed to achieve the desired angle of incidence. Furthermore, silicon wafers with thicknesses on the order of hundreds of micrometers allow multiple internal reflections, thereby increasing analytical sensitivity. Although silicon exhibits absorption in the lower wavenumber region of the spectrum (<1000 $cm^{-1}$) due to multiphonon lattice vibrations, the majority of the fingerprint region remains accessible. Disposable silicon IREs are also available, offering hygienic advantages for clinical analysis of biological fluids [19].

Another rapidly developing technology is polycrystalline diamond (PCD) waveguides. Diamond is one of the most durable optical materials and is well suited for applications requiring mechanical robustness, chemical stability, and biocompatibility. PCD waveguides are typically fabricated by chemical vapor deposition (CVD) of diamond films onto silicon substrates, making them more cost-effective than conventional monolithic single-bounce diamond ATR crystals while retaining robustness and non-toxicity. Additionally, reducing the geometrical dimensions of the waveguide enhances the strength of the evanescent field, improving signal-to-noise ratio (SNR). Once the waveguide thickness approaches the order of the guided wavelength, single-mode behaviour can be achieved. In this configuration, a continuous evanescent field is generated along the propagation path, as opposed to discrete interaction regions characteristic of macroscopic ATR elements [20].

### Concluding remarks

MIR optical fibers and waveguides have evolved into powerful tools for flexible light delivery and molecular sensing, enabling applications across laser technology, industrial process control, medical diagnostics, and environmental monitoring. With improving transmission performance, better coupling efficiency, and growing compatibility with compact QCL, supercontinuum, micro-electro-mechanical systems (MEMS), and upconversion-based systems, MIR fiber technologies are steadily overcoming historical limitations. As these developments converge, they promise increasingly robust, efficient, and versatile solutions for next-generation sensing and laser-delivery systems in the 2-20 μm range.

Mid-infrared ATR waveguides play a critical role in biomedical spectroscopy; however, their performance and applicability strongly depend on material properties. Traditional ATR materials such as ZnSe and GaAs offer high sensitivity but suffer from mechanical fragility, chemical instability, and toxicity, limiting their suitability for *in situ* medical use. Emerging alternatives—including silicon and polycrystalline diamond waveguides—address many of these limitations. Silicon provides a cost-effective, non-toxic platform compatible with disposable formats, while advances in CVD-grown polycrystalline diamond enable robust, biocompatible waveguides capable of enhanced field confinement and improved signal-to-noise performance. Together, these developments are paving the way for more reliable, hygienic, and clinically deployable MIR diagnostic systems.

### Acknowledgements

This study has been supported by the Programmes for Female Researchers, Office for Gender Equality @ Ulm University, Germany.

# 31. Nanophotonic metasurfaces for enhanced mid-IR bio-chemical sensing

**Ivan Sinev[1], Nikita Glebov[1], Berkay Dagli[1] and Hatice Altug[1]**

[1] Institute of Bioengineering, EPFL, Lausanne, Switzerland

E-mail: hatice.altug@epfl.ch

### Status

Surface-enhanced infrared absorption spectroscopy (SEIRA) has emerged as a powerful strategy to overcome the limitations of conventional mid-IR spectroscopy. By employing engineered photonic nanostructures – metasurfaces - that generate strong local electromagnetic fields at sub-wavelength volumes, SEIRA dramatically enhances the interaction between light and molecular vibrations resonantly at selected wavelengths [1]. The near-field amplification that it provides can increase absorption signals by several orders of magnitude, enabling label-free molecular detection at extremely low concentrations from small sample volumes without requiring long acquisition times. In addition, nano-scale confinements and on-chip format of metasurfaces provide a natural path forward for integrated and miniaturized mid-IR spectroscopy. Over the past years, SEIRA has been evolving into a versatile platform to enable numerous applications ranging from biochemical sensing and molecular discrimination [1-4] to tailoring strong light–matter coupling [3,5] and generation of superchiral fields [6]. Currently, the field of metasurfaces is undergoing a paradigm shift from fundamental studies toward technological realization [7]. The central challenge for SEIRA lies in translating its exceptional properties into impactful applications.

### Current and future challenges

SEIRA aim to overcome the fundamental and practical limitations of mid-IR spectroscopy and introduce new functionalities by engineering tailored metasurfaces and incorporating them into novel device architectures. For instance, for accurate molecular identification and structural analysis, there is an inherent trade-off between broad spectral coverage and high spectral resolution. Broadband operation is essential to capture multiple vibrational bands of complex analytes, while high resolution is necessary to resolve closely spaced spectral features and subtle conformational changes. Designing metasurfaces that support these two often mutually exclusive features is a challenging task, which is further aggravated by limited material options and fabrication imperfections. In addition, strong coupling between metasurface resonances and molecular vibrational modes, while beneficial for signal enhancement, can distort native spectral profiles and complicate the extraction of molecular information. Furthermore, most existing metasurfaces are static, exhibiting fixed optical responses that cannot be actively tuned or reconfigured after fabrication. Finally, their realization still depends mostly on low-throughput and costly nanofabrication techniques such as electron-beam lithography, which limits scalable manufacturing and thus translation of SEIRA technologies.

In the context of bioanalytical and medical applications, future efforts in SEIRA field need to involve increasingly advanced lab-on-chip, microarray and optofluidic integration designs for high-throughput screening capabilities and proper handling of delicate biological samples. Water is essential to study living cells and maintain native conformational state of the proteins. But at the same time, strong water absorption drastically attenuates the already weak mid-IR signals. SEIRA is well positioned to overcome this problem by increasing the sensitivity near the sensor surface through the confined fields and reducing the contribution of bulk liquid volume [1,4] (Fig. 1a,b). However, as SEIRA restricts the detection to small surface areas, it requires carefully designed bioassays to bind the analytes and microfluidic systems to ensure efficient sample delivery, reproducibility, and stability of the biological interface.

Detection of molecular conformations and structural differences represents another critical yet highly demanding application area. The ability to reliably detect conformation-specific mid-IR signatures would be invaluable in biomedical diagnostics, such as the early detection of neurodegenerative diseases [8], or in pharmaceutical research, for discrimination of chiral enantiomers [6]. On the other hand, these signals are extremely weak and require not only strong local field enhancement provided by SEIRA but also precise optical mode engineering to support and amplify selective coupling to conformation-specific vibrational modes [9]. Finally, conventional mid-IR systems rely on expensive and bulky instrumentation such as Fourier transform infrared (FTIR) spectrometers and, more recently, quantum cascade laser (QCL) based IR microscopes, which limit their applicability in compact, low-cost sensors and point-of-care diagnostic devices. Novel device schemes that allow miniaturization and system-level integration of mid-IR components and SEIRA devices in a compact footprint are needed for the realization of miniaturized and portable mid-IR sensors that harness the unique strengths of nanophotonics in real-world settings.

**Advances in science and technology to meet challenges**

Metasurfaces with novel designs have been the main driving force behind the progress in SEIRA. Over the recent years, the field has been undergoing remarkable evolution, with new architectures enabling broadband operation, high-Q resonances, active tuning, and compatibility with large-scale manufacturing. Plasmonic-based SEIRA platforms (Fig. 1a) have remained the dominant solution for bioanalytical measurements, particularly in aqueous environments. Novel plasmonic designs now offer broadband response [4,5,10] and significantly enhanced sensitivity, sufficient to resolve even subtlest spectral signatures, such as those associated with different molecular conformations [8,11] (Fig. 1b). These developments have also enabled SEIRA measurements on living cells [12] (Fig. 1c), facilitating real-time monitoring of drug delivery processes, protein translocations, and other dynamic biological phenomena at the molecular level. Two-dimensional materials provide an alternative route to support plasmonic resonances and have been used for tunable mid-IR biosensing [13] and chemical detection [14].

In parallel, novel SEIRA platforms based on high-Q dielectric metasurfaces developed rapidly. Contrary to plasmonics, such structures exhibit low optical losses and narrowband resonances, which makes them particularly well-suited for measurements in dry conditions, such as the detection of thin films and surface contaminants, or tracing gas molecules. To address the downside of limited spectral coverage of high-Q resonators, new design strategies, such as pixelated arrays [2] (Fig. 1d) and gradient metasurfaces [3,15], have been introduced. These approaches, especially when combined with quantum cascade laser-based imaging spectroscopy, enable efficient multiplexed measurements and rapid spectral mapping. Unparalleled design flexibility of all-dielectric metasurfaces also allows advanced engineering of optical modes, opening new avenues for chiral mid-IR spectroscopy, where local superchiral fields promise exceptional sensitivity to molecular handedness.

From the component perspective, a significant breakthrough was the development of quantum cascade lasers (QCLs), which provide narrowband yet extremely bright emission in the mid-IR [16]. Broadband coverage of these sources needed for spectroscopy applications is implemented via external cavity tuning or through the generation of frequency combs that offer unprecedented spectral resolution. The superior power and coherence of QCLs have already drastically reduced measurement times and enabled alternative, spectrometer-free measurement modes[2]. Concurrent efforts in the on-chip integration of QCL [17] and developments in wafer-scale fabrication of SEIRA chips and Mid-IR flat optics [10,18] (Fig. 1e) propel the field toward compact and highly functional sensing platforms.

The complexity of molecular dynamics and biological processes probed by SEIRA also motivates the development of hybrid approaches that combine mid-IR spectroscopy with complementary modalities, e.g. Raman spectroscopy and fluorescence. Such multimodal techniques provide more comprehensive insights into chemical and biological systems [19,20]. Furthermore, the vast datasets generated by mid-IR hyperspectral imaging and multimodal spectroscopy are perfectly suited for machine learning and AI-assisted data analysis. The incorporation of these tools has already improved diagnostic capabilities[8] and promises the possibility of autonomous, data-driven discovery in mid-IR sensing.

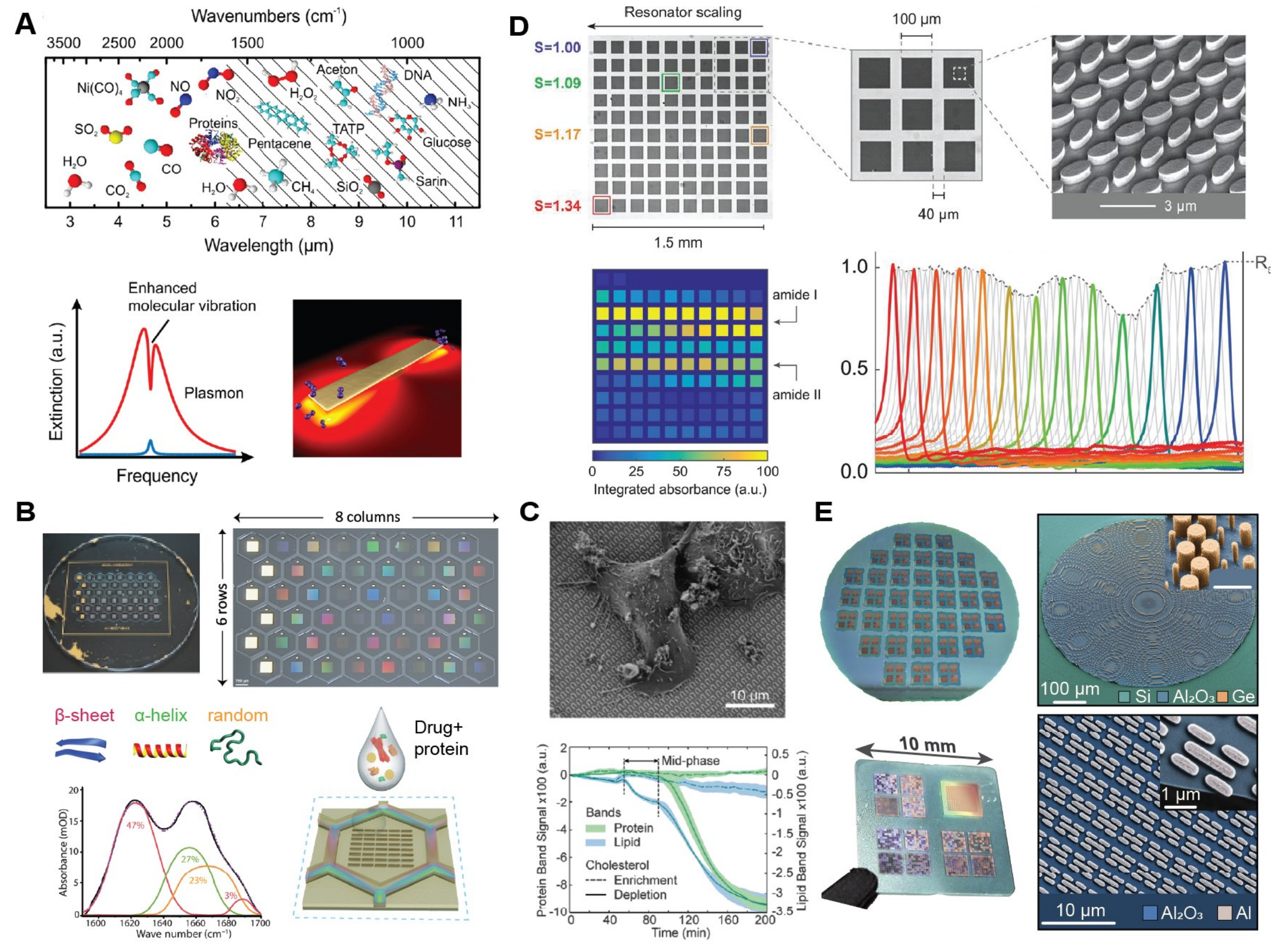

**Figure 1. (a)** Characteristic mid-IR vibrations of molecular species and illustration of the principle of surface-enhanced infrared absorption (SEIRA) with resonant nanoantennas. Adapted with permission from [1]. **(b)** Nanoplasmonic mid-IR microarrays for screening of structural the effect of drugs on the conformation of proteins. Adapted with permission from [8] and [11]. **(c)** Metasurface-based SEIRA for monitoring the effects of chemical stimuli on live cells. Adapted with permission from [12] **(d)** Pixelated dielectric metasurfaces for imaging-based mid-IR molecular barcode sensing using QCL microscopy. Adapted with permission from [2]. **(e)** Wafer-scale fabrication of mid-IR flat optics and CMOS-compatible SEIRA metasurfaces integrated with microfluidics. Adapted with permission from [18].

## Concluding remarks

With the recent advances in both technology and fundamental nanophotonics, metasurface-enhanced mid-IR spectroscopy is well-positioned for translation from laboratory innovation to practical technology. Crucially, the critical dimensions of many mid-IR metasurface designs are already compatible with wafer-scale fabrication, which puts them at the forefront of scalable nanophotonics and enables realistic pathways towards compact integrated devices. The ability to deliver label-free biosensing, chemical and conformational specific detection places SEIRA among the most promising platforms for next-generation diagnostics, environmental monitoring and pharmaceutical quality control. At the same time, the field remains open and rapidly evolving: exploration of nanoscale and quantum phenomena, nanomaterials, machine-learning–assisted analysis, multimodal detection schemes highlight its both fundamental and practical perspectives. Advances in active tuning, high-Q dielectric architectures, and chip-scale mid-IR components further support this momentum. Together, these developments indicate that SEIRA will play a central role in shaping future bioanalytical technologies.

# 32. Closing remarks

**Lukasz Sterczewski[1,*,&], Werner Mäntele[2,3,*,-], Borislav Hinkov[4,*,+], Johannes Kunsch[5,*,#],**

[1]Wrocław University of Science and Technology, Wrocław, Poland
[2] Uni Frankfurt, Frankfurt, Germany
[3] Diamontech, Berlin, Germany
[4] Silicon Austria Labs, Villach, Austria
[5] Laser Components, Olching, Germany

*these authors contributed equally
E-mail: [&]lukasz.sterczewski@pwr.edu.pl, [-]maentele@biophysik.uni-frankfurt.de
[+]borislav.hinkov@silicon-austria.com , [#]j.kunsch@lasercomponents.com

### Status

In the scientific community, it is widely accepted that infrared (IR) spectroscopy has matured from a laboratory-grade sample characterization technique to a promising tool for health analysis and monitoring. Several decades of IR spectrometer development (in various forms) accompanied by progress in suitable light sources and detectors, IR-compatible materials (mainly semiconductors & polymers), nanofabrication routines, and computationally-driven simulation tools and measurement analysis protocols, have equipped us with the necessary tools to drive the next photonic revolution. This becomes evident from the collection of papers in this IR Roadmap. IR spectroscopy must no longer be confined to a laboratory environment.

Unfortunately, this message loses its strength when communicating to laymen – general population and future patients. Infrared radiation is invisible to the naked eye and has been, even in the IR community, historically associated with high equipment cost, usability by specialists, and lack of portability. Fortunately, this notion can be changed when referring to well-known literature. Although this interpretation goes beyond what Antoine de Saint-Exupéry wanted to convey in his celebrated book The Little Prince, the quote "*What is essential is invisible to the eye (L'essentiel est invisible pour les yeux)*" also means that the most valuable information about the analyte may be carried in the IR region rather than the visible part of the electromagnetic spectrum. In other words, the technical difficulties of reaching the IR pay off in specificity and selectivity are both of paramount importance in the context of human health monitoring.

### Current and future challenges

If the role of IR spectroscopy as an indispensable tool for photonics-assisted medicine is secured and set as a priority, it is difficult to say which application will be the most widespread. From today's perspective, one can expect the most mature ones with the strongest reliance on physical principles and approval from authorities realized in the example of non-invasive glucose monitoring. Obviously, much hope is in the protecting health Global Initiative led by Nobel Laureate Prof. Ferenc Krausz to detect noncommunicable diseases at an early stage using blood plasma. As shown in Fig. 1, this, however, can be greatly facilitated if additional effort is made to involve more people in this endeavor, even if indirectly. *Education and popularization*

From the authors' perspective, IR physics and spectroscopy are a dying field at the university level with literally little to no coverage in current academic curricula. Some existing courses focus on gas-phase spectroscopy of simple molecules, but even this is rather scarce. More emphasis on the "academic rejuvenation" of this field, particularly when supported by modern AI-enhanced computational chemistry [1] or spectral databases [2], will yield a new generation of specialists who can greatly contribute to further advancing the field to transfer existing knowledge to novices and into new concepts for real world applications. With the ability to accurately model and predict MIR and far-infrared spectra of molecules from data available in other spectral regions, the practical broadband analysis of samples becomes a reality without the need for a "PhD in molecular spectroscopy".

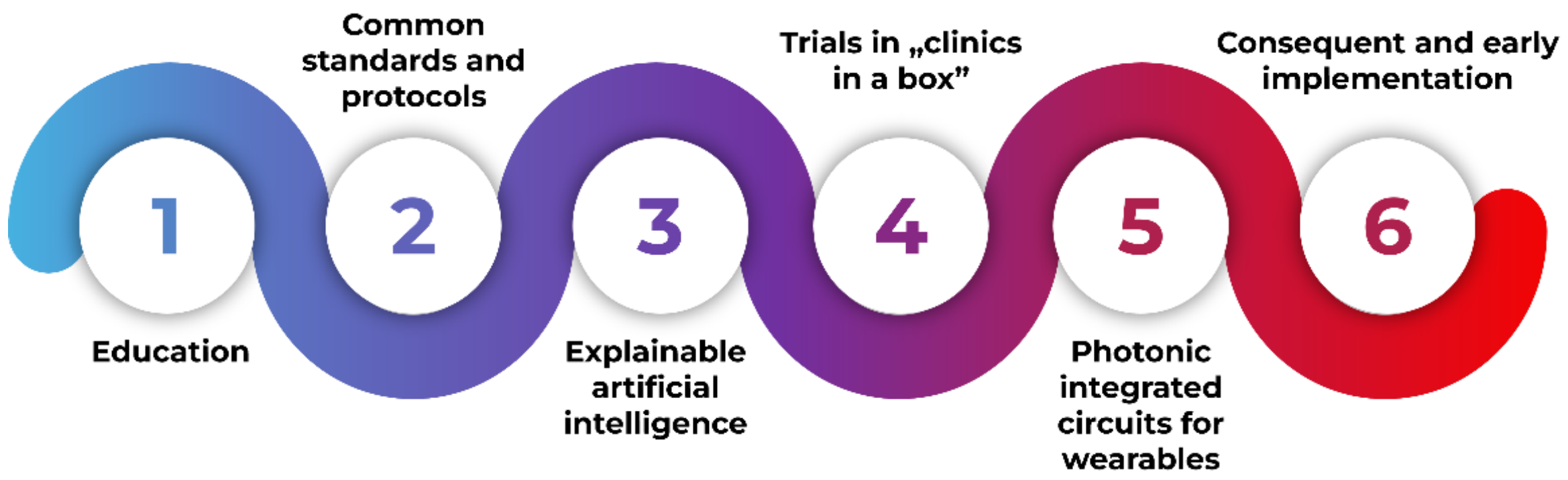

**Figure 1.** Boosters of the infrared photonic revolution for health-related applications.

*Common standards and protocols*

Similar to the open-source software revolution, we are obliged to define protocols and common frameworks for massive collaboration settings and to ensure mutual compatibility between different IR approaches and devices. Otherwise, software tools or spectroscopic instruments will rely too much on a proprietary sample handling routine preventing us from building a reliable computer reference model and eventually a "spectroscopic twin". Currently, standards are typically defined within a laboratory setting but not as global values. This change is obviously also, e.g., accompanied by the need for instrument calibration protocols concerning linearity and measurement reproducibility. While this certainly poses a fundamental metrological challenge, the current approach of mostly isolated laboratories with their own (incompatible) standards will fundamentally prevent further progress into medical use and applications.

In practice, the establishment of permanent application-, data-, and design centers would enable them serving as hubs for advancing metrology in medical infrared IR technologies. Their core functions would include the coordination and maintenance of international reference standards, ensuring consistency and reliability in measurements across borders and beyond individual routines. They would also be responsible for training personnel in best practices of sample handling, measurement techniques, and data interpretation, thereby raising the overall standard of expertise in the field. Furthermore, these centers would manage centralized data repositories, facilitating secure storage, sharing, and analysis of medical measurement data. By offering expert consultation and supporting collaborative research, these centers would foster innovation and help resolve discrepancies in measurement protocols and reference materials. Ultimately, such infrastructure would enhance the precision, repeatability, and global interoperability of medical infrared IR metrology, accelerating progress towards robust and, scalable applications.

*Explainable artificial intelligence*

The immense complexity of IR spectra and the multi-species nature of any medical sample often requires the use of machine learning or artificial intelligence algorithms for deconvolution. However, in medical applications, these face a challenge imposed by government regulation. In the European Union, the primary requirement for such an algorithm is that it has to be explainable rather than constituting a "black box approach". According to the 32024 Regulation (EU) 2024/1689 of the European Parliament and of the Council of 13 June 2024: "High-risk AI systems should be designed and developed in such a way that natural persons can oversee their functioning… High-risk AI systems should perform consistently throughout their lifecycle and meet an appropriate level of accuracy, robustness, and cybersecurity…". Both conditions are still a subject of intensive research and discussion, and obviously strongly dependent on the standardization requirements mentioned above. Without the latter, how can one ensure "consistent performance throughout the algorithm's lifetime"?"

*Trials in remote doctor's office (care stations)*

In response to the problem of "medical deserts" (i.e. regions with hindered access to healthcare) introduced in Chapter 1, and the growing costs of healthcare in general, an interesting solution has emerged from the OnMed company [4] also referred to as a "clinic-in-a-box". A soundproof booth with a large display remotely connects to a physician to perform a routine medical examination using established instruments like a pulse oximeter, blood pressure monitor, stethoscope, and a high-resolution video camera. Although all are sufficient to provide basic examination of vitals, they lack the capabilities that IR spectroscopy instruments can offer. For instance, exhaled air or urine analysis can potentially be used to detect diabetes or some metabolic conditions, all without any samples to be stored and sent for remote laboratory testing. The cost efficiency of these instruments can be greatly improved when they use largely-scalable technological concepts such as photonic integrated circuits (PICs) rather than bulky free-space optical setups.

*Photonic integrated circuits (PICs) for portable devices*

Much like highly-integrated semiconductor chips in electronics ("ICs") have led us to the smartphone revolution in terms of access to information, one can expect that IR PICs will soon enable real-time health monitoring using smart and cheap wearables. While their performance will not always be on the same level as mid- or high-end instruments deployed in point-of-care diagnostics locations, their low costs resulting from large technological scaling-effects will allow them to be used, e.g., as individual, private pre-screening devices. This is likely to open up new opportunities, as currently, the economic aspect is a significant limiting factor for existing solutions. To prepare the ground for future mass-scale use of PIC-based medical devices, it is imperative to define specific interfaces for their interaction with the human body. Also the legal and social aspects: what devices are needed for what purpose, how they will be used, and how sensor data will be handled and stored securely are of increasing importance.

*Consequent and early implementation*

Infrared omics (integrated spectroscopic signature of a system as a whole as defined in Chapter 16) is data-driven, and the benefit will strongly depend on broad and frequent usage. Let us imagine that this technology had already been implemented prior to the COVID pandemic, and personal data had been regularly taken before. The strength of infrared omics would not have been so much to detect the acute disease but to monitor and better understand post COVID phenomena by looking at existing co-indications. In addition, it would enable even retro integration of new scientific knowledge, demystifying vaccination effects, and develop a dynamic and personalized health concept.

## Concluding remarks

A well-known response to any crisis is the famous call to urgency, "The time to act is now" (before it is too late). If the price one pays for procrastination is human health, there are no excuses any more. The authors believe in defining common goals for the whole IR community, all unified by common standards and protocols to collectively achieve this via basic research and engineering efforts. It is critical to provide the first successful medical use case of IR spectroscopy on a mass scale, which cannot be easily replaced by competing technologies. Once this is fulfilled, the adoption of IR technologies for health monitoring and management will come in naturally. Our time starts now.